\documentclass[%
 reprint,
preprintnumbers,
nofootinbib,
 amsmath,amssymb,
 aps,
]{revtex4-1}

\usepackage{graphicx}
\usepackage{dcolumn}
\usepackage{bm}
\usepackage{amsmath,amssymb,amsfonts}
\usepackage{siunitx}
\usepackage{color}
\usepackage{booktabs}
\usepackage{mathtools}
\usepackage{multirow}
\usepackage{placeins}
\usepackage{bm}
\usepackage{xspace}
\usepackage{comment}
\usepackage{hyperref}
\usepackage{ulem}

\newcommand{\mnras}{Mon.~Not.~Roy.~Astron.~Soc.}
\newcommand{\physrep}{Phys.~Rep.}
\newcommand{\jcap}{JCAP}
\newcommand{\pasj}{Publ.~Astron.~Soc.~Japan}

\newcommand{\aap}{Astronomy~\&~Astrophysics}
\newcommand{\aj}{The~Astronomical~J.}

\usepackage[T1]{fontenc} 

\newcommand{\mtrv}[1]{{\textcolor{black}{#1}}}
\newcommand{\ssrv}[1]{{\textcolor{black}{#1}}}

\newcommand{\diff}{{\rm d}}
\newcommand{\diffrac}[2]{\frac{\diff #1}{\diff #2}}

\newcommand{\phaseint}[1]{\int\frac{\diff\vec{#1}}{{(2\pi)^3}}}
\newcommand{\LCDM}{$\Lambda$CDM\xspace}
\newcommand{\Sigmacrit}{\Sigma_{\rm crit}}

\newcommand{\Omegam}{\Omega_{\rm m}}
\newcommand{\dSigma}{\Delta\!\Sigma}
\newcommand{\wproj}{{w_{\rm p}}}

\newcommand{\PL}{P_{\rm L}}

\newcommand{\hiMpc}{h^{-1}{\rm Mpc}}
\newcommand{\hiGpc}{h^{-1}{\rm Gpc}}
\newcommand{\hMpci}{h{\rm Mpc}^{-1}}
\newcommand{\hiMsun}{h^{-1}M_\odot}
\newcommand{\bk}{{\bf k}}
\newcommand{\br}{{\bf r}}
\renewcommand{\eqref}[1]{Eq.~\ref{#1}}
\newcommand{\apdxref}[1]{Appendix~\ref{#1}}
\newcommand{\tableref}[1]{Table~\ref{#1}}
\newcommand{\sectref}[1]{Sec.~\ref{#1}}

\newcommand{\figref}[1]{Fig.~\ref{#1}}
\newcommand{\setup}[1]{``#1''\xspace}
\newcommand{\mock}[1]{{\tt #1}\xspace}
\newcommand{\code}[1]{{\sf #1}\xspace}
\newcommand{\survey}[1]{{#1}\xspace}
\newcommand{\avrg}[1]{\left\langle#1\right\rangle}

\begin{document}

\preprint{IPMU20-0063,YITP-20-80}

\title{Validating a minimal galaxy bias method for cosmological parameter inference\\
using HSC-SDSS mock catalogs}

\author{Sunao~Sugiyama${}^{1,2}$}
\email{sunao.sugiyama@ipmu.jp}
\author{Masahiro~Takada${}^{2}$}
\email{masahiro.takada@ipmu.jp}
\author{Yosuke~Kobayashi${}^{1,2}$}
\author{Hironao~Miyatake${}^{3,4}$}
\author{Masato~Shirasaki${}^{5}$}
\author{Takahiro~Nishimichi${}^{6,2}$}
\author{Youngsoo~Park${}^{2}$}


\affiliation{
    ${}^{1}$Department of Physics, The University of Tokyo, Bunkyo, Tokyo 113-0031, Japan}
\affiliation{
    ${}^{2}$Kavli Institute for the Physics and Mathematics of the Universe (WPI),The University of Tokyo Institutes for Advanced Study (UTIAS), The University of Tokyo, Chiba 277-8583, Japan}
\affiliation{
    ${}^{3}$Institute for Advanced Research, Nagoya University, Nagoya 464-8601, Japan}
\affiliation{
    ${}^{4}$Division of Physics and Astrophysical Science, Graduate School of Science, Nagoya University, Nagoya 464-8602, Japan}
\affiliation{
    ${}^{5}$Division of Science, National Astronomical Observatory of Japan, Mitaka, Tokyo 181-8588, Japan}
\affiliation{
    ${}^{6}$Center for Gravitational Physics, Yukawa Institute for Theoretical Physics, Kyoto University, Kyoto 606-8502, Japan
}


\date{\today}

\begin{abstract}
	We assess the performance of a perturbation theory  inspired method for inferring cosmological parameters from the joint measurements of galaxy-galaxy weak lensing ($\Delta\Sigma$) and the projected galaxy clustering ($w_{\rm p}$). 
	To do this, we use a wide variety of mock galaxy catalogs constructed based on a large set of $N$-body simulations that mimic the Subaru HSC-Y1 and SDSS galaxies, and apply the method to the mock signals to address whether to recover the underlying true cosmological parameters in the mocks.
	We find that, as long as the appropriate scale cuts, $12$ and $8~h^{-1}{\rm Mpc}$ for $\Delta\Sigma$ and $w_{\rm p}$ respectively, are adopted, a ``minimal-bias'' model using the linear bias parameter $b_1$ alone and the nonlinear matter power spectrum can recover the true cosmological parameters (here focused on $\Omega_{\rm m}$ and $\sigma_8$) to within the 68\% credible interval, for all the mocks we study including one in which an assembly bias effect is implemented. 
	This is as expected if physical processes inherent in galaxy formation/evolution are confined to local, small scales below the scale cut, and thus implies that real-space observables have an advantage in filtering out the impact of small-scale nonlinear effects in parameter estimation, compared to their Fourier-space counterparts. 
	In addition, we find that a theoretical template including the higher-order bias contributions such as nonlinear bias parameter $(b_2)$ does not improve the cosmological constraints, but rather leads to a larger parameter bias compared to the baseline $b_1$-method. Another non-trivial finding is that the cosmological parameters are not necessarily recovered, even when the model prediction is used as the input mock signals, as a consequence of marginalization or projection of  asymmetric posterior distributions in a multidimensional parameter space, such as the case of the ``banana''-shaped distribution in the $(\Omega_{\rm m}, \sigma_8)$ plane.
	We also study the performance of alternative observables, $\Upsilon$ or $Y$ statistic, where the {\it same} scale cut for both the weak lensing and the galaxy clustering can be employed thanks to their same sensitivity to the Fourier modes, but do not find a promising advantage of these statistics over the fiducial observables, $\{\Delta\Sigma,w_{\rm p}\}$.
\end{abstract}

\maketitle


\section{Introduction}
\label{sect:intro}

Wide-area imaging and spectroscopic galaxy surveys provide us with powerful tools for constraining the energy composition of the universe, the growth of cosmic structure formation over cosmic time, and properties of the primordial perturbations \citep{2013PhR...530...87W}. In particular, when combined with high-precision measurements of cosmic microwave background \citep{wmap5,planck-collaboration:2015fj}, galaxy surveys allow us to explore the origin of the late-time cosmic acceleration such as properties of dark energy or a possible breakdown of General Relativity on cosmological scales. There are many ongoing and upcoming galaxy surveys aimed at advancing our understanding of these fundamental questions; e.g.,  the SDSS-III Baryon Acoustic Oscillation Spectroscopic Survey (BOSS) \citep{dawsonBaryonOscillationSpectroscopic2013}, the Subaru Hyper Suprime-Cam (HSC) survey \cite{aiharaHyperSuprimeCamSSP2018}, the Dark Energy Survey (DES)\footnote{https://www.darkenergysurvey.org}, the Kilo-Degree Survey (KiDS)\footnote{http://kids.strw.leidenuniv.nl}, the Subaru Prime Focus Spectrograph (PFS) survey \citep{takadaExtragalacticScienceCosmology2014}, the Dark Energy Spectrograph Instrument (DESI) survey\footnote{https://www.desi.lbl.gov}, and then ultimately the Rubin Observatory Legacy Survey of Space and Time (LSST)\footnote{https://www.lsst.org}, the ESA Euclid\footnote{https://sci.esa.int/web/euclid}, and the NASA Roman Space Telescope\footnote{https://roman.gsfc.nasa.gov}.

A main challenge in the use of galaxy surveys for precision cosmology lies in uncertainties in the relation between matter and galaxy distributions in large-scale structure, i.e. the so-called galaxy bias uncertainties \citep{kaiser84} \citep[also see][for a thorough review]{Desjacques18}. Since physical processes inherent in galaxy formation and evolution are still very challenging to accurately model from first principles, we need to empirically model the galaxy bias and/or observationally lift the uncertainties. One promising way, as an observational approach, is a joint probes cosmological analysis, done by combining galaxy-galaxy weak lensing and galaxy clustering \citep{2005PhRvD..71d3511S,2009MNRAS.394..929C,mandelbaumCosmologicalParameterConstraints2013,cacciatoCombiningGalaxyClustering2012,hikage:2013kx,moreWeakLensingSignal2015,krauseDarkEnergySurvey2017,2018PhRvD..98d3526A,2018MNRAS.476.4662V,wibkingCosmologyGalaxygalaxyLensing2020,2020JCAP...03..044N}. 
The two-point correlation function, $\xi_{\rm gg}$, is the standard tool to characterize the clustering properties of galaxy distribution in large-scale structure \citep{peacock01,2004ApJ...606..702T}. For a cold dark matter dominated universe with adiabatic initial conditions, the galaxy correlation function is related to the two-point correlation function of the underlying matter distribution, $\xi_{\rm mm}$, at large scales via a linear bias parameter as $\xi_{\rm gg}\simeq b_1^2\xi_{\rm mm}$, where $b_1$ is a {\it constant} whose value varies with types of galaxies \citep{dalal08}. Cross-correlating the positions of galaxies with shapes of background galaxies as a function of their separations on the sky enables one to probe the average matter distribution around the foreground (lensing) galaxies -- the so-called galaxy-galaxy weak lensing \citep{mandelbaumSystematicErrorsWeak2005,mandelbaum06}. The galaxy-galaxy weak lensing arises from the galaxy-matter cross-correlation, $\xi_{\rm gm}$, that is given, at large scales, as $\xi_{\rm gm}\simeq b_1\xi_{\rm mm}$. Hence, combining the two clustering observables allows one to observationally lift the galaxy bias uncertainty for the foreground galaxy sample, at least on large scales. 

On the theory side, there is still a difficulty in attaining the full potential of galaxy surveys; we need sufficiently accurate theoretical templates for extracting cosmological information from the clustering observables. There are two competing requirements on an analysis method -- ``robustness'' and ``precision'' \citep[also see][for a study based on similar motivation]{nishimichiBlindedChallengePrecision2020}. For robustness, we want to minimize any possible bias or shift in the estimated value of cosmological parameter(s) from the true value(s). On the other hand, for precision,  we want to obtain as small credible interval (error bars) in cosmological parameters as possible from the given observables. Obviously it is not straightforward to achieve these two requirements simultaneously. 
For example, since the galaxy clustering observables have a higher signal-to-noise ratio at smaller scales, which are affected by nonlinear structure formation and galaxy physics, achieving a higher precision (smaller error bars) in cosmological parameters requires one to use the clustering observables down to smaller scales in the nonlinear regime. If a theoretical model is not accurate enough on these small scales, it can easily lead to a large bias in the estimated parameters. The worst case scenario is that one could claim a wrong cosmology that is away from the underlying true cosmology, at an apparent high significance for given datasets. 

Hence the purpose of this paper is to assess the performance of a joint probes analysis for extracting cosmological information. For a theoretical template we use a method motivated by the cosmological perturbation theory (PT) of structure formation \citep{bernardeau02}, where we introduce a set of bias parameters  and then allow the bias parameters to freely vary in cosmological parameter estimation \citep[also see][for a similar study]{maccrannY1ResultsValidating2018}. 
However, we note in advance that our baseline model is a ``minimal-bias'' model, where we include the linear bias parameter alone and  use the fully nonlinear matter power spectrum to model $\xi_{\rm gg}$ and $\xi_{\rm gm}$, following the method used in the DES cosmology analysis 
\citep{2018PhRvD..98d3526A}.
For the hypothetical clustering observables, we consider the galaxy-galaxy weak lensing, $\dSigma(R)$, and the projected galaxy clustering correlation function, $\wproj(R)$, mimicking those measured from the Subaru HSC-Y1 data \cite{2018PASJ...70S...8A,hikageCosmologyCosmicShear2019,hamanaCosmologicalConstraintsCosmic2019a} and the DR11 SDSS BOSS data \citep{dawsonBaryonOscillationSpectroscopic2013,2015ApJ...806....1M}. 
For this purpose, we generate mock catalogs of the HSC and SDSS galaxies using a suite of high-resolution $N$-body simulations in Refs.~\cite{nishimichiDarkQuestFast2018,kobayashiCosmologicalInformationContent2019}. Then we apply the 
method 
to the mock signals of $\dSigma$ and $\wproj$ to perform cosmological parameter estimation properly taking into account the error covariance matrices. We quantify the performance of the method in terms of ``robustness'' and ``precision''; represented by the size of parameter bias (i.e. the difference between the estimated parameter and the true value) compared to the 68\% credible interval, and the size of the 68\% credible interval, respectively. 
To do this, we pay particular attention to two questions. First, we estimate a proper ``scale'' cut in the clustering observables. Since the linear bias or PT-based method is valid only on large scales, we need to define in terms of scale cuts the minimum scale above which we can safely use the clustering observables for parameter inference. An advantage of the bias-expansion based method is that, as long as we treat the bias parameter(s) as a free parameter(s), it is expected to accurately model the clustering observables on large scales without detailed modeling of galaxy physics, irrespective of any specific galaxy sample. 
To test this advantage, we use a variety of mock catalogs including an assembly bias mock where the dependence of galaxy bias on inner structures of host halos, in addition to halo mass, is included. 
We then assess whether the method can successfully recover the cosmological parameters from each of the different mock catalogs. In this work we do not ask whether the bias parameter is recovered by the method, and we focus on the cosmological parameters ($\Omegam$ and $\sigma_8$). Thus our work gives a validation of the minimal-bias or PT-based method that we are planning to apply to the real HSC-Y1 and SDSS data in our forthcoming paper. 
This work also stands in contrast to our companion work (Miyatake et al. in prep.), where a more specific model of galaxy bias based on the halo model is employed in cosmological parameter estimation using exactly the same set of mock galaxy catalogs \citep[also see][]{vandenboschCosmologicalConstraintsCombination2013,wibkingCosmologyGalaxygalaxyLensing2019}.

This paper is organized as follows. In Sec.~\ref{sect:observables}, we first define the clustering observables, $\dSigma$ and $\wproj$, and then describe the PT-based method to model the clustering observables for a given cosmological model. In Sec.~\ref{sect:GalaxyMock}, we describe the details of $N$-body simulations, the mock galaxy catalogs of HSC and SDSS galaxies, the mock clustering signals, and also the error covariance matrices. 
\ssrv{In Sec.~\ref{sect:methods}, we describe the strategies of our validation,} i.e. \mtrv{validations for the analysis setups and for the minimal galaxy bias 
method.} 
In Sec.~\ref{sect:result}, we show the main results for cosmological parameter estimations, based on the Bayesian inference method, and especially discuss the performance and validation against different analysis setups and different mock galaxy catalogs. Finally we conclude in \sectref{sect:conclusion}. In appendix we also give discussion on the different methods and supplementary discussion.
Unless explicitly stated, we assume a flat-geometry $\Lambda$CDM cosmology that is consistent with the {\it Planck} CMB data \citep[hereafter {\it Planck} cosmology][]{planck-collaboration:2015fj}. 

\section{Observables and Models}\label{sect:observables}

\subsection{Definitions of clustering observables: $\dSigma$ and $\wproj$}
\label{sec:observables}

In this section we define the two clustering observables, galaxy-galaxy weak lensing and galaxy auto-clustering, which we use in this paper.

Galaxy-galaxy weak lensing can be measured by cross-correlating positions of foreground galaxies with shapes of background galaxies \citep{mandelbaum06,2015ApJ...806....1M}. It probes the excess surface mass density profile, $\Delta\!\Sigma$, around the lensing galaxies that is given in terms of the surface mass density profile $\Sigma$ as
\begin{align}
	\dSigma(R; \mtrv{z_{\rm l}}) &=\bar{\Sigma}(R)-\Sigma(R) \nonumber\\
  &= \left.\Sigmacrit(z_{\rm l},z_{\rm s})\gamma_{\rm t}(R;z_{\rm s})\right|_{R=\chi_{\rm l}\Delta\theta}, \label{eq:dSigma-def}
\end{align} 
where $\gamma_{\rm t}$ is the averaged tangential shear of background galaxies in the annulus of projected centric radius $R$ from the foreground galaxies, and $\chi_{\rm l}$ is the comoving angular diameter distance to each foreground galaxy. Note that the average of background galaxy shapes needs to be done for all the \mtrv{pairs of lens and source} galaxies in the same projected separation $R$, not the angular separation $\Delta\theta$ (see below), \mtrv{even if the foreground galaxies have a redshift distribution.} 
$\Sigma_{\rm crit}$ is the lensing critical surface density defined for an observer-lens-source system as
\begin{align}
	\Sigma_{\rm crit}(z_{\rm l},z_{\rm s}) = \frac{c^2 \chi_{\rm s}(z_{\rm s})}{4\pi G \chi_{\rm ls}(z_{\rm l},z_{\rm s})\chi_{\rm l}(z_{\rm l})
	(1+z_{\rm l})},
\end{align}
where $\chi_{\rm s}$
and $\chi_{\rm ls}$ are the comoving angular diameter distances to the source galaxy, 
and between the lens and the source, respectively. 
The use of the comoving angular diameter distance and the factor of $1/(1+z_{\rm l})$ is due to our use of the comoving coordinates. Here we consider a single lens redshift and a single source redshift for simplicity, but it is straightforward to take into account the redshift distributions of foreground (lensing) and source (lensed) galaxies, e.g. by following the method in Ref.~\cite{2015ApJ...806....1M}. 
Eq.~(\ref{eq:dSigma-def}) indicates that an estimation of $\dSigma$ depends on the assumed or ``reference'' cosmology, which is needed to estimate $\Sigma_{\rm crit}$ and $R$ via $R=\chi_{\rm l}\Delta\theta$ for each foreground-background galaxy pair. However, the reference cosmology generally differs from the underlying true cosmology. Hence, in parameter estimation, we need to properly take into account the cosmological dependences of the estimated $\dSigma$ \citep{moreCosmologicalDependenceMeasurements2013}.
\mtrv{Eq.~(\ref{eq:dSigma-def}) also means that, as long as accurate photometric redshifts of source galaxies are available (as we consider spectroscopic galaxies
for lensing galaxies), the lensing profile $\dSigma(R)$ depends only on the matter distribution at lens redshift or equivalently does not depend on 
source redshift. 
Throughout this paper we do not consider the source redshift dependence in the lensing signal, but do include the effect in the covariance matrix as we will describe below. On the other hand, if we use $\gamma_t$ as an observable, we need to properly take into account both the 
redshift dependences of $z_{\rm l}$ and $z_{\rm s}$.
Hereafter we  omit $z_{\rm l}$ in the argument of the lensing profile $\dSigma(R;z_{\rm l})$ for notational simplicity.}

The surface mass density profile $\Sigma(R)$ is given by a line-of-sight projection of the three-dimensional galaxy-matter cross-correlation function,
$\xi_{\rm gm}(r)$, as
\begin{align}
	\Sigma(R) = \Omega_{\rm m}\rho_{\rm cr0}
	\int_{-\infty}^{\infty}\!\diff\Pi~ \left[1+\xi_{\rm gm}\left(\sqrt{R^2+\Pi^2}\right)\right],
\end{align}
where $\rho_{\rm cr0}$ is the critical mass density today, defined as $\rho_{\rm cr0}=3H_0^2/8\pi G$. The first term on the r.h.s. of Eq.~(\ref{eq:dSigma-def}), $\bar{\Sigma}(R)$, is the averaged surface mass density defined within a circular aperture of radius $R$ with respect to the lensing galaxies at the center: $\bar{\Sigma}(R) = (2/R^2) \int_0^R\diff R' R'~ \Sigma(R')$.
Here we ignore redshift evolution of the matter-galaxy cross-correlation within a redshift slice of lensing galaxies for simplicity. Using the cross-power spectrum of matter and galaxy, denoted as $P_{\rm gm}(k)$, the excess surface density profile is expressed under the Limber approximation \cite{1953ApJ...117..134L} as 
\begin{align}
	\Delta\!\Sigma(R)=\bar{\rho}_{\rm m0}\int_0^\infty\!\frac{k\mathrm{d}k}{2\pi}~ P_{\rm gm}(k)J_2(kR),\label{eq:dsigma_fourier2}
\end{align}
where $J_2(x)$ is the 2nd-order Bessel function. Note $\xi_{\rm gm}(r)=\int_0^\infty\!k^2\mathrm{d}k/(2\pi^2)~P_{\rm gm}(k)j_0(kr)$, where
$j_0(x)$ is the zeroth order spherical Bessel function. 

Another clustering observable we use is the projected auto-correlation function for spectroscopic galaxy sample that is taken as foreground galaxies for the weak lensing analysis. The projected correlation function is defined by a line-of-sight projection of the three-dimensional auto-correlation function, $\xi_{\rm gg}(r)$, as
\begin{align}
	\wproj(R;z) = 2\int_{0}^{\pi_{\rm max}}\!\diff\Pi~ \xi_{\rm gg}\left(\sqrt{R^2 + \Pi^2};z\right),\label{eq:wp-def}
\end{align}
where $\pi_{\rm max}$ is the length of the line-of-sight projection. Note that the projected correlation function has 
a dimension of $[h^{-1}{\rm Mpc}]$. \mtrv{In our setting $z=z_{\rm l}$.}
Throughout this paper, unless explicitly stated, we use $\pi_{\rm max}=100~\hiMpc$ as our default choice. Also note that $\xi_{\rm gg}(r)$ is given in terms of the auto-power spectrum of the galaxy number density field, $P_{\rm gg}$, as $\xi_{\rm gg}(r)=\int_0^\infty\!\!k^2\mathrm{d}k/(2\pi^2)~P_{\rm gg}(k)j_0(kr)$. The projected correlation function is not sensitive to the redshift-space distortion (RSD) effect due to peculiar velocities of galaxies, if a sufficiently large projection length ($\pi_{\rm max}$) is taken \citep[see Fig.~6 in Ref.][]{vandenboschCosmologicalConstraintsCombination2013}. The RSD effect itself is a useful cosmological probe, but its use requires an accurate modeling \citep{kobayashiCosmologicalInformationContent2019,2020arXiv200506122K}, which is not straightforward. Hence, the projected correlation function makes it somewhat easier to compare with theory in a cosmological analysis. In the following, we ignore the RSD effect in most cases of our cosmology challenges, but will separately study the impact of the RSD effect on the parameter estimation. 
For convenience of our discussion, let us also consider the case of an infinite projection length, $\pi_{\rm max}=\infty$. In this case, the projected correlation function is rewritten under the Limber approximation as
\begin{align}
	\wproj(R) = \int_0^\infty\!\frac{k\mathrm{d}k}{2\pi}~ P_{\rm gg}(k) J_0(kR), \label{eq:wp_fourier2}
\end{align}
where $J_0(x)$ is the zero-th order Bessel function. We again ignore a possible redshift evolution of the auto galaxy power spectrum within a given redshift slice for simplicity. 

Comparing Eqs.~(\ref{eq:dsigma_fourier2}) and (\ref{eq:wp_fourier2}) manifests that $\dSigma(R)$ and $\wproj(R)$ at a particular projected radius $R$ arise from different Fourier modes due to the different kernels, $J_2$ and $J_0$, in the Fourier-space integral. 
Here $\dSigma$ at a particular radius is more sensitive to the larger $k$ modes than $\wproj$ is, reflecting the fact that $\dSigma$ is a non-local quantity arising from the tidal field due to the mass distribution around lensing galaxies; e.g., even a point mass lens causes an extended profile of $\dSigma$ around the lens on the sky.
On the other hand, $\wproj$ is a local quantity, and satisfies the integral constraint $\int_0^\infty\!R\mathrm{d}R~\wproj(R)=0$ (under the flat-sky approximation). 
In order to use the perturbation theory for parameter inference, we can use only the large-scale information in $\dSigma$ and $\wproj$, i.e. at scales greater than a certain scale cut. However, due to the different Fourier kernels, we need to employ different scale cuts in the two observables. This is not an obvious question, and is one of main scopes that we address in this paper.

Regarding the different dependencies of $\dSigma$ and $\wproj$ on Fourier modes, several works proposed alternative statistical quantities,  reconstructed from the observables, in that they turn out to be sensitive to the same Fourier modes and one can impose the same scale cut 
for both the quantities. Such examples are the Annular Differential Surface Density (ADSD) statistic ($\Upsilon$) \cite{baldaufAlgorithmDirectReconstruction2010,mandelbaumCosmologicalParameterConstraints2013}, and the reconstructed surface mass density profile ($Y$) \cite{parkLocalizingTransformationsGalaxyGalaxy2020}
%
\begin{align}
	&\left\{ \Upsilon_{\rm gm}(R| R_0),\Upsilon_{\rm gg}(R| R_0)\right\},\\
	&\left\{ Y\!(R),\wproj(R)\right\}.\label{eq:ADSD}
\end{align}
where $R_0$ is the scale that a user needs to introduce in order to filter out the small-scale information in the observable $\Upsilon$. $\Upsilon_{\rm gm}$ and $\Upsilon_{\rm gg}$ are the ADSD profiles for the galaxy-matter cross-correlation and the galaxy auto-correlation, which are constructed from $\dSigma$ and $\wproj$, respectively. Here $Y$ is a local quantity reconstructed from $\dSigma$, as in $\Sigma(R)$, and is designed to be sensitive to the same Fourier modes as $\wproj$, for a particular $R$.
In Appendix~\ref{apdx:sigma-upsilon}, we describe in detail the definitions and properties of these observables, and will study the performance of these observables for cosmological parameter estimations, in comparison with our default observables $\{\dSigma(R),\wproj(R)\}$.

\subsection{Standard Perturbation Theory}
\label{sec:pt}

\begin{figure}[h]
	\begin{center}
		\includegraphics[clip,width=0.95\columnwidth]{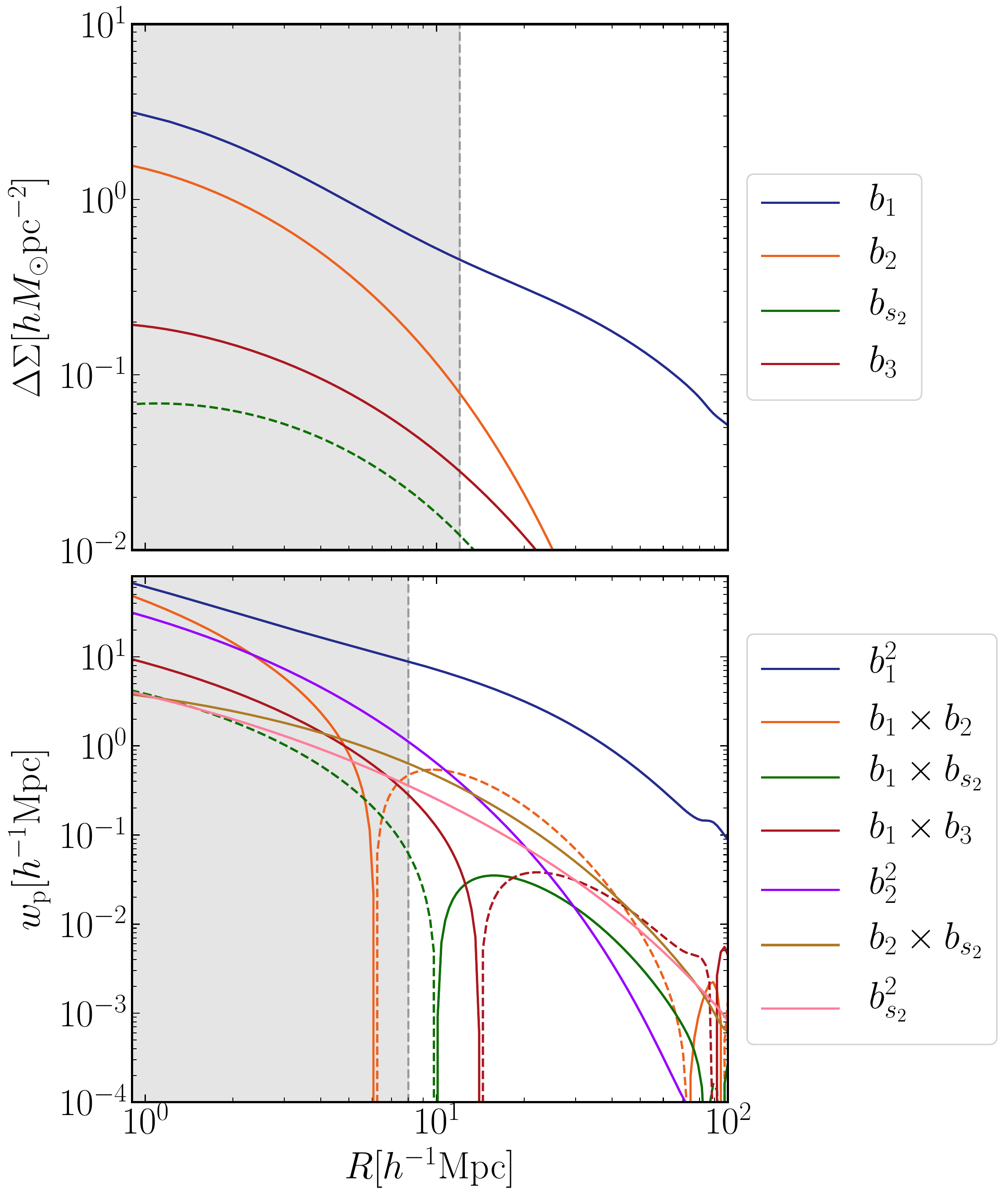}
		\caption{
			The different lines show how each term involving the different combinations of bias parameters, $b_1, b_2, b_3$ and $b_{s_2}$, in $P_{\rm gm}(k)$ and $P_{\rm gg}(k)$ (Eqs.~\ref{eq:Pgm} and \ref{eq:Pgg}) affects the galaxy-galaxy weak lensing $\dSigma(R)$ and the projected auto-correlation function $\wproj(R)$. Here, for illustration purposes, we set all the bias parameters to unity: $b_1=b_2=\cdots=1$. 
			The solid and dashed parts of each line show positive and negative parts of the corresponding term. For our default analysis of the parameter estimation, we use only the large-scale information in $\dSigma$ and $\wproj$ over the range of $R$ above the scale cuts, as denoted by the unshaded regions. We use the scale cuts  of $R_{\rm cut}=12$ and $8~\hiMpc$ for $\dSigma$ and $\wproj$, respectively, for the default choice. 
		}
		\label{fig:pt-bias-terms}
	\end{center}
\end{figure}

To compare measurements against theory, we need to model the observables in terms of cosmological parameters within a flat-geometry, $\Lambda$CDM model with adiabatic initial conditions. As can be found from Eqs.~(\ref{eq:dsigma_fourier2}) and (\ref{eq:wp-def}), once the power spectra $P_{\rm gm}$ and $P_{\rm gg}$ are provided for a given cosmological model, we can compute $\dSigma$ and $\wproj$. We adopt the following PT approach for large-scale structure formation and galaxy bias expansion \citep{fry:1993lr,bernardeauLargeScaleStructureUniverse2002,Desjacques18} to model the power spectra. 

The number density contrast field of galaxies, $\delta_{\rm g}({\bf x})$, is formally expressed in terms of the underlying matter field, $\delta_{\rm m}({\bf x})$, as
\begin{align}
	\delta_{\rm g} = \tilde{b}_1\delta_{m} + \frac{1}{2}\tilde{b}_2 \delta_m^2+\cdots.
\end{align}
The coefficients appearing in this expansion, $\tilde{b}_{i}~(i=1,2,\cdots)$, are bare bias parameters. By using the Eulerian cosmological perturbation theory, we can express the matter-galaxy cross-power spectrum and the galaxy auto-power spectrum in a series of the underlying matter power spectra \citep{mcdonaldClusteringDarkMatter2006,mcdonaldClusteringDarkMatter2009}:
\begin{align}
	P_{\rm gm}(k) &= b_1P_{\rm NL}(k) + b_2 P_{b_2}(k) + b_{s_2} P_{b_{s_2}}(k) + b_3P_{b_3}(k)\label{eq:Pgm}\\
	P_{\rm gg}(k) &= b_1^2P_{\rm NL}(k) + 2b_1\left[b_2P_{b_2}(k) + b_{s_2}P_{b_{s_2}}(k)+b_3P_{b_3}(k)\right]\nonumber\\
	&\hspace{2em}+ b_2^2 P_{b_2^2}(k) + b_{s_2}^2P_{b_{s_2}^2}(k) + 2b_2b_{s_2}P_{b_2b_{s_2}}(k)\label{eq:Pgg}
\end{align}
where $P_{\rm NL}$ is the nonlinear matter power spectrum, and the functions, $P_{b_2}$, $P_{b_3}$, $P_{b_{s_2}}$, $P_{b_2^2}$, $P_{b_{s_2}^2}$, and $P_{b_2b_{s_2}}$, are given in \apdxref{sect:power-spectra}. 
The bias parameters, $b_{X}~(X=1,2,s_2,3)$, are the so-called renormalized bias parameters and are different from the bare bias parameters. For $P_{\rm NL}$, we use the fitting formula developed in Ref.~\citep{takahashiRevisingHalofitModel2012}. The PT functions are given as a function of the linear power spectrum, for input cosmological parameters and redshift.
Strictly speaking, the expansion above is not self-consistent, because we use the fully nonlinear matter power spectrum that formally includes all the higher-order contributions including the non-perturbative effects after the shell-crossing (galaxy/halo formation), and is not at the same order as the other terms in the above equations\footnote{For comprehensiveness we will also consider the self-consistent PT-based method using the PT model including the one-loop corrections for the matter power spectrum, $P_{\rm SPT}$,
instead of the full nonlinear model, $P_{\rm NL}$.}.
All other terms are at the next-to-leading order (one-loop corrections) in the PT expansion.
We Fourier-transform back the above power spectra to obtain the configuration-space correlation functions, $\xi_{\rm gm}(r)$ and $\xi_{\rm gg}(r)$. However, the one-loop correction terms decay with the wavenumber $k$ rather slowly, making the inverse Fourier transformation unstable.
Therefore we put an artificial cutoff in the Fourier space, by multiplying a Gaussian function $\exp(-k^2/k_{\rm off}^2)$, in the Fourier integral. We employ $k_{\rm off}=10~\hMpci$ for our default choice. 
The PT modeling and the galaxy bias expansion are valid only on large scales, $k\lesssim k_{\rm NL}$, and break down in the nonlinear regime, $k\gtrsim k_{\rm NL}$. Hence it is necessary to adopt scale cuts in $\dSigma$ and $\wproj$ to properly use the PT method in parameter estimations. 
We also note that, although one might think that a residual shot noise term needs to be included in $P_{\rm gg}$ for self-consistency of the PT modeling, it contributes only to the correlation function at zero lag, and we ignore it in the following.

To compute the nonlinear matter power spectrum  $P_{\rm NL}(k)$ for a given cosmological model, we use the improved \code{halofit} in Ref.~\citep{takahashiRevisingHalofitModel2012} for the input linear matter power spectrum that is computed using \code{CAMB} for the assumed cosmological model. To compute the higher-order PT terms, we use the \code{FAST-PT} algorithm \cite{mcewenFASTPTNovelAlgorithm2016}. For $\dSigma$, we use \code{FFTLog}\footnote{\url{https://jila.colorado.edu/~ajsh/FFTLog/index.html}} to obtain $\dSigma$ to compute an infinite-range line-of-sight projection of the computed $P_{\rm gm}$, based on Eq.~(\ref{eq:dsigma_fourier2}). 
For $\wproj$, we first compute the three-dimensional correlation function $\xi_{\rm gg}(r)$ using \code{FFTLog} from the computed $P_{\rm gg}$, and then numerically compute the line-of-sight integration of $\xi_{\rm gg}(\sqrt{R^2+\Pi^2})$ over $\Pi=[0,\pi_{\rm max}]$ to obtain $\wproj(R)$.

For our baseline method we employ the model involving only the $b_1$ terms, without other terms in Eqs.~(\ref{eq:Pgm}) and (\ref{eq:Pgg}). 
Hereafter we will often call this model the ``minimal-bias'' model.
This method is similar to what was used in the DES-Y1 cosmological analysis \cite{2018PhRvD..98d3526A} \citep[also see][for a similar discussion]{nishizawa13}. 
This baseline model, by definition, satisfies the following identity for the cross-correlation coefficient over all the scales, irrespective of galaxy types:
\begin{align}
r^{(\xi)}_{\rm cc}(r)\equiv \frac{\xi_{\rm gm}(r)}{\sqrt{\xi_{\rm gg}(r)\xi_{\rm mm}(r)}}=1, 
\label{eq:def_r_coefficient}
\end{align}
where $\xi_{\rm mm}$ is the two-point correlation function of matter. As shown in Fig.~31 of Ref.~\cite{nishimichiDarkQuestFast2018}, the clustering of halos measured in $N$-body simulations 
fairly well satisfies $r\simeq 1$ on scales $r\gtrsim 10~\hiMpc$.
The PT picture generally predicts $r_{\rm cc}^{(\xi)}\ne 1$ if the higher-order bias parameters are non-zero. One of the main purposes of this paper is to assess the performance of the baseline \mtrv{model.}

In~\figref{fig:pt-bias-terms}, we show how different PT terms in $P_{\rm gm}$ and $P_{\rm gg}$ alter the observables, $\dSigma$ and $\wproj$. 
For illustrative purpose, we here set all the bias parameters to unity, $b_1=b_2=\cdots = 1$, and employed the fiducial {\it Planck} cosmology to compute the PT terms. 
It is clear that the different terms lead to complicated scale-dependent modifications in the theoretical templates. 
In our baseline analysis, we use the large-scale clustering information, denoted by the unshaded regions, to perform parameter estimation. Even on large scales, some of the PT terms affect the theoretical templates, and we below address whether including these PT terms can improve the performance of parameter estimations.

\section{$N$-body simulations and mock catalogs of HSC and SDSS galaxies}\label{sect:GalaxyMock}

\subsection{Galaxy mock catalogs}\label{subsect:galaxy-mock-catalog}

We use galaxy mock catalogs that are built using a suite of high-resolution $N$-body simulations, generated in Ref.~\citep{nishimichiDarkQuestFast2018}, for the flat-geometry $\Lambda$CDM cosmology that is consistent with the {\it Planck} CMB data \citep[hereafter {\it Planck} cosmology; Ref.][]{planck-collaboration:2015fj}.
The {\it Planck} cosmology is characterized by the parameters, $(\omega_{\rm b},\omega_{\rm c},\Omega_{\rm de},\ln(10^{10}A_{\rm s}),n_{\rm s},w)=(0.02225,0.1198,0.6844,3.094,0.9645,-1)$, where $\omega_{\rm b}\equiv \Omega_{\rm b}h^2$ and $\omega_{\rm c}\equiv \Omega_{\rm c}h^2$ are the physical density parameters of baryon and cold dark matter with $h=H_0/(100~{\rm km~s}^{-1}~{\rm Mpc}^{-1})$ being the Hubble parameter, $\Omega_{\rm de}\equiv 1-(\omega_{\rm b}+\omega_{\rm c}+\omega_\nu)/h^2$ is the dark energy density parameter for a flat-geometry universe, $A_{\rm s}$ and $n_{\rm s}$ are the amplitude and tilt parameters for the primordial curvature power spectrum normalized at $0.05~{\rm Mpc}^{-1}$, and $w$ is the equation of state parameter for dark energy. 
As for the neutrino density $\omega_\nu\equiv \Omega_\nu h^2$, we fix to $0.00064$, corresponding to $0.06~{\rm eV}$ for the total mass of three neutrino species. 
We include the neutrino effect only in the initial linear power spectrum and ignore the dynamical effect of massive neutrinos in the $N$-body simulations \citep[see][for details]{nishimichiDarkQuestFast2018}. 
The fiducial {\it Planck} cosmology gives, as derived parameters, $\Omega_{\rm m}\equiv 1-\Omega_{\rm de}=0.3156$ (the present-day matter density parameter), $h=0.672$ and $\sigma_8=0.831$ (the rms linear mass density fluctuations within a top-hat sphere of radius $8~\hiMpc$).
The initial conditions of the $N$-body simulations are generated using a numerical code developed in \citep{nishimichi09,Valageas11a} based on the second-order Lagrangian perturbation theory \citep{scoccimarro98,crocce06a}. The initial linear power spectrum is computed by \code{CAMB} \citep{camb}, and then the initial Gaussian random fields are generated from the input power spectrum.

All the $N$-body simulations used in this paper are performed with $2048^3$ dark matter particles in comoving cubes with side length $1~h^{-1}\mathrm{Gpc}$. The mass of the simulation particle is $m = 1.02\times10^{10}~\hiMsun$ for the {\it Planck} cosmology. We use halos identified in each simulation output using the halo finder algorithm in phase space, \code{ROCKSTAR}\citep{behrooziRockstarPhaseSpaceTemporal2013}.
Throughout this paper, we adopt $M\equiv M_{200} = (4\pi/3)R_{200}^3\times200\bar{\rho}_{\rm m0}$ for the halo mass definition, where $R_{200}$ is the spherical halo boundary within which the mean mass density is $200\bar{\rho}_\mathrm{m0}$. 
The center of each halo is determined by \code{ROCKSTAR}. After identifying halo candidates, we determine whether they are central or satellite halos. When the separation between the centers of different
halos is closer than $R_{200}$ of any neighboring halo with a larger mass, we mark it as a satellite halo and those unmarked are finally identified as central halos. We kept only the central halos with mass $M\ge 10^{12}~\hiMsun$.

\begin{table}
	\centering
    \caption{
		Specifications of the mock galaxy catalogs that resemble the LOWZ and CMASS galaxies for the SDSS BOSS survey. For the CMASS sample we consider two subsamples divided into two redshift ranges. We give the redshift range and the comoving volume for each sample, assuming $8300$~deg$^2$ for the area coverage. \mtrv{The column denoted as ``representative redshift'' is a representative redshift of each sample which we use to make the model predictions.}
    The lower columns, after double lines, denote the HOD parameters used to build the mock catalogs of each sample from $N$-body simulations for the {\it Planck} cosmology.
	}
    \label{tab:HOD_parameters}
    \begin{tabular}{c|ccc} \hline \hline
        parameters  & \multicolumn{3}{c}{sample} \\
        /quantities & LOWZ & CMASS1 & CMASS2 \\ 
		\hline
        redshift range  & $[0.15,0.35]$ & $[0.47,0.55]$ & $[0.55,0.70]$ \\ 
        \mtrv{representative} redshift  & $0.251$ & $0.484$ & $0.617$ \\ 
		volume [$(\hiGpc)^3$] & 0.67 & 0.81 & 2.00 \\
	    \hline \hline
		& \multicolumn{3}{c}{fiducial values}\\ \hline
		$\log M_{\rm min}$ & $13.62$   & $13.94$   & $14.19$  \\
		$\sigma_{\log{M}}$ & $0.6915$ & $0.8860$ & $0.7919$  \\
		$\log M_1$         & $14.42$   & $14.46$   & $14.85$  \\
		$\log M_{\rm sat}$ & $13.33$   & $13.72$   & $13.01$ \\
		$\alpha_{\rm sat}$           & $0.9168$  & $1.192$   & $0.9826$ \\ \hline 
    \end{tabular}
\end{table}
\begin{table}[htb]
	\centering
	\caption{
		The cumulative signal-to-noise (S/N) ratios of $\dSigma$, $\wproj$ and the joint measurements for the LOWZ, CMASS1 and CMASS2 samples, which are estimated using the mock signals and the covariance matrices. 
		Here we define the ``cumulative'' S/N over the ranges of $R=[12,70]$ and $[8,70]$ for $\dSigma$ and $\wproj$, respectively, which are our baseline choices of the radial range (see text for details). For the ``total'' S/N of $\dSigma$ we take into account the cross-covariances between $\dSigma$'s of different galaxy samples. 
    \mtrv{We assume that}
    the $\wproj$-signals for the three samples are independent from each other, and ignore the cross-covariances between $\dSigma$ and $\wproj$.
	}
	\begin{tabular}{c|ccc|c} \hline\hline
		& LOWZ & CMASS1 & CMASS2 & total \\
		\hline
		$\dSigma$ & 5.12 & 5.51 & 5.39 & 8.74 \\
		$\wproj$  & 23.1 & 23.2 & 22.7 & 39.8 \\
		joint ($\dSigma+\wproj$) & 23.7 & 23.8 & 23.3 & 40.7 \\ \hline
	\end{tabular}
	\label{table:signal-to-noise-ratio}
\end{table}

To build mock catalogs of galaxies that resemble the \survey{SDSS} spectroscopic galaxy samples, we employ the halo occupation distribution (HOD) prescription \citep{jingSpatialCorrelationFunction1998,seljak:2000uq,scoccimarro:2001fj,zhengTheoreticalModelsHalo2005}. More precisely we follow the method described in Ref.~\citep{kobayashiCosmologicalInformationContent2019,2020arXiv200506122K}. 
Here the HOD model gives the expected number of galaxies in host halos as a function of halo mass. 
The HOD consists of two contributions, the HODs for central and satellite galaxies, respectively:
\begin{align}
	\avrg{N_{\rm g}}\!(M)=\avrg{N_{\rm c}}\!(M)+\avrg{N_{\rm s}}\!(M).\label{eq:hod}
\end{align}
We assume that the mean number of central galaxy is given by
\begin{align}
	\langle N_{\rm c}\rangle(M) = \frac{1}{2}\left[1+{\rm erf}\left(\frac{\log M-\log M_{\rm min}}{\sigma_{\log M}}\right)\right],\label{eq:fiducial-Ncen}
\end{align}
where $\mathrm{erf}(x)$ is the error function, and $M_{\rm min}$ and $\sigma_{\rm \log M}$ are the model parameters. 
The central HOD has the following properties: $\langle N_{\rm c}\rangle(M)\rightarrow0$ for halos with $M\ll M_{\rm min}$, while $\langle N_{\rm c}\rangle(M)\rightarrow1$ for halos with $M\gg M_{\rm min}$. 
We populate $N_{\rm c}$ central galaxy into each halo, where $N_{\rm c}$ is a random number drawn from the Bernoulli distribution with mean, $\langle N_{\rm c}\rangle(M)$, solely determined by the halo mass of each halo.
When $\langle N_c\rangle(M)=1$, all halos in the mass bin host central galaxies.
For the default HOD method, we populate satellite galaxy(ies) into halos that already host a central galaxy.
To determine how many satellite galaxies reside in a halo, we employ the satellite HOD given by $\langle N_{\rm s}\rangle(M) = \langle N_{\rm c}\rangle(M)\lambda_{\rm s}(M)$, where 
\begin{align}
    \lambda_{\rm s}(M) \equiv \left(\frac{M-M_{\rm sat}}{M_1}\right)^{\alpha_{\rm sat}},\label{eq:hod_ns}
\end{align}
where $M_{\rm sat},M_1$ and $\alpha_{\rm sat}$ are model parameters. 
We populate $N_{\rm s}$ satellite galaxies into each halo, where $N_{\rm s}$ is a random number drawn from the Poisson distribution with mean $\lambda_{\rm s}(M)$.
For the default model, we assume that the spatial distribution of satellite galaxies follows the Navarro-Frenk-White (NFW) profile with the concentration-mass relation given in \cite{diemerUNIVERSALMODELHALO2015,Diemer_2019} for the {\it Planck} cosmology. Thus, in each of the galaxy mock catalogs, we have the 3-dimensional distributions of dark matter and galaxies at a given redshift. 

We then \mtrv{construct mock catalogs of galaxies}
that resemble spectroscopic galaxies in the SDSS-III BOSS DR11 data covering about 8,300~deg$^2$
\citep{dawsonBaryonOscillationSpectroscopic2013}. We consider three galaxy samples at three redshifts: the ``LOWZ'' galaxies in the redshift range $z=[0.15,0.30]$, and two subsamples of ``CMASS'' galaxies that are divided into two redshift bins of $z=[0.47,0.55]$ and $z=[0.55,0.70] $. Here we consider luminosity-limited samples rather than flux-limited samples for the SDSS-like galaxies, following the method in Refs.~\cite{2015ApJ...806....1M,2015ApJ...806....2M}. 
With this selection, each sample is considered as nearly volume-limited in the  redshift range, and we expect that properties of galaxies do not strongly evolve within the redshift range. Table~\ref{tab:HOD_parameters} gives the characterizations and HOD parameters for each of LOWZ- and CMASS-like galaxies we consider. 
The overlapping area between the HSC-Y1 data and the BOSS survey is only about 140~deg$^2$, so the comoving volumes covered by the overlapping area is smaller than the number in Table~\ref{tab:HOD_parameters}\footnote{Table~\ref{tab:HOD_parameters} is the same as Tables~II and III of Ref.~\cite{kobayashiCosmologicalInformationContent2019}. However, the volumes in Table~II of the paper were typos, and the volumes in Table~\ref{tab:HOD_parameters} are correct, which is the volume for each redshift range with the area coverage of $8300$~sq. deg.}, by a factor of 0.017. Although the overlapping region has such a small volume coverage, the galaxy-galaxy weak lensing, measured from the HSC-Y1 data, plays a crucial role in the parameter estimation, especially by breaking degeneracies between the cosmological parameters and the galaxy bias parameters, as we will show below.

We use the outputs of the $N$-body simulations at $z=0.251$, $0.484$ and $0.617$ to build mock catalogs for each of the LOWZ and the two CMASS galaxies, respectively. \mtrv{We take these redshifts as}
representative redshifts of each sample. 

In addition to the default mock catalogs described above, we also build the mock catalog including the RSD effect of galaxies. As described in Ref.~\cite{kobayashiCosmologicalInformationContent2019}, each mock galaxy has its own peculiar velocity: the total velocity of each galaxy is given by the sum of the bulk velocity of its host halo and the internal virial velocity inside the host halo. 
Assuming the distant observer approximation, we first make a mapping of each galaxy from the real to redshift space according to the RSD effect, and then use the same procedures to measure $\wproj$. 
To test the impact of RSD effect on the parameter estimation we prepare a mock catalog where we include the RSD effect using the the same halo-galaxy connection method as in the \mock{fiducial} mock, and will use the RSD mock for the test.

\subsection{Measurements from the mock catalogs}
\label{subsect:measument-of-signal-from-catalog}

\begin{figure*}
	\centering
	\includegraphics[clip,width=0.95\textwidth]{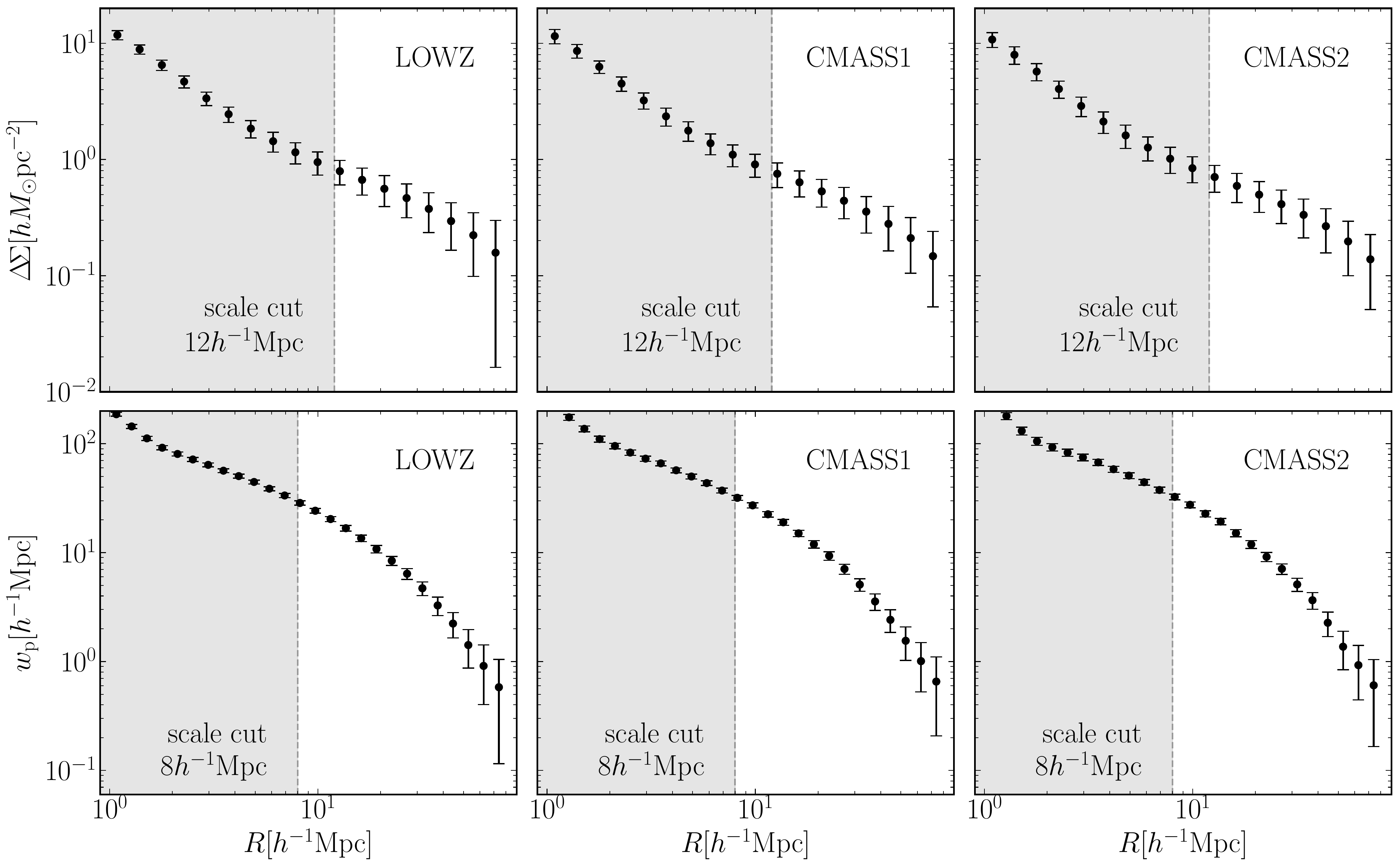} 
    \caption{
		The simulated signals of galaxy-galaxy weak lensing, $\dSigma$, and the galaxy auto-correlation function, $\wproj$, for the \survey{SDSS} spectroscopic data and the 1st-year \survey{HSC} data, which are computed from the mock catalogs (see main text for the details). Here we consider the three samples, LOWZ, CMASS1 and CMASS2, centered at different redshifts, z=0.251, 0.484, and 0.617, respectively. 
		The error bars around each data point denotes the statistical errors that are computed from the mock catalogs based on the light-cone, full-sky simulations (see text for details). Since we use the mock catalogs with much larger volume than those of \survey{SDSS} and \survey{HSC} Y1 data, the scatters in the signals are negligible or at least much smaller than the statistical errors. We will use the mock signals for tests and validations of the analysis methods for the parameter inference.
		}
    \label{fig:fiducial-signal}
\end{figure*}

To obtain the simulated signals, we measure the projected correlation function ($\wproj$) and the galaxy-galaxy weak lensing ($\dSigma$) from each of the mock catalogs. \mtrv{As we discussed in Sec.~\ref{sec:observables}, both the signals depend only on large scale structures at lens redshift, 
$z_{\rm l}$ (each of the three SDSS redshifts), where the lensing profile does not depend on source redshift because we assumed that 
the dependence of the lensing critical density, $\Sigma_{\rm crit}$, is corrected for in the measurement. Hence, to generate the mock signals, we 
use the $N$-body simulation outputs \ssrv{and galaxy mock catalogs} at each of the SDSS redshifts mentioned above. However, we will also use the different mock catalogs of HSC source galaxies to compute 
the covarinace matrix of lensing profile, as we will describe later.}

For $\wproj$ we first generate grid-based data for the three-dimensional number density fields of galaxies using the Nearest Grid Point (NGP) interpolation kernel. Then we obtain the Fourier-transformed quantities of the density field using the Fast Fourier Transformation (FFT) method with $1024^3$ grids. 
By inverse-Fourier transforming $|\tilde{\delta}_{\rm g}(\bk)|^2$, we numerically obtain the three-dimensional auto-correlation function $\xi_{\rm gg}(\br)$ via the relation $\xi_{\rm gg}(\br)=\int\!\mathrm{d}^3\bk/(2\pi)^3~
|\tilde{\delta}_{\rm g}(\bk)|^2e^{i\bk\cdot\br}$. Here we take the $x^3$-axis in the simulation to be along the line-of-sight direction, assuming the distant observer approximation. We then take the azimuthal angle average of $\xi_{\rm gg}(\br)$ in the $(x^1,x^2)$-plane for a fixed projected radius $R$, project it along the line-of-sight ($x^3$) direction over the range of $x_3=[0,\pi_{\rm max}]$, and then obtain the simulated signal for the projected correlation function, $\wproj(R)$, for an assumed $\pi_{\rm max}$ ($\pi_{\rm max}=100~\hiMpc$ for our default choice). 

For the lensing profile $\dSigma(R)$, we employ a similar method to $\wproj$, but in this case we first project both the density fields of galaxies and matter ($N$-body particles) along the $x^3$-direction over the entire simulation box to obtain the projected density fields. Then we use the two-dimensional FFT to obtain the Fourier coefficients, using the $46,332^2$ FFT grid points, from which we compute the projected cross-correlation function between galaxies and matter and take the azimuthal average to obtain $\Sigma_{\rm gm}(R)$. Finally, based on Eq.~(\ref{eq:dSigma-def}), we numerically compute the simulated signals for $\dSigma_{\rm gm}(R)$ in each realization. 

We use 22 and 19 realizations of the galaxy mock catalogs to compute $\wproj$ and $\dSigma_{\rm gm}$ at each redshift for each SDSS-like galaxy sample in Table~\ref{tab:HOD_parameters}\footnote{The reason we used slightly different number of realizations for $\wproj$ and $\dSigma$ is because the $N$-body simulation data for the 4 realizations were lost in the middle of this work. However, the difference is small compared to the statistical errors we need, and the main results of this paper are not changed.}. 
These correspond to 22~ and 19~$(\hiGpc)^3$ volumes in total for $\wproj$ and $\dSigma$, respectively. The simulated volume is much greater than the volume of SDSS survey (for each galaxy sample) by at least a factor of 11, as can be found from Table~\ref{tab:HOD_parameters}, and the volume is greater than the volume of SDSS-HSC overlapping region by a factor of 
650.
We use the averaged signals among all the realizations as simulated signals used in cosmology challenges in order to minimize any unwanted bias in estimated parameters due to sample variance.
In this way, we can properly evaluate the performance of each method or theoretical template, i.e. whether to recover the true cosmological parameters, without being affected by the sample variance.

\subsection{Error covariance matrices}
\label{subsect:measurement-of-covariance}

The error covariance matrices of the observables ($\wproj$ and $\dSigma$)  characterize statistical uncertainties in the observables for the assumed surveys, HSC-Y1 and SDSS surveys. 

To accurately model the covariance matrix of $\dSigma$,  
we employ the methods in Refs.~\cite{shirasakiMockGalaxyShape2019,shirasakiRobustCovarianceEstimation2017}. 
We use the 108 realizations of {\it full-sky} simulations in Ref.~\cite{takahashiFullskyGravitationalLensing2017}, each of which includes the halo distribution and the lensing fields in each of source redshift bins in the light-cone volume. 
Using the real HSC-Y1 catalog of source galaxies that contains the information on the position (RA and dec), shape, photometric redshift, and the lensing weight for individual galaxies, we generate the mock catalog of HSC galaxies as follows:
\begin{itemize}
	\item[(i)] Assign the HSC survey footprints (RA and dec regions) to each of the full-sky realizations.
	\item[(ii)] Populate each source galaxy into each realization of the light-cone simulations according to its angular position and photometric redshift.
	\item[(iii)] Randomly rotate the shape of each galaxy to erase the real lensing signal. 
	\item[(iv)] Simulate the lensing distortion effect on each source galaxy according to the lensing information of the full-sky simulations for the source redshift 
	\item[(v)] Repeat the steps (i)--(iv) for all the source galaxies (about 4.3 millions galaxies in total after the photo-$z$ cut to select background galaxies).
\end{itemize}
We cut out 21 mocks of the HSC galaxies from each of the full-sky simulations, where the HSC survey footprints consist of 6 distinct fields (about 170~sq.~deg. when not including the masks) and the different HSC mocks in the same simulation are taken using the rotation in RA and dec directions to avoid an overlap between the different HSC mocks \citep[see Fig.~2 in Ref.][]{shirasakiMockGalaxyShape2019}. Hence we have 2268$(=21\times 108)$~HSC mocks in total. The HSC mocks constructed in this way include observational effects such as the survey geometry, masks, the distributions of positions, shapes and redshifts for source galaxies, and the galaxy weights. 

Then we populate SDSS-like (lensing) galaxies into halos in each full-sky, light-cone realization using an HOD model that is chosen to fairly reproduce the number density and the projected correlation function of LOWZ and CMASS galaxies in the redshift range of $0.15<z<0.35$ (LOWZ), $0.43<z<0.55$ (CMASS1), and $0.55<z<0.70$ (CMASS2), respectively. 
Here we assume the NFW profile \cite{NFW} to model the spatial distribution of satellite galaxies in the host halo. In addition we include the redshift-space distortion effect; for central galaxies we assume the same peculiar velocities as those of the host halos, while we assign the random virial velocities to satellite galaxies assuming a Gaussian distribution with the virial velocity dispersion \citep{kobayashiCosmologicalInformationContent2019}. 
Note that we include the SDSS-like galaxies outside the HSC survey footprints because the galaxy-galaxy weak lensing signals can be measured for such pairs including the foreground SDSS galaxies outside the HSC survey footprints for a given projected separation as we did for actual measurements. 

Then we measure the galaxy-galaxy lensing signal, $\dSigma$, from each mock using the same measurement pipeline as we used in the actual measurements.
We estimate the covariance matrix from the scatters among the signals in the 2268 realizations. We also estimated the cross covariance between different redshift bins.
The covariance matrix estimated in this way includes all the contributions: the shape noise, the Gaussian covariance contribution, and the non-Gaussian covariance including contributions of the connected 4-point correlation function of sub-survey modes and the super-sample covariance \citep{takadaPowerSpectrumSupersample2013}. 
With this method, Ref.~\cite{shirasakiRobustCovarianceEstimation2017} shows that the shape noise of actual galaxies and the Gaussian covariance matrix give a dominant contribution to the total covariance over the scales we consider.

For the covariance matrix of the projected correlation function ($\wproj$), we use the same mock catalogs of SDSS galaxies in the light-cone simulations we described above. 
Similarly we define the SDSS DR11 footprints into each of 108 simulations, and then estimate the covariance matrix of $\wproj$ from each mock using the jackknife method in Ref.~\cite{2015ApJ...806....1M}. 
Then we take the average of the covariance matrices among 108 realizations, and use the averaged covariance for our cosmology challenges.
For the galaxy clustering, the shot noise and the Gaussian covariance give a dominant contribution \citep{2019MNRAS.482.4253T}, and the jackknife is a good approximation to estimate the genuine covariance matrix. 

The overlapping region between the SDSS and HSC-Y1 data is only 140~sq.~deg., which is a tiny region of the SDSS footprints of $8300$~sq.~deg. 
Hence, we can safely ignore the cross-covariance matrix between $\dSigma$ and $\wproj$, because $\wproj$ is mainly from the SDSS region 
\mtrv{(the non-overlapping region).}
Or for actual cosmological analysis, we can check whether the cosmological constraints are significantly changed if we use $\wproj$ measured only from the non-overlapping SDSS region with HSC, because we can safely ignore the cross-covariance in this case. 

Fig.~\ref{fig:fiducial-signal} shows the simulated signals for $\dSigma$ and $\wproj$ for each of LOWZ-, CMASS1-, and CMASS2-like samples. 
For the simulated signals, we used the mock catalogs for the {\it Planck} cosmology as described in Sections~\ref{subsect:galaxy-mock-catalog} and \ref{subsect:measument-of-signal-from-catalog}. 
The error bar at each radial bin denotes the statistical error expected for measurements of $\wproj$ and $\dSigma$ for the HSC-Y1 and SDSS surveys, which is computed from the diagonal components of the covariance matrix based on the method we described in this section. The figure shows that the HSC and SDSS data allows for a significant detection of these clustering signals in each bin of separations, for each of the three galaxy samples. 
To be more quantitative, Table~\ref{table:signal-to-noise-ratio} gives the cumulative signal-to-noise ratio for each sample, which is estimated by $({\rm S/N})^2=\sum_{ij}d_i{\rm Cov}^{-1}_{ij}d_j$, where $d_i=\dSigma(R_i)$, $\wproj(R_i)$ or the joint data, ${\rm Cov}^{-1}$ is the inverse of the covariance matrix, and the summation is done over the range of radial bins $R_{i}$. Here we consider, as our baseline choices, the ranges of radial bins, $R\simeq [12,70]~\hiMpc$ or $R\simeq [8,70]~\hiMpc$ for $\dSigma$ and $\wproj$, respectively. 
Even if the S/N value for $\dSigma$ is smaller than that for $\wproj$ for each sample, the use of the weak lensing information is crucial to break the parameter degeneracies with the galaxy bias parameters in $\wproj$, as we will show below.

\section{Method}
\label{sect:methods}

\subsection{Parameter estimation methodology}

We assume that the likelihood of the mock signals (data vector) follows a multivariate Gaussian distribution:
\begin{align}
	\ln{\cal L}(\bm{d}|\bm{\theta})=
	-\frac{1}{2}\sum_{i,j} [d_i - t_i(\bm{\theta})] {\rm Cov}_{ij}^{-1} [d_j - t_j(\bm{\theta})] + {\rm const.},
	\label{eq:likelihood}
\end{align}
where $\bm{d}$ is the data vector, $\bm{t}$ is the theoretical templates of the observables, and $\bm{\theta}$ is the set of parameters. We consider the following model parameters and the data vector for parameter inference with the joint probes ($\dSigma$ and $\wproj$): 
\begin{align}
    \bm{d}&=\left\{\dSigma(R_1; z_1),\dots,\dSigma(R_{N_{\dSigma}}; z_1),\right.\nonumber\\
    &\hspace{3em} \wproj(R_1;z_1), \dots, \wproj(R_{N_{\wproj}};z_1), \nonumber\\
    &\hspace{5em} \dots,\nonumber\\
    &\hspace{3em} \dSigma(R_1; z_3),\dots,\dSigma(R_{N_{\dSigma}}; z_3),\nonumber\\
    &\hspace{3em} \left.
    \wproj(R_1;z_3),\dots,\wproj(R_{N_{\wproj}};z_3)\right\},\nonumber\\
    \bm{\theta}&=\left\{\Omega_{\rm m},\sigma_8, b_1(z_1), b_2(z_1), \right.\nonumber\\
    &\hspace{3em}\left.\dots, b_1(z_2), b_2(z_2), \dots, b_1(z_3), b_2(z_3), \dots \right\},
\end{align}
where $N_{\dSigma}$ and $N_{\wproj}$ are the number of radial bins for $\dSigma$ and $\wproj$ (see below), respectively.
For the cosmological parameters we consider only $\Omega_{\rm m}$ and $\sigma_8$ for simplicity, as those are the primary parameters which the clustering observables can constrain. 
We introduce each bias parameter for each of the galaxy samples at three redshift bins and treat the bias parameters independently in parameter estimation. 
For the baseline ``minimal-bias'' method in which we assume a linear bias model, we have 5 parameters: $\Omega_{\rm m}, \sigma_8$ and 3 linear bias parameters $b_1(z_i)$ ($i=1,2,3$). For the data vector, we include both $\dSigma$ and $\wproj$ for each of the three galaxy samples; 
for the baseline setup (see below), we employ 8 $R$-bins and $14$ $R$-bins in the ranges of $R\simeq [12, 71.2]$ and $[8,73.9]~\hiMpc$ for $\dSigma$ and $\wproj$, respectively, where the radial bins are evenly spaced in the logarithmic space; we have 66 data points in total, $66= 3\times (8+14)$. 
For the covariance, we use the fixed covariance matrix that is obtained from the mock catalogs described in Sec.~\ref{subsect:measurement-of-covariance}, and in other words we do not include the cosmology dependences of the covariance matrix following the discussion in Ref.~\citep{2019OJAp....2E...3K}. This is suitable for our study, because we can avoid any difference in parameter inference arising from the cosmology-dependent covariances. 

\begin{table}
	\centering
	\caption{
		Prior ranges of model parameters. 
		The redshifts of galaxy biases, $z_i(i=1,2,3)$, correspond to the LOWZ, CMASS1 and CMASS2 galaxy samples. The prior ranges of derived parameter, $\sigma_8$ and $\Omegam$ are also shown, which are computed from the prior ranges of $10^{10}A_\mathrm{s}$ and $\Omega_\mathrm{de}$.
	}
    \label{tab:prior-range}
    \begin{tabular}{cc} \hline \hline
      \multirow{2}{*}{Parameters} & Prior range \\
	  & [minimum, maximum] \\ 
	  \hline\hline
	  $10^{10}A_\mathrm{s}$ & $[0,100]$ \\
	  $\Omega_\mathrm{de}$ & $[0.4594,0.9094]$ \\
	  $b_1(z_i)$ & $[0,5]$ \\
	  $b_2(z_i)$ & $[-100,100]$\\
	  $b_{s_2}(z_i)$ & $[-100,100]$\\
	  $b_3(z_i)$ & $[-100,100]$\\\hline \hline 
	  & varying range \\ \hline
	  $\sigma_8$ & $[0.163, 3.833]$\\
	  $\Omega_\mathrm{m0}$ & $[0.0906, 0.5406]$ \\ \hline\hline
    \end{tabular}
\end{table}
We then perform cosmology challenges (parameter estimation) based on the Bayesian inference:
\begin{align}
	{\cal P}(\bm{\theta}|\bm{d}) \propto {\cal L}(\bm{d}|\bm{\theta}) {\Pi}(\bm{\theta}) \label{eq:Bayesian-posterior},
\end{align}
where ${\cal P}(\bm{\theta}|\bm{d})$ is the posterior distribution of model parameters and $\Pi(\bm{\theta})$ is the prior distribution of the parameters. Throughout this paper, we use a flat prior on each of model parameters as given in Table~\ref{tab:prior-range}. 
Exactly speaking, we use the cosmological parameters, $\Omega_{\rm de}$ and $10^{10}A_{\rm s}$, in the parameter estimation following the design in Ref.~\cite{nishimichiDarkQuestFast2018}, and employ their flat priors given by $\Omega_{\rm de}=[0.4594, 0.9094]$ and $10^{10}A_{\rm s}=[1.0, 400]$. Other cosmological parameters such as $\omega_{\rm b}$, $\omega_{\rm c}$ and $n_{\rm s}$ are fixed to their fiducial values for the {\it Planck} cosmology. 
In the following, we show the posterior distributions for $\Omega_{\rm m}$ and $\sigma_8$ instead of $\Omega_{\rm de}$ and $10^{10}A_{\rm s}$, which are derived parameters for a flat-geometry $\Lambda$CDM model. The varying ranges of $\Omegam$ and $\sigma_8$ in Table~\ref{tab:prior-range} are computed from the prior ranges of $\Omega_{\rm de}$ and $10^{10}A_{\rm s}$. \ssrv{The prior ranges for the higher-order galaxy bias parameters are chosen to be wide enough so that the resultant posterior distribution is not bounded by the prior range\footnote{\ssrv{We found that, if the prior range for} \ssrv{the higher-order galaxy bias parameters are taken to be as narrow as that of the linear galaxy bias parameter,  the posterior distribution is bounded by the prior range, meaning that the result depends on the choice of prior range. \mtrv{However, we checked that 
the following results for the cosmological parameters remain almost unchanged.}}}.}

For a given dataset $\bm{d}$, we make use of the likelihood package \code{MultiNest}\cite{ferozMultiNestEfficientRobust2009} implemented in \code{MontePython} to estimate the posterior distribution, which is based on the nested sampling method.
We also use the public code \code{emcee}\cite{foreman-mackeyEmceeMCMCHammer2013}, which is based on the Markov Chain Monte Carlo method, to check the robustness of our results against different samplers (see \apdxref{apdx:sampler-check-emcee} for details). 
Sampling of parameters is done for $10^{10}A_\mathrm{s}$, $\Omega_\mathrm{de}$ and galaxy biases, where we use the flat priors in Table~\ref{tab:prior-range}.
Once a parameter chain is obtained after the sampling is converged, we compile
the chain to obtain flat priors for $\sigma_8$ and $\Omegam$, by multiplying each sample by the weight given by the conversion factor $\sigma_8/A_\mathrm{s}$
\footnote{The correction factor to change the priors of $(\Omegam,\sigma_8)$ 
from the flat priors of  ($10^{10}A_\mathrm{s}$ and $\Omegam$) can be obtained from conservation of the probability density. Because the likelihood part is identical, conservation of the probability density
\begin{align}
	\mathcal{P}(10^{10}A_\mathrm{s}, \Omega_\mathrm{de})\diff (10^{10}A_\mathrm{s})\diff\Omega_\mathrm{de}
	=
	\mathcal{P}(\sigma_8, \Omegam)\diff\sigma_8\diff\Omegam,
	\nonumber
\end{align}
leads to conversion for the priors $\Pi(10^{10}A_\mathrm{s},\Omegam)\diff (10^{10}A_\mathrm{s})\diff\Omega_\mathrm{de}=\Pi(\sigma_8, \Omegam)\diff\sigma_8\diff\Omegam$, yielding 
$\Pi(A_\mathrm{s},\Omega_{\rm de})\propto|\partial(\sigma_8,\Omegam)/\partial(10^{10}A_\mathrm{s},\Omega_{\rm de})|\propto \sigma_8(A_\mathrm{s},\Omega_{\rm de})/A_\mathrm{s}$
for the flat $\Lambda$CDM model if assuming the flat priors for $\sigma_8$ and $\Omegam$ ($\Pi(\sigma_8,\Omegam)={\rm const.})$. 
Here we do not care about the constant factor which is irrelevant to the weighting of chain as far as we focus only on the posterior distribution.}. 
Hence the results we show below are equivalent to the results obtained by employing flat priors for $\sigma_8$ and $\Omegam$.

In this paper, we adopt the mode value and highest density interval \ssrv{(HDI)} to infer the central value and the credible interval of parameter(s), \mtrv{respectively, both of which} 
are obtained from the 
\ssrv{marginalized}
posterior distribution. 
We also report the best-fit parameters \mtrv{at the maximum posterior in a multi-dimensional parameter space, before marginalization.} 
The important difference between the parameter value of the best-fit model and the central (mode) value is that the former is \mtrv{a point estimate in the posterior distribution of the full parameter space,}
while
the latter is estimated from the posterior distribution that is obtained by marginalizing over other parameters.
We note that since our parameter estimation is based on an interval estimate rather than a point estimate, we assess the robustness of each method by asking whether the true value of each cosmological parameter is included within the credible interval.
Hence the choice of central parameter is not important for our purpose.
Nevertheless, we will also infer the point estimate of a parameter, e.g. the central (mode) value and/or the value of the best-fit model, only when we discuss how the difference between the two values arises, e.g. after marginalization of the posterior distribution over other parameters.

\subsection{Validation Methodology}
\label{sec:validation}

The purpose of this paper is to assess two aspects of the cosmology inference method. First, we want to find proper scale cuts  so that the PT-based theoretical templates are safely applied to the observables above the scale, for the $\Lambda$CDM model. 
Second, we want to assess the robustness of each method to correctly estimate cosmological parameters against properties of galaxy physics, which, if confirmed, would be a main advantage of the PT-based method compared to a more restrictive model such as the HOD-based method, where halo bias is specified by halo mass for a given cosmology. 
In the HOD-based method, the 1-halo term signal of the galaxy-galaxy lensing gives a strong constraint on the (average) mass of host halos for a galaxy sample, and in turn helps break degeneracies between galaxy bias and cosmological parameters. 
However, if this scaling relation of galaxy bias with the host halo mass is broken, e.g. in a case of the assembly bias, the HOD-based method could cause a bias in the estimated cosmological parameters.
On the other hand, in the PT-based method we treat the galaxy bias parameter(s) as completely free parameter(s), which can be considered as a more flexible bias model.
In this section, we describe our strategy and setups that we use to address the above two questions.

\subsubsection{Validation strategy of the measurement effects: scale cuts, the RSD effect and the covariance matrix}
\label{fiducial-mock-validation}

In \tableref{table:fiducial-mock-setup}, we summarize the setups of our analysis to assess performance of each method. 
As we described, it is a bit tricky to properly estimate scale cuts  for $\dSigma$ and $\wproj$ because the two observables are sensitive to different range of Fourier modes in the underlying matter power spectrum (see Eqs.~\ref{eq:dsigma_fourier2} and \ref{eq:wp_fourier2}). 
As our baseline 
setup, we employ the scale cuts of $12$ and $8~\hiMpc$ for $\dSigma$ and $\wproj$, respectively, and use the linear bias parameter, $b_1(z_i)$ for each of the three galaxy samples, denoted as \setup{$(\Delta\!\Sigma_{12},w_\mathrm{p,8})$}, following the method used in the DES analysis \cite{krauseDarkEnergySurvey2017,maccrannY1ResultsValidating2018}. 
To be more explicit, we use the information in $\dSigma(R)$ and $\wproj(R)$ over the ranges of $12\le R/[\hiMpc]\le 71.2$ and $8\le R/[\hiMpc]\le 73.9$, respectively, in the parameter estimation. 
Note that, throughout this paper, we fix the maximum separation cut to $R_{\rm max}\simeq 70~\hiMpc$ for all the results. Thus we do not include the information at BAO scales. 
To validate the baseline choice of scale cuts, we will also study the performance of different cuts as given in Table~\ref{table:fiducial-mock-setup}: 
$(R^{\dSigma}_{\rm cut},R^{\wproj}_{\rm cut})=(6,4), (9,6), (18, 12)$ and $(24,16)$ in units of $\hiMpc$, respectively, where the scale cuts for $\dSigma$ and $\wproj$ are chosen to have the same ratio as that in the \setup{baseline} setup. 
We quantify the performance of each setup by robustness and precision in estimations of the cosmological parameters, 
$\Omega_{\rm m}$ and $\sigma_8$. 
Here we quantify the ``robustness'' of each setup by the degree to recover the input parameter; we check whether the true value of cosmological parameter is included within the credible interval. We do not so much care about the central value of estimated parameter, because our inference is not a point estimate but rather an interval estimate.
We quantify the ``precision'' by the size of credible interval in a given parameter; a smaller credible interval is better. 
What we most worry about is a situation that we have a bias in any cosmological parameter larger than the credible  interval, which could lead us to claim a wrong cosmology at a high significance.
\mtrv{For our assessment}
we do not care about an accuracy to recover
the galaxy bias parameters.

By including the higher-order bias parameters, we hope that the model can more accurately describe the clustering observables down to smaller scales, i.e. more nonlinear scales. 
This could in turn improve the parameter estimation due to the increased information content, since the  clustering signals at smaller scales has a higher signal-to-noise ratio.
To assess the performance for the method using the higher-order bias parameters, we first consider different setups using different combinations of the bias parameters as listed in Table~\ref{table:sigma-upsilon-validation}, where we fix the scale cuts to the baseline choice, $(12,8)~\hiMpc$.

\ssrv{\mtrv{As shown in Fig.~\ref{fig:pt-bias-terms}, the terms involving the higher-order bias parameters alter the observables ($\dSigma$ and 
$\wproj$) on relatively small scales in the quasi nonlinear regime.} 
Thus we will also investigate how the accuracy 
and robustness could be 
improved by \mtrv{applying the higher-order bias model to the mock signals using the smaller scale cuts,}
compared to the fiducial method. The setups are denoted as ``$(\Delta\Sigma_{9},w_{\rm p,6})$ w/ $b_1,b_2$'' and ``$(\Delta\Sigma_{6},w_{\rm p,4})$ w/ $b_1,b_2$''.}

For the method labeled as ``self-consistent PT-method'' we use the PT model for nonlinear matter power spectrum including the next-leading-order (one-loop) correction, $P_{\rm SPT}$, instead of the fitting formula $P_{\rm NL}$, to compute the model predictions of $P_{\rm gm}$ and $P_{\rm gg}$ (Eqs.~\ref{eq:Pgm} and \ref{eq:Pgg}). 
For the bias parameters, in addition to $b_1$, we include the high-order bias parameters that are at the same order as $P_{\rm SPT}$. 
This model is considered a self-consistent model under the PT framework.
We consider models using different combinations of the bias parameters, $(b_1,b_2)$, $(b_1,b_2,b_{s^2})$ or $(b_1,b_2,b_{s^2},b_3)$, respectively. 
Here the model of $(b_1,b_2,b_{s^2},b_3)$ is a most thorough model including all possible terms up to the one-loop order.
We apply each of the methods to the fiducial mock catalog to test whether the method can recover the cosmological parameters. However, note that this model generally 
predicts $r_{\rm cc}\neq 1$ for the cross-correlation coefficient (Eq.~\ref{eq:def_r_coefficient}) in the presence of the non-zero higher-order bias parameter(s).

We also use the mock catalog including the RSD effects of individual galaxies to assess the impact of the RSD effect on the parameter estimation. For this test, we employ the baseline analysis setup. 

To quantify the complementarity between $\dSigma$ and $\wproj$ in the parameter estimation, we also consider the setups using either of $\dSigma$ or $\wproj$ alone. 
The setups of \setup{$(\Delta\!\Sigma_{12},w_\mathrm{p,8})$ w/ $0.1\mathrm{Cov}_{\Delta\!\Sigma}$} are for the parameter forecast when using the full dataset of the HSC survey, which has a larger area coverage, by a factor of 10, than that of HSC-Y1 data.
To mimic the full HSC information, we use the reduced covariance matrix, by multiplying the default covariance of $\dSigma$ by a factor of 1/10, as denoted as ``$0.1\mathrm{Cov}_{\dSigma}$''. 

\begin{table*}[htb]
	\centering
	\caption{Analysis setups in the parameter inference.}
	\begin{tabular}{l| c l l}
		\hline \hline
		\multirow{2}{*}{Setup label} & Scale cuts & \multirow{2}{*}{Sampled parameters} & \multirow{2}{*}{Notes}\\
        & in $\mathrm{Mpc}/h$ && \\
		\hline
		$(\Delta\!\Sigma_{12},w_\mathrm{p,8})$ & (12,8) & $\sigma_8,\Omegam,b_1(z_i)$ & baseline analysis\\
		\hline
		$(\Delta\!\Sigma_{6},w_\mathrm{p,4})$ & (6,4) & \multirow{4}{*}{$\sigma_8,\Omegam,b_1(z_i)$} & \multirow{4}{*}{tests for scale cuts}\\
		$(\Delta\!\Sigma_{9},w_\mathrm{p,6})$ & (9,6) & & \\
		$(\Delta\!\Sigma_{18},w_\mathrm{p,12})$ & (18,12) & & \\
		$(\Delta\!\Sigma_{24},w_\mathrm{p,16})$ & (24,16) & & \\
		\hline
        $(\Delta\!\Sigma_{12},w_\mathrm{p,8})~\mathrm{w/}~b_1,b_2$ & \multirow{3}{*}{(12,8)} & $\sigma_8,\Omegam,b_1(z_i),b_2(z_i)$& \multirow{3}{*}{tests for galaxy bias}\\
        $(\Delta\!\Sigma_{12},w_\mathrm{p,8})~\mathrm{w/}~b_1,b_2,b_{s_2}$ && $\sigma_8,\Omegam,b_1(z_i),b_2(z_i),b_{s_2}(z_i)$&\\
        $(\Delta\!\Sigma_{12},w_\mathrm{p,8})~\mathrm{w/}~b_1,b_2,b_{s_2},b_{3}$ && $\sigma_8,\Omegam,b_1(z_i),b_2(z_i),b_{s_2}(z_i),b_{3}(z_i)$&\\
		\hline
		\ssrv{$(\Delta\!\Sigma_{9},w_\mathrm{p,6})~\mathrm{w/}~b_1,b_2$} & \ssrv{(9,6)} & \multirow{2}{*}{\ssrv{$\sigma_8,\Omegam,b_1(z_i),b_2(z_i)$}} & \ssrv{tests for the smaller scale cuts}\\
		\ssrv{$(\Delta\!\Sigma_{6},w_\mathrm{p,4})~\mathrm{w/}~b_1,b_2$} & \ssrv{(6,4)} &&\ssrv{with the higher-order galaxy bias}\\
		\hline
		$(\Delta\!\Sigma_{12},w_\mathrm{p,8})~\mathrm{w/}~P_{\rm SPT}, b_1,b_2$ & 
        (12,8) & $\sigma_8,\Omegam,b_1(z_i),b_2(z_i)$ & \multirow{3}{*}{self-consistent PT method}\\
        $(\Delta\!\Sigma_{12},w_\mathrm{p,8})~\mathrm{w/}~P_{\rm SPT}, b_1,b_2,b_{s_2}$ & 
        (12,8) & $\sigma_8,\Omegam,b_1(z_i),b_2(z_i),b_{s_2}(z_i)$ &\\
        $(\Delta\!\Sigma_{12},w_\mathrm{p,8})~\mathrm{w/}~P_{\rm SPT}, b_1,b_2,b_{s_2},b_3$ & 
		(12,8) & $\sigma_8,\Omegam,b_1(z_i),b_2(z_i),b_{s_2}(z_i),b_{3}(z_i)$ &\\
		\hline
		$\dSigma$ alone & 12 & \multirow{2}{*}{$\sigma_8,\Omegam,b_1(z_i)$} & \multirow{2}{*}{check of degeneracy breaking}\\
		$\wproj$ alone & 8 & & \\
		\hline
		RSD & (12,8) & $\sigma_8,\Omegam,b_1(z_i)$ & impact of RSD\\
		\hline
		$(\Delta\!\Sigma_{12},w_\mathrm{p,8})~\mathrm{w/}~0.1\mathrm{Cov}_{\Delta\!\Sigma}$ & (12,8) & $\sigma_8,\Omegam,b_1(z_i)$ & impact of a change in the $\dSigma$-covariance \\
		\hline\hline
	\end{tabular}
	\label{table:fiducial-mock-setup}
\end{table*}

\subsubsection{Validation strategy against various galaxy mocks}
\label{sec:differnt_mocks}

\begin{figure*}
	\centering
    \includegraphics[clip,width=1.0\textwidth]{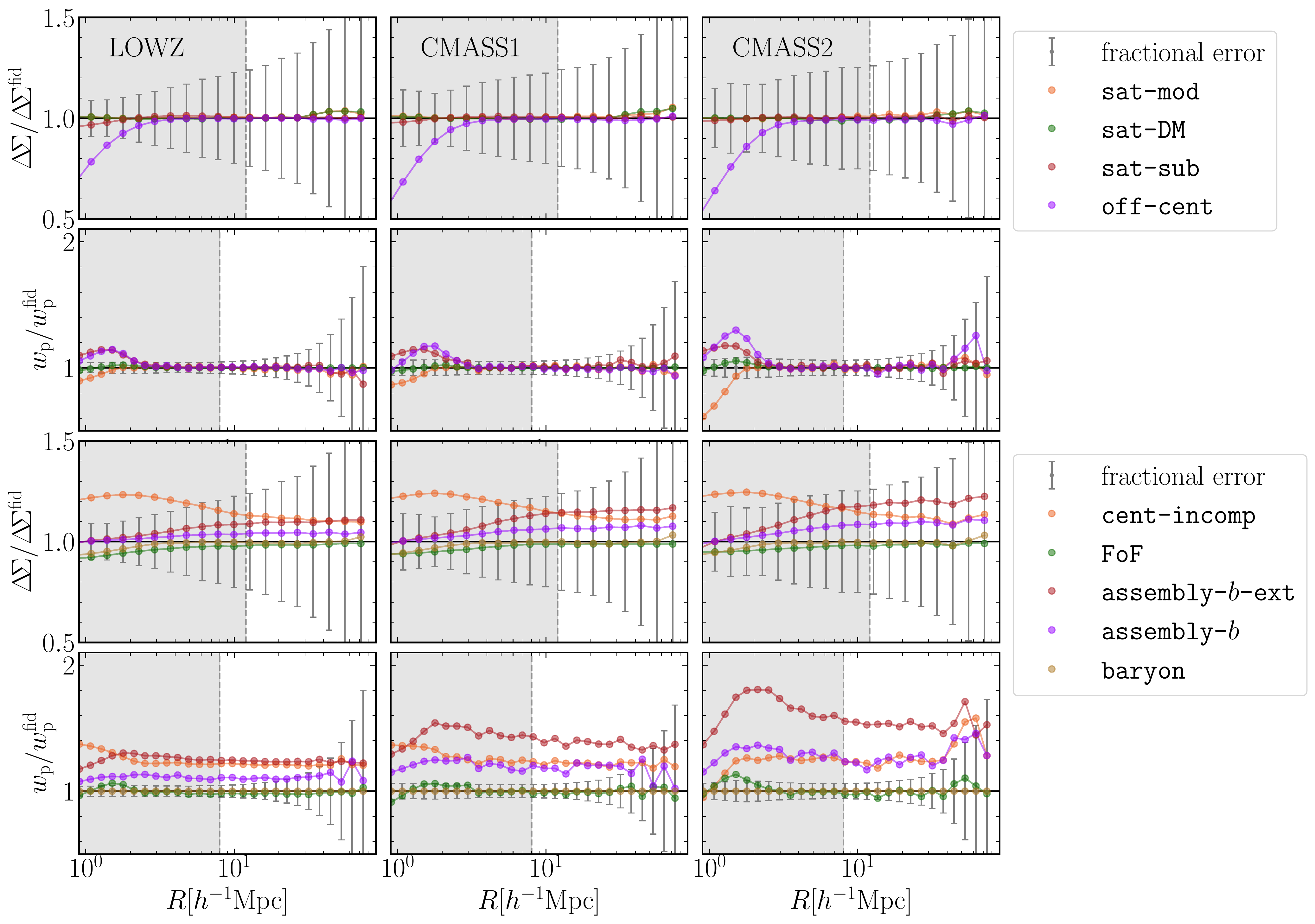}
	\caption{
		Comparison of the mock signals, $\dSigma$ and $\wproj$, for different mock catalogs relative to those for the fiducial mock (see Sec.~\ref{sec:differnt_mocks}). We generate the different mocks based on the consideration to model various effects of galaxy properties, and 
		then measure the mock signals from \mtrv{each mock catalog.}  All the mocks are based on the same simulations for the {\it Planck} cosmology. 
		Except for the \mock{incompleteness} and \mock{FoF halo} mocks, all the mocks have the same HOD, but differ in ways of populating galaxies into halos in each simulation \mtrv{realization.}
		For comparison, we show the fractional error bars around the fiducial mock, which are computed 
		from the diagonal components of the covariance matrices for the SDSS and HSC-Y1 data.
		}
	\label{fig:galaxy-mock-signals}
\end{figure*}
\begin{figure*}
	\centering
    \includegraphics[clip,width=1.0\textwidth]{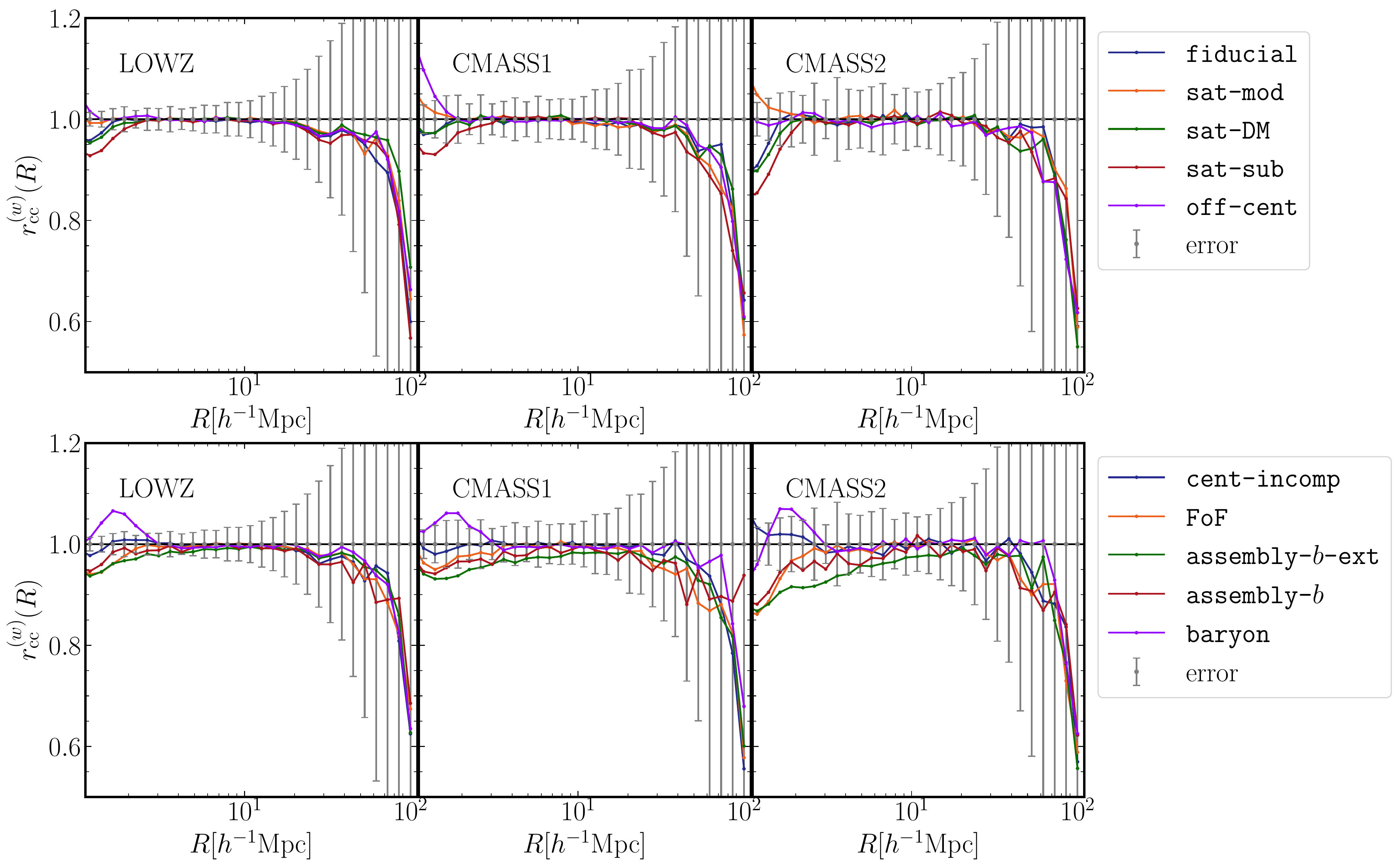}
	\caption{
		Shown is the cross-correlation coefficient, defined from the projected correlation functions as 
		$r^{(w)}_{\rm cc}(R)\equiv w_{\rm gm}(R)/\sqrt{w_{\rm gg}(R)w_{\rm mm}(R)}$, for each mock in the previous figure. 
		Note that we use $\pi_{\rm max}=100~\hiMpc$ for the line-of-sight projection \mtrv{in the computation of}
    $w_{\rm gg}$ and $w_{\rm mm}$ (see Eq.~\ref{eq:wp-def}), while we use the infinity length for $w_{\rm gm}$ that \mtrv{is relevant for}
    the lensing profile. 
		Even if the different mocks predict different signals of $\dSigma$ and $\wproj$ as shown in the previous figure, all the lines are close to unity at $R\gtrsim 10~\hiMpc$. 
		The deviation from unity at $R\gtrsim 40~\hiMpc$ is not physical, is rather due to the different projection lengths, and can be well reproduced in the theoretical templates if using the different projection lengths.
		}
	\label{fig:galaxy-mock-r}
\end{figure*}

An advantage of the PT-based method is that it is expected to be robust against properties of galaxies. As long as we allow the bias parameter(s) to freely vary in the parameter estimation, it captures the effect of galaxy properties on the clustering signals on large scales without modeling galaxy physics in detail, because the clustering signals on large scales, far beyond the scale of galaxy physics, are governed by gravitational interaction and properties of the primordial fluctuations. 
To test the robustness of the PT-based method, we use different types of mock galaxy catalogs generated in Miyatake et al. (in preparation) \citep[also see][for details of the similar mocks]{kobayashiCosmologicalInformationContent2019}. 
In the following, we briefly describe each mock catalog. \mtrv{Readers who are interested in the results can skip this subsection and directly go to Sec.~\ref{sect:result}.}
\begin{itemize}
	\item \mock{fiducial} --  This is the fiducial mock catalog, built using the HOD model in Sec.~\ref{sect:GalaxyMock} and Table~\ref{tab:HOD_parameters}.
	\item \mock{sat-mod} -- This mock is based on a slight modification from the fiducial mock. In this mock we populate satellite galaxies in halos irrespective of whether the halos already host central galaxies. We keep the same form of the satellite HOD, 
	$\avrg{N_{\rm s}}\!(M)$,  as that in the fiducial mock (Eqs.~\ref{eq:hod} and \ref{eq:hod_ns}).
	\item \mock{sat-DM} -- In this mock we populate satellite galaxy(ies) in each host halo by randomly assigning each satellite to dark matter particle($N$-body particles) in the halo, in contrast to the NFW profile used in the fiducial mock.
	\item \mock{sat-subhalo} -- In this mock  we populate satellite galaxy(ies) in each host halo by randomly assigning each satellite to subhalo in the host halo, where we use subhalos in the \code{ROCKSTAR} output.
	\item \mock{FoF halo} --  Instead of using spherical-overdensity halos in each simulation realization (see Sec.~\ref{sect:GalaxyMock}), we use halos that are identified by the friends-of-friends (FoF) method with linking length $b_{\rm FoF}=0.2$. 
	The FoF halos do not have one-to-one correspondence to the fiducial \code{ROCKSTAR} halos, and their halo masses are also different on an individual halo basis, even for the matched halos. Nevertheless, we use the same HOD to populate galaxies into FoF halos by treating the FoF halo mass as halo mass in the HOD. 
	\item \mock{off-cent} -- Since there is no unique definition of halo center,  central galaxies might have an offset from the ``center'' (mass density maximum) in the host halos as a result of complicated assembly history of host halos such as merger and  dynamical friction \citep{hikage:2013kx,masaki13}. Based on this consideration, we include the off-centering effects of central galaxies in this mock. 
	As an extreme case, in this mock we assume that all central galaxies are off-centered, and that the spatial profile of off-centered galaxies follows a Gaussian profile with width given by ${\cal R}_{\rm off}=2.2$, where ${\cal R}_{\rm off}$ is a parameter to characterize the off-centering radius in units of the scale radius of NFW profile of each host halo \cite{moreWeakLensingSignal2015}.
	\item \mock{assembly-$b$} -- Numerical simulations \citep{gao:2005fk,wechslerDependenceHaloClustering2006} as well as analytical considerations \citep{dalalHaloAssemblyBias2008} predict that halo bias could additionally depend on parameter(s) other than halo mass, depending on the details of mass assembly history in each halo, which is referred to as ``assembly bias''. This assembly bias is one of the most dangerous, physical systematic effects in the parameter inference, if we want to use a prior on galaxy bias inferred from the scaling relation of halo bias with halo mass.
	However, assembly bias effects have not yet been detected from real data at a high significance \cite{2016ApJ...819..119L}. 
	The ``\mock{assembly-$b$-ext}'' mock is intended to study the impact for an extreme case. To generate this mock, we preferentially populate galaxies into halos in the order of halo mass concentration from the lowest one.  We generate this mock based on the following method \citep[also see][for details]{kobayashiCosmologicalInformationContent2019}. 
	First, using the member $N$-body particles of each halo, we compute the mass enclosed within the sphere of radius of $0.5R_{\rm 200}$ as a proxy of the mass concentration. We then make a ranked list of all the halos in the ascending order of the inner-mass fraction in each of halo mass bins, and populate central galaxies into halos from the top of the list (from the lowest-concentration halo) in the mass bin, according to the number fraction specified by the central HOD $\avrg{N_{\rm c}}\!(M)$ in the mass bin. 
	Then we populate satellite galaxies into halos that already host a central galaxy using the satellite HOD. In this way, we can have a maximum boost in the large-scale clustering amplitude for all the three galaxy samples in this mock; $\wproj$ in this mock have larger amplitudes at large separations, by up to a factor of 1.6, than that of $\wproj$ in the fiducial mock. 
	We call this extreme-case mock ``\mock{assembly-$b$-ext}''. For the ``\mock{assembly-$b$}'' mock, we introduce a scatter to populate central galaxies to halos in the ranked list in the above procedures so that the large-scale amplitude of $\wproj$ is smaller than that of the ``\mock{assembly-$b$-ext}'' mock.
	The $\wproj$ of the ``\mock{assembly-$b$}'' mock, however, still possesses the greater amplitudes than the \mock{fiducial} mocks by up to a factor of 1.3. 
	The latter is a more realistic case for host halos of the \survey{SDSS} galaxies ($\sim 10^{13}~\hiMsun$), even if the assembly bias exists, as indicated by the simulation based study for the \LCDM model in Ref.~\cite{wechslerDependenceHaloClustering2006}.
	\item \mock{baryon} -- The baryonic physics inherent in galaxy formation/evolution is another important effect that needs to be considered. 
	Due to the complexity in the physical processes, it is still difficult to accurately model the effects from first principles. An encouraging fact is that the large-scale bias relation of galaxy distribution to matter distribution is not largely affected by the baryonic physics as shown in 
	hydrodynamical simulations 
	\cite{2014MNRAS.440.2997V,2018MNRAS.475..676S}.
  	This is ascribed to the fact that galaxies form from the same density peaks in the initial mass density fields in a CDM-dominated structure formation scenario, even in the presence of baryonic physics. 
	This would be the case for the BOSS galaxies that are early-type galaxies. Based on this consideration we do not change the mock 
	signals of $\wproj$. 
	On the other hand, the baryonic physics causes a redistribution of matter around each halo (galaxy) as a consequence of dissipation effects from star formation and feedback effects from supernovae or AGN. 
	We adopt the method in Refs.~\citep{schneiderNewMethodQuantify2015,2019JCAP...03..020S} to model the baryonic effects on the matter distribution around halos, which in turn alter the weak lensing profile. 
	We tuned the model parameters so that the model predictions reproduce the baryonic effects on the weak lensing profile in the {\tt Illustris} hydrosimulations \citep{2014Natur.509..177V} that implemented too large baryonic effects than implied by observations. Hence the \mock{{\tt baryon}} mocks are considered as a worst case scenario for the baryonic physics. 
	\item \mock{cent-incomp} -- This mock is intended to model an incomplete selection of galaxies against host halos. We follow the method in Ref.~\cite{moreWeakLensingSignal2015} (Eq.~5 in the paper) to implement this effect. This model includes a possibility that some fraction of even very massive halos might not host a SDSS galaxy due to an incomplete selection after the specific color and magnitude cuts. 
	We adopt $\alpha_{\rm inc}=0.5$ and $\log M_{\rm inc}=13.9$ ($M_{\rm inc}$ in units of $\hiMsun$) following the results in Ref.~\cite{moreWeakLensingSignal2015} (see Table~1 in the paper) for all the three samples. 
    This effect modifies the HOD shape, and therefore alters the numbers of central and satellite galaxies. 
\end{itemize}
Except for the \mock{cent-incomp} and \mock{FoF halo} mocks, all the mocks have exactly the same HOD and differ only in ways of populating galaxies into halos in each simulation realization. 

In Fig.~\ref{fig:galaxy-mock-signals} we show variations in $\dSigma$ and $\wproj$ for different mocks, relative to those for the \mock{fiducial} mocks.
It can be found that different mocks lead to quite different $\dSigma$ and $\wproj$ in the amplitudes and shapes compared to those of the \mock{fiducial} mock, even when employing the same HOD (except for the \mock{cent-incomp} and \mock{FoF halo} mocks, which essentially does not follow the same HOD as the \mock{fiducial} one). 
The changes for some of the mocks are found to be larger than the statistical errors for the HSC-Y1 and SDSS data.
We use these mocks to assess robustness of the PT-based method against the effects of uncertainties in the galaxy-halo connection. 

Fig.~\ref{fig:galaxy-mock-r} shows the cross-correlation coefficient for each mock, defined similar to Eq.~(\ref{eq:def_r_coefficient}) as $r_{\rm cc}^{(w)}(R)\equiv w_{\rm gm}/\sqrt{w_{\rm gg}w_{\rm mm}}$. 
Here $w_{\rm gm}(R)$ and $w_{\rm mm}(R)$ are the projected cross-correlation function between galaxies and matter and the projected auto-correlation of matter, which are defined similarly to Eq.~(\ref{eq:wp-def}) by using $P_{\rm gm}(k)$ and $P_{\rm mm}(k)$ instead of $P_{\rm gg}(k)$. 
However, note that we use the infinite projection length for $w_{\rm gm}$ as used in the lensing profile, while we use $100~\hiMpc$ for the projection length for $w_{\rm gg}$ and $w_{\rm mm}$. 
We measure these projected correlation functions from each of the different mocks, while $w_{\rm mm}$ is the same for all the mocks as it is measured from the original $N$-body simulations \mtrv{for the fiducial {\it Planck} cosmology.} 
Although the different mocks lead to different clustering signals in their amplitudes and shapes as shown in Fig.~\ref{fig:galaxy-mock-signals}, the cross-correlation coefficients are close to unity at large scales, $R\gtrsim 10~\hiMpc$. 
Note that deviation from unity at $R\gtrsim 40~\hiMpc$ is due to the different projection lengths as stated above, and is a geometrical effect; the theoretical template accounting for the different projection lengths can well reproduce the behavior. 
Thus, as long as modifications in the halo-galaxy connection are confined to local scales of galaxy formation in each host halo, a few Mpc at most corresponding to a size of most massive halos, they cannot impact the cross-correlation coefficients at scales sufficiently larger than the local scale. In other words, the large-scale clustering signals are governed solely by gravitational interactions, given the initial conditions that we assume to be adiabatic and Gaussian. These results imply that, as long as we use $\dSigma$ and $\wproj$ on large scales, we can recover 
the underlying matter correlation function $w_{\rm mm}$ by reconciling the galaxy bias uncertainty, and then use the reconstructed $w_{\rm mm}$ to estimate cosmological parameters, as we study more quantitatively below. 

\section{Results: validation and performance of minimal galaxy bias method for cosmology parameter inference}
\label{sect:result}

In this section, we show the main results; the robustness and precision of the PT-based method for estimating cosmological parameters from the clustering observables, $\dSigma$ and $\wproj$.

\subsection{Cosmological parameter dependence of observables}
\label{subsect:result:fiducial-mock}

\begin{figure}[h]
	\centering
    \includegraphics[clip,width=0.95\columnwidth]{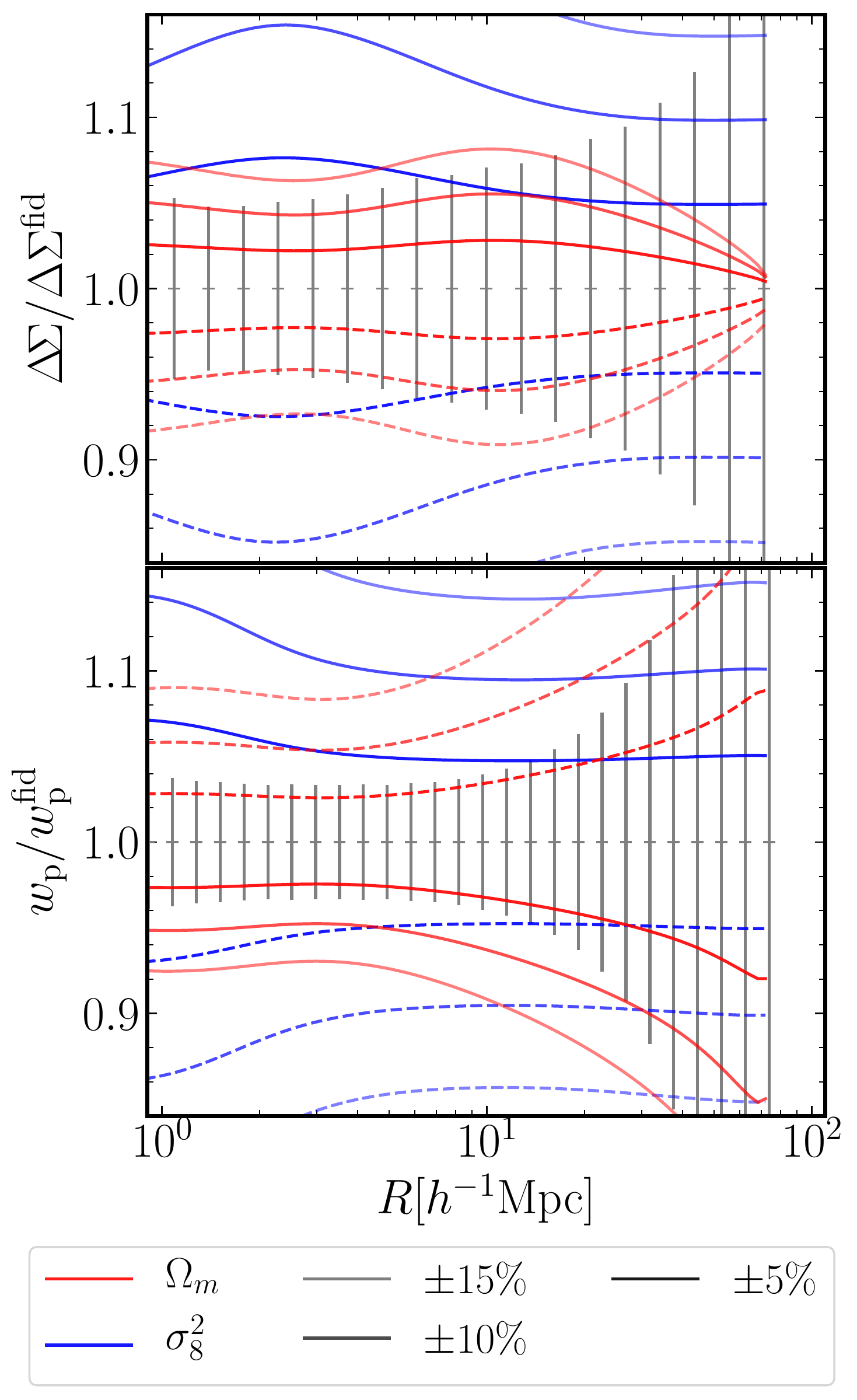}
	\caption{
		Dependences of $\dSigma$ and $\wproj$ on the cosmological parameters, $\Omegam$ (red) and $\sigma_8$ (blue). Shown is the ratio of the model prediction relative to that for the fiducial model. When we change one parameter, we kept the other parameter fixed to the fiducial value. 
		Here we show the observables at $z=0.251$ corresponding to the representative redshift of LOWZ sample. Note that we consider the linear bias parameter (baseline model), but  do not vary the bias parameter for different models; the $b_1$-dependence cancels out in the ratio.
		The solid or dashed lines of each color show the fractional changes in $\dSigma$ and $\wproj$ when increasing or decreasing either parameter from the fiducial value. 
		From thicker to thinner respective lines, we change \ssrv{the cosmological parameter, $\Omegam$ or $\sigma_8^2$,} by $\pm 5$, 10 and 15\% in fraction from the fiducial value, respectively.
		For comparison we also show the error bars that are computed from the covariance matrices, where we scaled the covariance matrix of $\dSigma$ by 
		0.1 for illustrative purpose.
		}
	\label{fig:cosmo-dep-dswp}
\end{figure}
Before going to the main results, we show the cosmological parameter dependences of $\dSigma$ and $\wproj$.
In Fig.~\ref{fig:cosmo-dep-dswp} we study how changes in the cosmological parameters ($\sigma_8$ or $\Omega_{\rm m}$) alter the clustering observables at $z=0.251$ (the redshift of the LOWZ sample) relative to those for the fiducial {\it Planck} cosmology. 
We vary either of \ssrv{$\sigma_8^2$} or $\Omega_{\rm m}$ alone by $\pm 5$, 10, or 15\%, respectively, and fix the other to its fiducial value. 
Here we consider the linear bias model ($b_1$ terms alone in Eqs.~\ref{eq:Pgm} and \ref{eq:Pgg}), and do not change its value so that the $b_1$-dependence cancels out in the ratio. 
The figure shows that a change in $\sigma_8$ causes an overall shift in the clustering amplitudes on large scales in the linear regime, but causes a scale-dependent change in the observables at nonlinear scales via the dependence on the nonlinear matter power spectrum $P_{\rm NL}$. 
On the other hand, a change in $\Omegam$ causes a characteristic scale-dependent change in the observables over all the scales we consider. 
For comparison, we show the statistical errors expected for the HSC-Y1 and SDSS data, which are the square root of the diagonal components of the covariance matrix for each observable. Note that the neighboring bins are highly correlated with each other, and we multiply the covariance matrix of $\dSigma$ by 0.1 for illustrative purpose. 

\begin{figure}[h]
	\centering
    \includegraphics[clip,width=0.95\columnwidth]{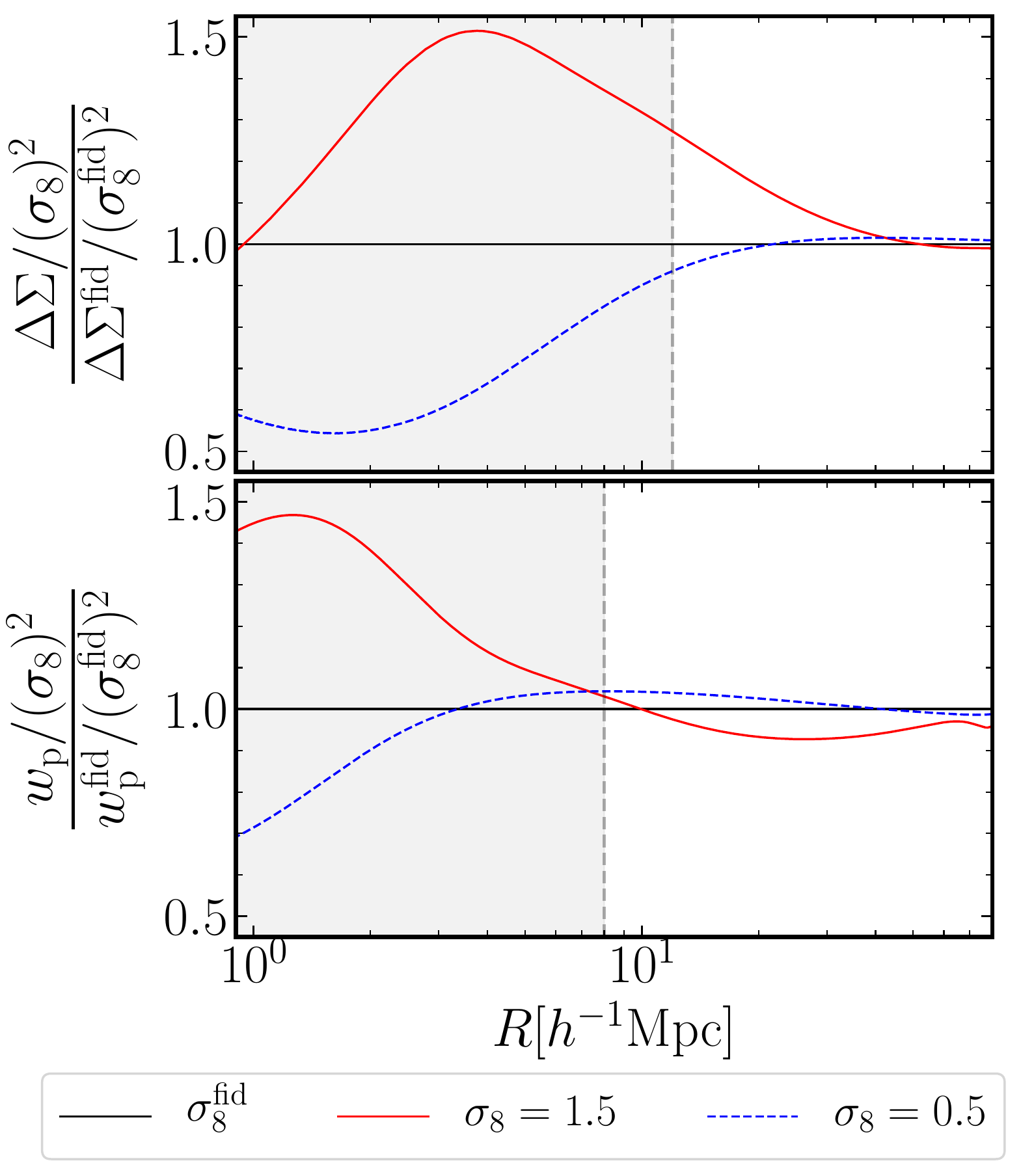}
	\caption{
		Dependence of $\dSigma$ and $\wproj$ on $\sigma_8$, as predicted by our baseline model. 
		Here we fixed $\Omegam$ to its fiducial value, and we do not vary the linear bias parameter as in the previous figure. 
        We plot the ratio in each observable relative to the linear theory dependence on $\sigma_8$ that is given by $\dSigma \propto \sigma_8^2$ and $\wproj \propto \sigma_8^2$. 
		That is, the linear theory predicts that all the lines 	give unity, and a deviation from unity is due to the nonlinear clustering effects (via $P_{\rm NL}$ in our model).
		An increase in $\sigma_8$ leads to greater amplitudes in $\dSigma$ an $\wproj$. More importantly, a change in $\sigma_8$ causes a scale-dependent change in these observables, even at scales greater than the scale cut (unshaded region), although the scale-independent change is naively expected at such large scales as predicted by the linear theory. Note that all curves got to unity at very large scales, far beyond 100~$\hiMpc$.
		}
    \label{fig:phalo-transition}
\end{figure}
\begin{figure*}
	\begin{center}
		\includegraphics[clip,width=0.95\textwidth]{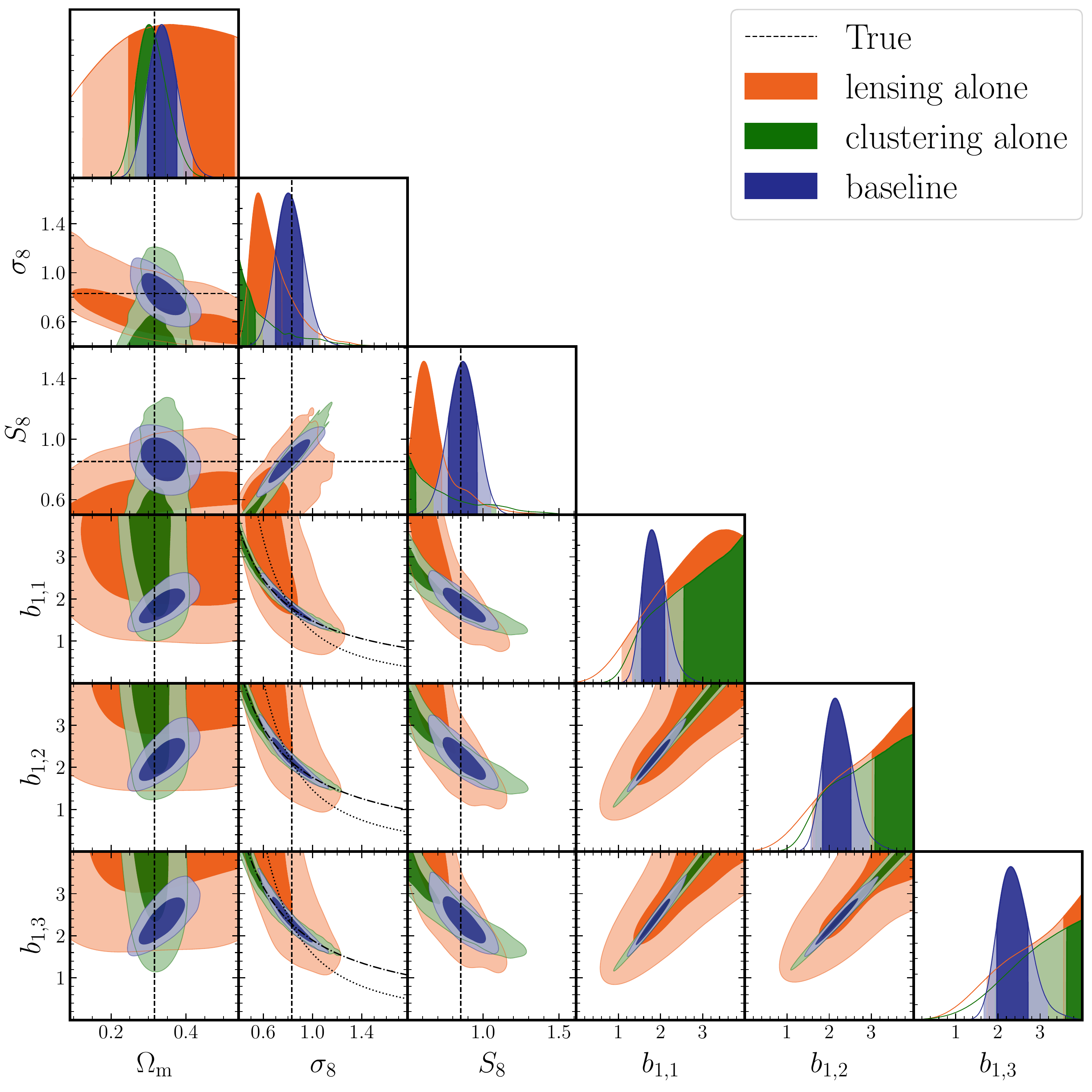}
		\caption{
			Marginalized posterior distributions for all the sampled parameters, obtained from the lensing information alone (\setup{$\dSigma$ alone}; 
			orange contours), the projected clustering information alone (\setup{$\wproj$ alone}; green), and the joint information (\setup{baseline}; blue), respectively. 
			We used the baseline setup for the scale cuts, $12$ and $8~\hiMpc$, for $\dSigma$ and $\wproj$, respectively.
            The dashed lines denote the input values of $\Omegam$ and $\sigma_8$, which correspond to the {\it Planck} cosmology used in the mock galaxy catalogs. The parameter, $S_8\equiv \sigma_8(\Omegam/0.3)^{0.5}$, is a derived parameter computed from the chains. 
            In the panels of $(\sigma_8,b_{1,i}) (i=1,2,3)$ for each of the three galaxy samples, 
            the dotted and dot-dashed lines display $b_{1,i}\sigma_8^2={\rm const.}$ and 
            $b_{1,i}^2\sigma_8^2={\rm const.}$, which denote the degeneracy directions in the amplitudes of 
            $\dSigma$ and $\wproj$, respectively, in the linear theory predictions. The lines are shown around the input value of $\sigma_8$
            for the {\it Planck} cosmology ($\sigma_8=0.831$) and $b_1=1.78$, 2.12 or $2.28$ for each of the three galaxy samples that are
            the linear bias parameters estimated from the same mock catalog in Ref.~\cite{2019arXiv190708515K} (Table~III of the paper).
		}
		\label{fig:comp-baseline-wp-dsigma}
	\end{center}
\end{figure*}
In Fig.~\ref{fig:phalo-transition}, we further investigate how a change in $\sigma_8$ causes scale-dependent changes in the observables.
\ssrv{In each plot for $\dSigma$ and $\wproj$, the red-solid and blue-dashed lines show the \mtrv{changes in the signals}}
when assuming quite small or large value of $\sigma_8$, $\sigma_8=0.5$ or 1.5, instead of the fiducial value $\sigma_{8,{\rm fid}}=0.83$, while the black-solid line is the result for the fiducial model. Note that we fix the other parameters to the fiducial values. 
The \ssrv{figure} shows non-trivial, interesting results. Recalling that linear theory predicts $\dSigma\propto \sigma_8^2$ and $\wproj\propto \sigma_8^2$, we show the ratios of the clustering observables, normalized by the linear-theory prediction, to those for the fiducial model. 
The linear theory predicts unity for the ratio, so a deviation from unity arises from the nonlinear dependence of the clustering observables on $\sigma_8$ via the dependence on $P_{\rm NL}$. 
We checked that the ratio goes to unity at very large scales beyond the plotting range in this figure. 
For reference, the gray-shaded region denotes the scale cuts of the \setup{baseline} setup (Table~\ref{table:fiducial-mock-setup}); we use the clustering information at scales in the non-shaded region for parameter estimation. 
The figure clearly shows that the increase or decrease in $\sigma_8$ from its fiducial value causes a scale-dependent change in the observables even at large scales greater than the scale cut, which is also confirmed if we use the PT prediction for the nonlinear matter power spectrum, instead of $P_{\rm NL}$, to compute the model predictions (see Fig.~\ref{fig:signal-plin-pnlin-pspt}).
Thus the transition scale\footnote{In Fourier space the transition scale is roughly given by $k_{\rm NL}$ satisfying the condition $\left.k^3P_{\rm L}(k)/2\pi^2\right|_{k=k_{\rm NL}}=1$. 
Roughly speaking, the transition scale in real space is given by $r_{\rm NL}\sim 1/k_{\rm NL}$.} to divide the linear and nonlinear scales is quite sensitive to the assumed $\sigma_8$.
For example, for $\sigma_8=0.5$, the transition scale moves to a smaller scale compared to that for the fiducial model, and consequently the clustering observables in the non-shaded region respond to $\sigma_8$ as predicted by the linear theory (a subtle scale dependence is due to slightly stronger nonlinear effect in the fiducial model, especially around the scale cut). 
On the other hand, the increase in $\sigma_8$ causes a  complicated scale-dependent change in the observables at these scales. This means that a large $\sigma_8$ causes additional scale-dependent changes in the observables due to more significant nonlinear effects.
Thus the scale cut depends on the genuine $\sigma_8$-value of the underlying true cosmology, and this should be kept in mind. 

In \figref{fig:comp-baseline-wp-dsigma} we show the marginalized posterior distribution in each of 1D or 2D parameter space when comparing the theoretical templates of the baseline setup (Table~\ref{table:fiducial-mock-setup}) with the mock signals measured from the fiducial mock catalogs. 
The orange and green distributions show the results when using either of $\dSigma$ or $\wproj$ alone, and both case display severe degeneracies. This can be understood as follows. 
If we use the linear theory prediction, instead of our baseline method using the nonlinear matter power spectrum, either of the two observables alone leaves perfect degeneracies given by $b_1\sigma_8^2$ or $b_1^2\sigma_8^2$, because different models keeping $b_1\sigma_8^2$ or $b_1^2\sigma_8^2$ fixed to the same value give the same amplitudes in $\dSigma$ and $\wproj$ (for a fixed $\Omegam$) and such models are not distinguishable by either of $\dSigma$ or $\wproj$ alone. 
The dotted and dot-dashed lines in each plane of $(\sigma_8,b_1)$ display $b_1\sigma_8^2={\rm const.}$ or $b_1\sigma_8^2={\rm const.}$, respectively, which nicely reproduce the degeneracy directions of the posterior distributions.
However, we adopt the fully nonlinear matter power spectrum, $P_{\rm NL}$, in the baseline method, and this model causes a stronger dependence on $\sigma_8$ even over the fitting range of scales above the scale cut, if the input value of $\sigma_8$ becomes sufficiently large; that is, for such large-$\sigma_8$ models, the transition scale to divide the linear and nonlinear scales enter into the fitting range of scales as explained in Fig.~\ref{fig:phalo-transition}.
This nonlinear dependence of $\sigma_8$ leads to a cutoff in the posterior distributions at models with large $\sigma_8$. For this reason, the marginalized 1D posterior distribution has a skewed distribution extending to the boundary of each parameter at the lower end of $\sigma_8$ (Table~\ref{tab:prior-range}), and an apparent peak of some 1D posterior distributions is artificial due to the boundary effect of the prior. 
In summary, either of $\dSigma$ and $\wproj$ alone cannot lead to any meaningful constraints on the model parameters.

Hence {\it only} if combining the two observables $\dSigma$ and $\wproj$, we can break the parameter degeneracies and simultaneously constrain the model parameters, as displayed by the blue contours/distributions.
For the current size of error bars in the observables, $\wproj$ has more constraining power than $\dSigma$ does. 
This can be found from the blue contour in the ($\sigma_8,b_1$) plane, which shows an elongated distribution along the degeneracy direction of $\wproj$, i.e. $b_1^2\sigma_8^2={\rm constant}$.
However, a closer look reveals that the posterior distribution for the joint constraints still display  an asymmetric distribution towards models having lower $\sigma_8$ values. 
This means that the input value is not necessarily perfectly recovered even if only the large-scale clustering information is used, after projection or marginalization of the asymmetric (``banana''-shaped)  posterior distribution in a multidimensional parameter space, as we will explicitly study this in the next section. 
We should keep in mind this caveat, and in the following we show only the posterior distribution for the joint constraints of $\dSigma$ and $\wproj$.

\subsection{Validation for the fiducial mocks and baseline setup}
\label{subsubsect:result:fiducial:baseline-analysis}

\begin{figure*}
	\centering
	\includegraphics[clip,width=0.95\textwidth]{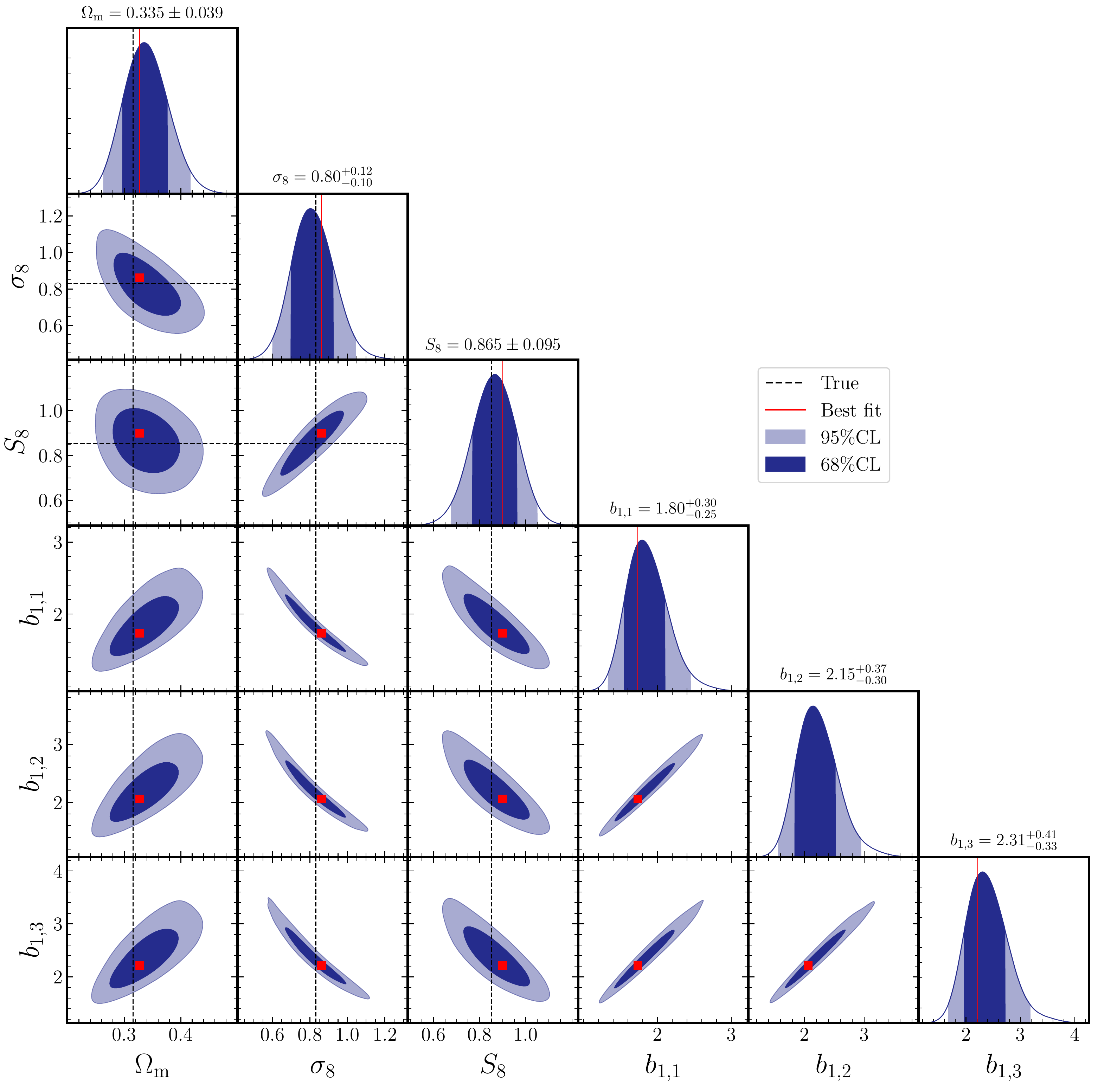}
	\caption{
		The 1D and 2D posterior distributions of the parameters for the baseline analysis, where the distribution includes marginalization over other parameters. 
		The number with error at the top of each of the upper panels denotes the central value and the 68\% credible interval for each parameter, respectively. Dotted line indicates the true value of each cosmological parameter, i.e. \mtrv{an input parameter used in}  the $N$-body simulations.
		The red points and lines show the best-fit parameters, \mtrv{corresponding to the model that has the highest likelihood value.} 
    We define $S_8\equiv\sigma_8(\Omegam/0.3)^{0.5}$.
		}
	\label{fig:corner-baseline}
\end{figure*}
\begin{figure*}
	\centering
	\includegraphics[clip,width=0.95\textwidth]{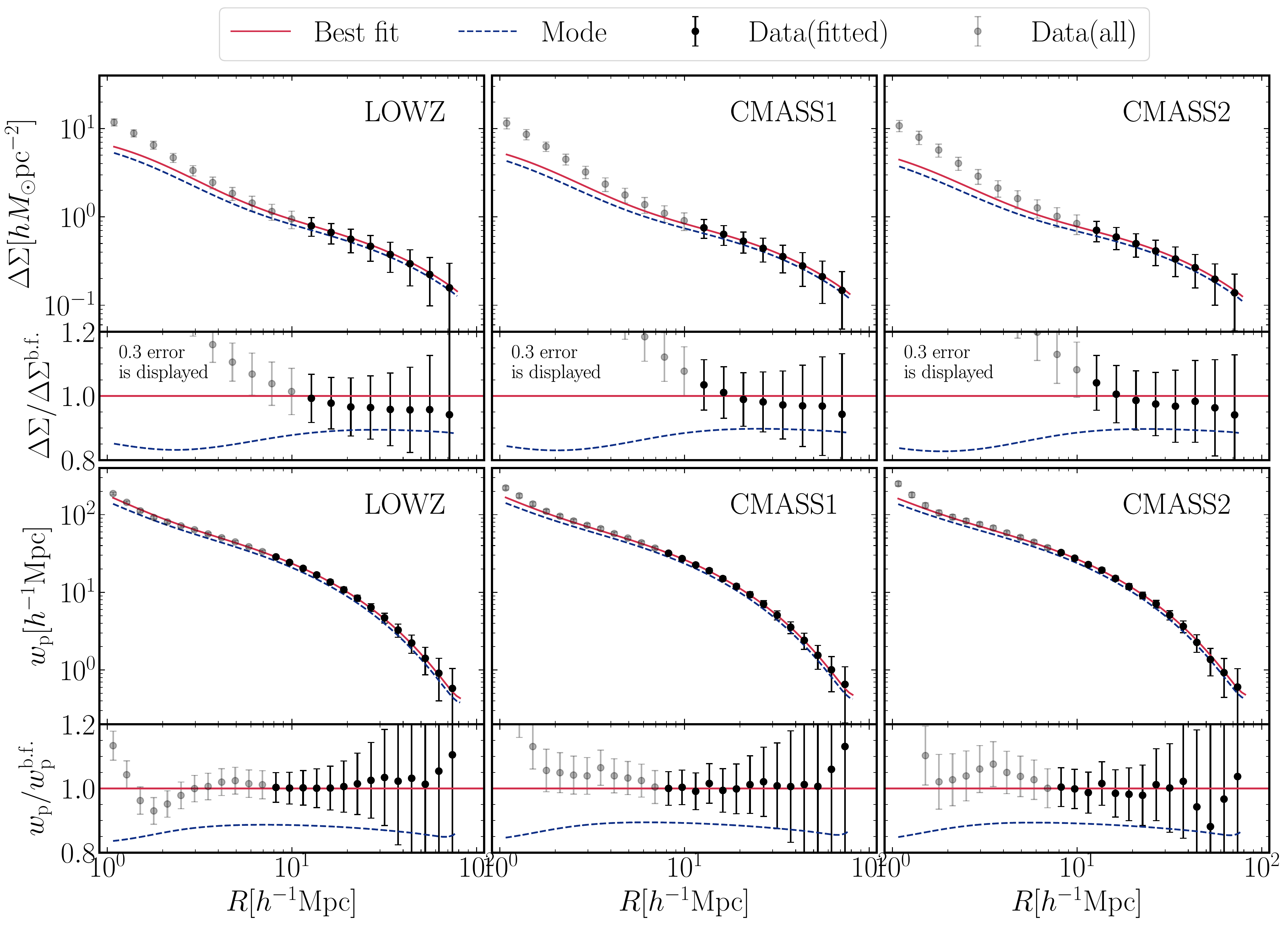}
	\caption{
		Comparison of the best-fit model predictions (red solid) with the mock signals (black points) for each of the galaxy samples, for the baseline analysis. 
		The black points are the mock data at scales greater than the scale cut, which are used in the parameter inference. The gray points are data points under the scale cuts.
		The error bars in each bin are the expected errors for the SDSS and HSC-Y1 data, but in the lower panel we show the smaller error bars computed by multiplying the covariance of $\dSigma$ by 0.1 for illustrative purpose. 
		For comparison, the blue-dashed line in each panel shows the model prediction, where we use the mode value of the marginalized 1D posterior distribution of each model parameter (the central value of each parameter dented in Fig.~\ref{fig:corner-baseline}).   
	}
	\label{fig:baseline-signal}
\end{figure*}
\begin{figure}[h]
	\centering
	\includegraphics[clip,width=0.95\columnwidth]{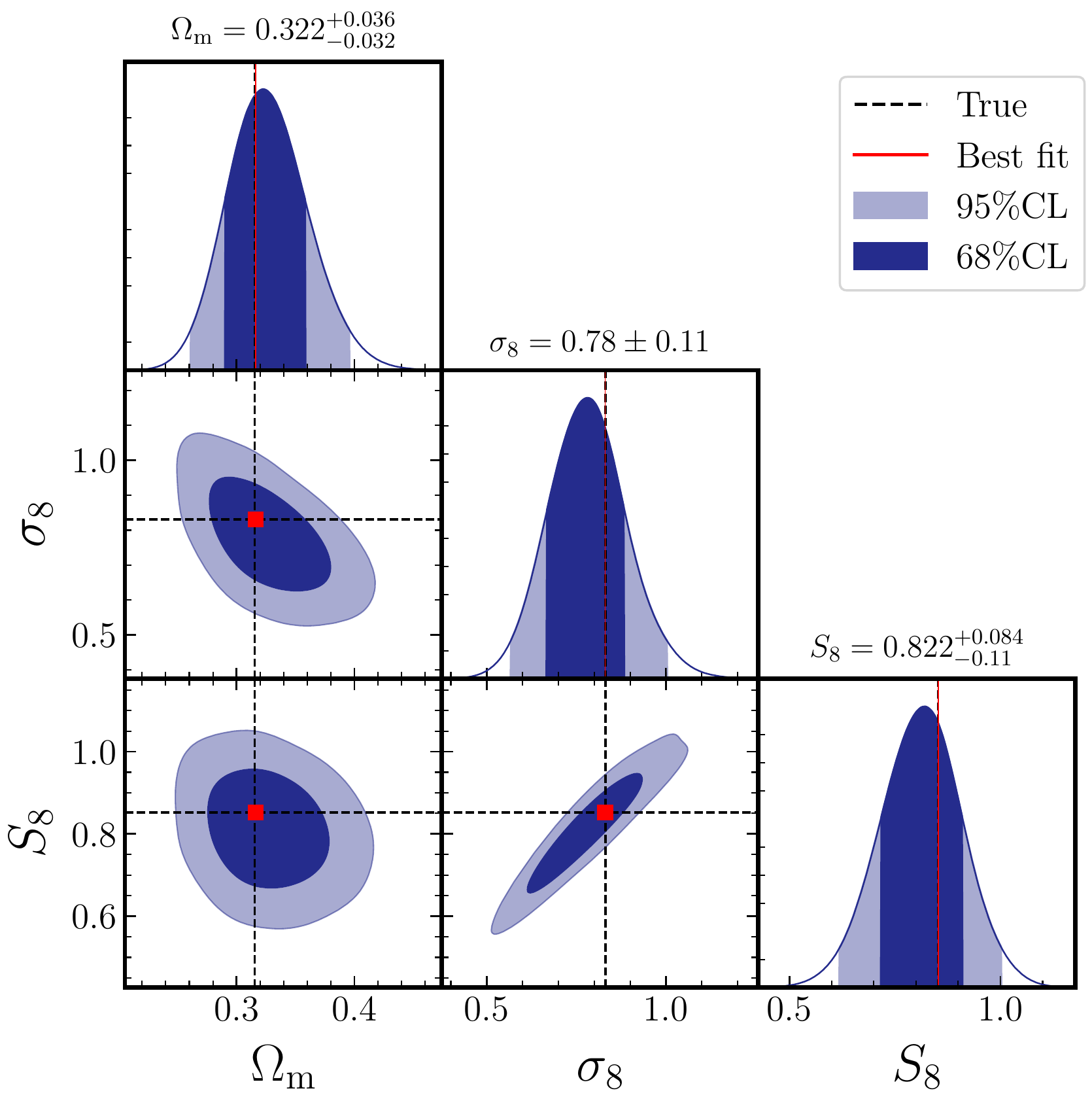}
	\caption{
        Similar to Fig.~\ref{fig:corner-baseline}, but the figure shows the results if we use the model predictions for the fiducial model as the input data vector in the parameter estimation, instead of the simulated signals from the mock catalogs. 
        Here we adopt the minimal-bias model in both the data vector and the model predictions; for the input data vector, we employ the model predictions  assuming the {\it Planck} cosmology and the linear bias parameters, $b_{1,1}=b_{1,2}=b_{1,3}=2$ for the three galaxy samples. 
        }
	\label{fig:corner-ptinput}
\end{figure}
\begin{figure}[h]
	\centering
	\includegraphics[clip,width=0.95\columnwidth]{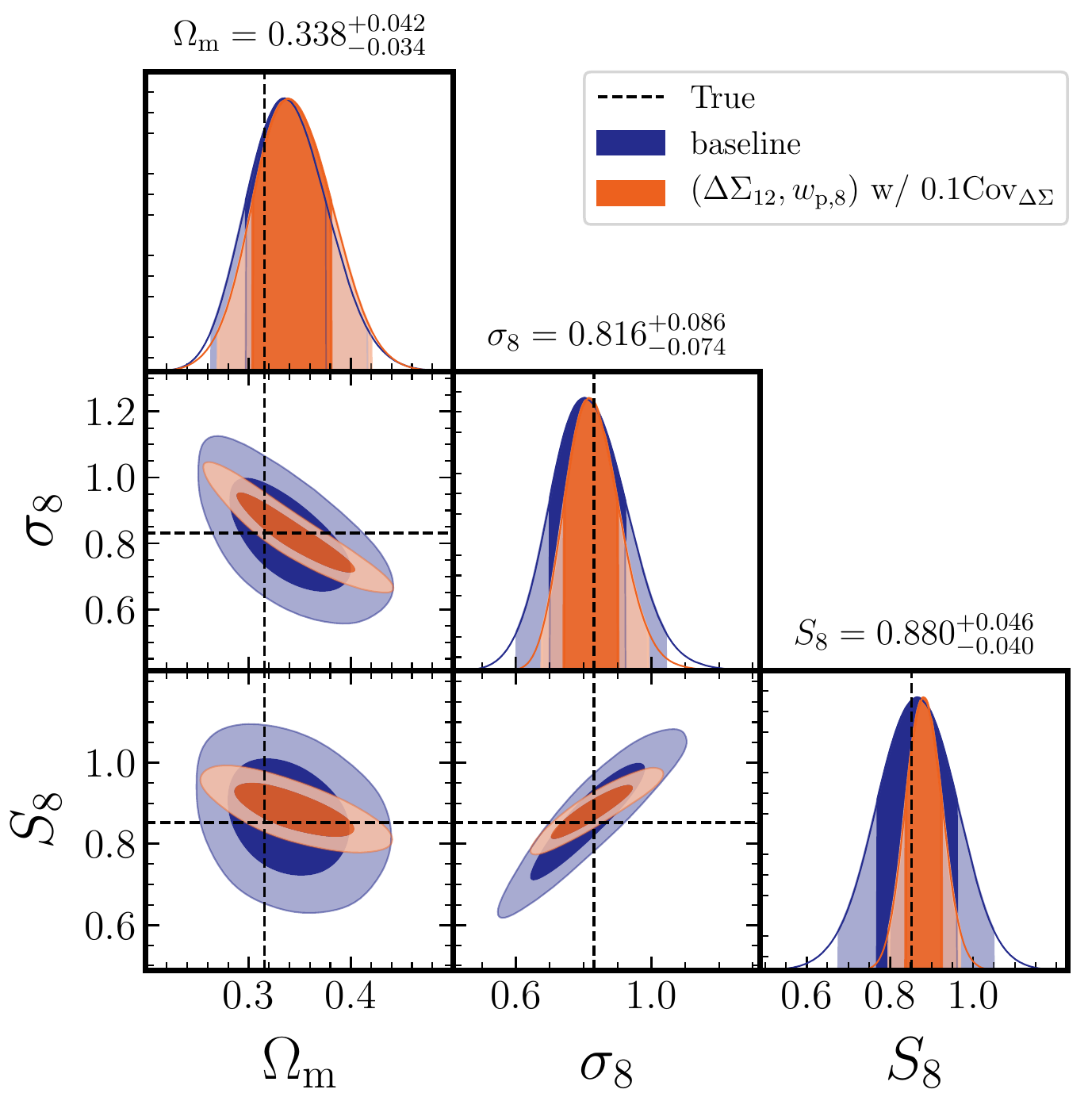}
	\caption{
		Similar to Fig.~\ref{fig:corner-baseline}, but the posterior distributions obtained by using the expected full HSC data that has a factor of 10 larger sky coverage than the HSC-Y1 data. Here we simply employ the reduced covariance matrix of $\dSigma$ by a factor of 10 in the parameter estimation. 
		}
	\label{fig:corner-01cov}
\end{figure}
\begin{figure*}
	\centering
    \includegraphics[clip,width=1.0\textwidth]{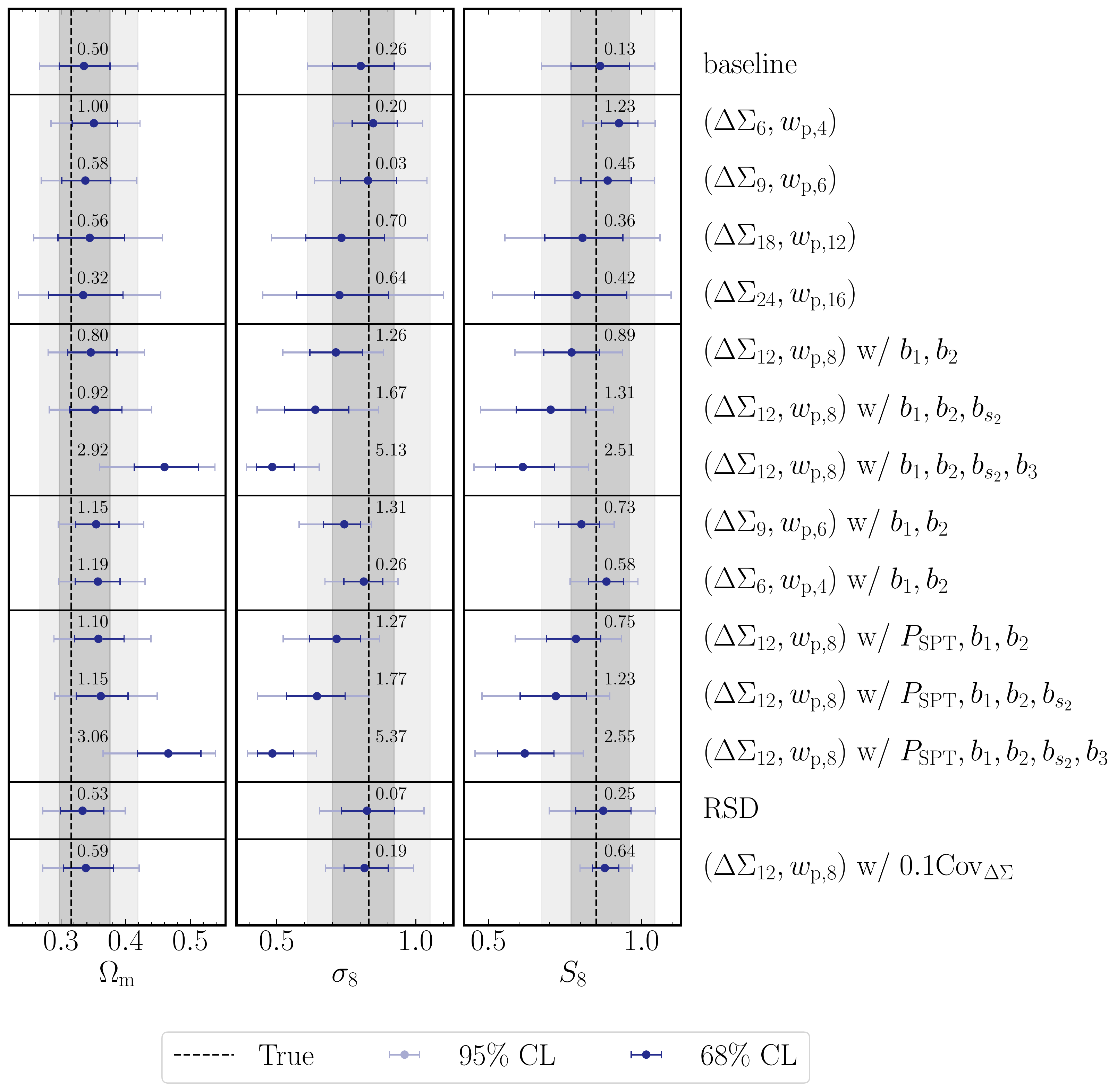}
	\caption{
		Constraints on the cosmological parameters for the SDSS and HSC-Y1 data, and the performance and robustness against the different analysis setups and the different bias models (see Table~\ref{table:fiducial-mock-setup}).
		For each column the circle symbol denotes the central value obtained from the marginalized posterior, and the thin and thick error bars denote the 68\% and 95\% credible intervals, respectively. 
		\ssrv{The numeric value shown above each error bar is the relative size of systematic bias, defined in Eq.~(\ref{eq:relative-bias}).}
		\ssrv{For comparison the dark- and light-gray shaded regions denote the 
    the 68\% and 95\% credible intervals for the ``baseline'' method.} 
		Note $S_8=\sigma_8(\Omegam/0.3)^{0.5}$. The vertical dashed line denotes the true value of each parameter used in the mock catalog. 
		For comparison, we also show the result of ``baseline'' analysis by gray shaded regions.
		}
	\label{fig:fiducial-parameter-constraint}
\end{figure*}

We now show the main results of this paper. 
In Fig.~\ref{fig:corner-baseline} we show the 1D and 2D posterior distributions of the parameters for the joint information of $\dSigma$ and $\wproj$, obtained by applying the baseline setup to the \mock{fiducial} mock. 
Throughout this paper, we adopt  the mode value and the highest density interval to infer the central value(s) and the marginalized credible interval, respectively, in both 1D and 2D posterior distributions. 
For comparison, we also display the best-fit parameters corresponding to the model that has the highest likelihood value among the obtained chains.
Note that the mode and best-fit models have some realization-by-realization scatters in each \code{Multinest} run, typically by about 0.02$\sigma$ and 0.1$\sigma$, respectively, where $\sigma$ is the 68\% credible interval.
Since the clustering probes are the most sensitive to the combination of $\sigma_8$ and $\Omegam$ \citep{hikageCosmologyCosmicShear2019}, we also show the result for $S_8\equiv \sigma_8(\Omegam/0.3)^{0.5}$, which is a derived parameter.
Here we simply follow Ref.~\cite{troxelDarkEnergySurvey2018} for the definition of $S_8$ to make it easier to compare our results with other existing results such as Refs.~\cite{troxelDarkEnergySurvey2018,Hikage19}, without considering e.g. the principal components of the parameter covariance drawn from the observables $\dSigma$ and $\wproj$.

First of all, the baseline setup recovers the true values of the cosmological parameters, used in the simulations/mock catalog, to within the 68\% credible intervals. 
Hence we conclude that the baseline setup, using the scale cuts of $12$ and $8~\hiMpc$ for $\wproj$ and $\dSigma$, respectively, can be safely applied to actual data, if the underlying cosmology is not far from the {\it Planck} cosmology. 
The HSC-Y1 and SDSS data allow for about 10\% precision in the fractional error of $S_8$. 
However, a closer look reveals that the central value for each of $\Omegam$ or $\sigma_8$ is slightly biased from the true value. 
The slight bias in $\Omegam$ can be understood as follows.  
Fig.~\ref{fig:baseline-signal} compares the model predictions for the best-fit model with the mock signals. Recalling that the cosmological constraints are mainly from the $\wproj$ information, the best-fit model is driven by the mock signal of $\wproj$ around the scale cut, $R\simeq 8~\hiMpc$ because of the highest signal-to-noise ratios at the scale.
Then the mock signal has a steeper slope at the scales just below the scale cut than the best-fit model predicts. This means that adopting smaller scale cuts leads to a higher $\Omegam$ than the input value to compensate this scale-dependent discrepancy as indicated by the $\Omegam$-dependence of $\wproj$ for a fixed $\sigma_8$ in Fig.~\ref{fig:cosmo-dep-dswp}.

The bias in $\sigma_8$ is due to the degeneracies of $\sigma_8$ with other parameters (mainly the $b_1$ parameters) or after marginalizing the asymmetric shape (``banana''-like shape) of posterior distribution in a multidimensional parameter space.
As explained in Fig.~\ref{fig:comp-baseline-wp-dsigma}, there are severe degeneracies between the $b_1$ parameters and $\sigma_8$ in the joint probes constraints, which are roughly given by $b_1^2\sigma_8^2={\rm constant}$. When we marginalize the posterior distribution over the $b_1$ parameters, the marginalized 1D posterior distribution of $\sigma_8$ is skewed towards lower $\sigma_8$, because the original ``banana-shaped'' 
posterior distribution has larger cross sections at the intersection of lower $\sigma_8$.

The positive and negative differences between the central and input values of $\Omegam$ and $\sigma_8$ cancel out to some extent in the constraint of $S_8$. 
However this cancellation does not always occur as we will show below. The degree of cancellation depends on the relative constraining power of $\dSigma$ and $\wproj$.

In Fig.~\ref{fig:corner-ptinput}, we show that
these biases in $\Omegam$ and $\sigma_8$ occur even when we use the model predictions as the data vector in the parameter estimation
\mtrv{(that is, a perfect case where the input mock signal is exactly the same as the model prediction for the input model).}
Here we generate the input data assuming the {\it Planck} cosmology and $b_{1,1}=b_{1,2}=b_{1,3}=2$ in the baseline model.
The figure shows that the central value of each parameter has an offset from the input value, in the same direction as that in Fig.~\ref{fig:corner-baseline}.
Thus, even if we consider an ideal situation for the parameter constraint, there is no guarantee that we can recover the input values due to the nonlinear parameter dependences.
For illustration, the blue-dotted line in each panel of Fig.~\ref{fig:baseline-signal} shows the model prediction where we use the central value (mode) of the marginalized 1D posterior of each model parameter. The model prediction is far off from the best-fit model and the mock signal, meaning that the central value is biased from the true value or the best-fit value after marginalization of the ``banana''-shaped posterior distribution in a multidimensional parameter space. 
For this reason, we should keep in mind that a point estimate of parameter such as the mode value in the marginalized posterior might not be useful. 
The credible interval is more important, and we give a validation of a given method if the credible interval includes the true value.

In Appendix~\ref{apdx:linear_treePT} we show the performance of the methods using the linear or PT theory predictions to model the matter power spectrum, instead of $P_{\rm NL}$ in the minimal bias model.
We do not find any strong advantage for these methods over our baseline method using $P_{\rm NL}$.
In fact, the baseline method gives the smallest bias in $S_8$. Hence we conclude that the baseline method is a reasonably good method, even though it is an empirical method.

We should note that the results of Fig.~\ref{fig:corner-baseline} are for a particular choice of the error covariance matrices that mimic the SDSS and the HSC-Y1 data, where the signal-to-noise ratio of $\dSigma$ is smaller than that of $\wproj$, by about a factor of 4--5 as shown in Table~\ref{table:signal-to-noise-ratio}. 
Since the sky coverage of \survey{HSC-Y1} is about 10\% of the area that we anticipate for the full \survey{HSC} survey \citep{2018PASJ...70S...4A}, we expect a huge improvement in the signal-to-noise ratio of $\dSigma$ in coming years. 
In Fig.~\ref{fig:corner-01cov}, we study how the full HSC dataset can improve the cosmological constraints. To study this, we use the lensing covariance matrix reduced by a factor of 0.1, but use the same setups for other quantities as those in the baseline analysis. 
The figure shows that the posterior distributions significantly shrink.
The bias in $\sigma_8$ is significantly reduced because the asymmetry of posterior distribution in multidimensional parameter space is mitigated, while the bias in $\Omegam$ still remains to be in the similar direction and amount compared to those in Fig.~\ref{fig:corner-baseline} (also see Fig.~\ref{fig:fiducial-parameter-constraint}). 
These results mean that the improved lensing signal of the expected HSC full dataset mainly improves the constraint on $\sigma_8$, and the constraint of $\Omegam$ is still mainly from $\wproj$. 
Consequently, we can achieve a $\sim 4\%$ constraint on $S_8$, corresponding to a factor of 2.5 improvement compared to that of the HSC-Y1 data. 
However, we note that $S_8$ suffers from a larger bias than that in the baseline setup, because only the bias in $\sigma_8$ is improved in the ``$(\Delta\Sigma_{12},w_\mathrm{p,8})$ w/ $0.1\mathrm{Cov}$'' setup compared to the baseline setup and biases in $\Omegam$ and $\sigma_8$ do not fully cancel out in $S_8$.

\begin{figure*}
	\centering
	\includegraphics[width=1.0\textwidth]{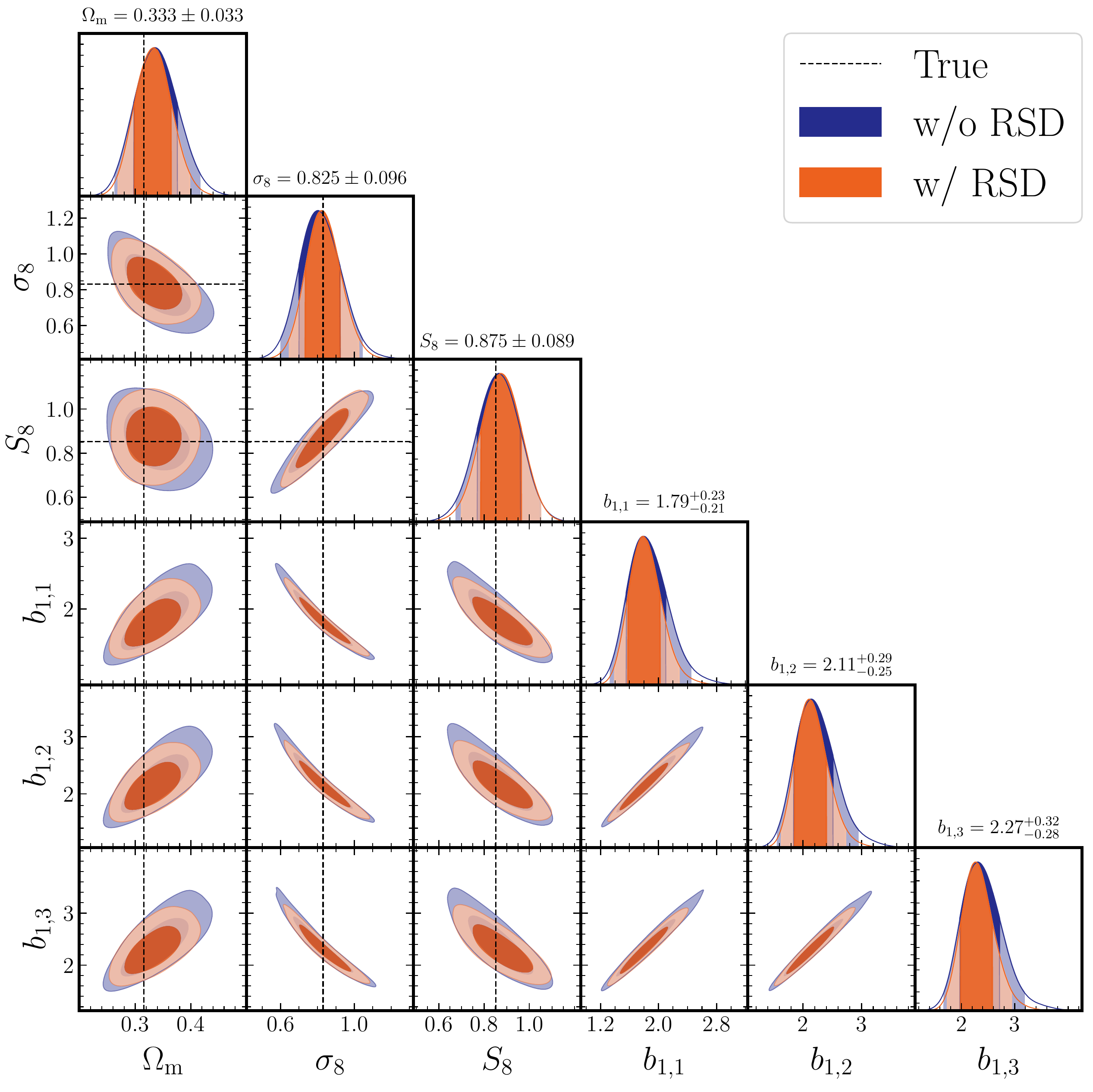}%
	\caption{
		The 1D and 2D posterior distributions, obtained by applying the baseline method to the mock catalog including the RSD effect. 
		Here we include the linear Kaiser factor to correct for the RSD effect in the theoretical template of $\wproj$, for a projection length of $100~\hiMpc$ (see text for details). Note that $\dSigma$ is not changed. 
		For comparison, the blue-color distributions are the same as those in Fig.~\ref{fig:corner-baseline}, i.e. the results for the mock without the RSD effect.
		The central value and 68\% interval shown in the top of each diagonal panel are the result for the mock with RSD effect.
	}
	\label{fig:corner-RSD}
\end{figure*}
\begin{figure*}
	\centering
	\includegraphics[width=1.0\textwidth]{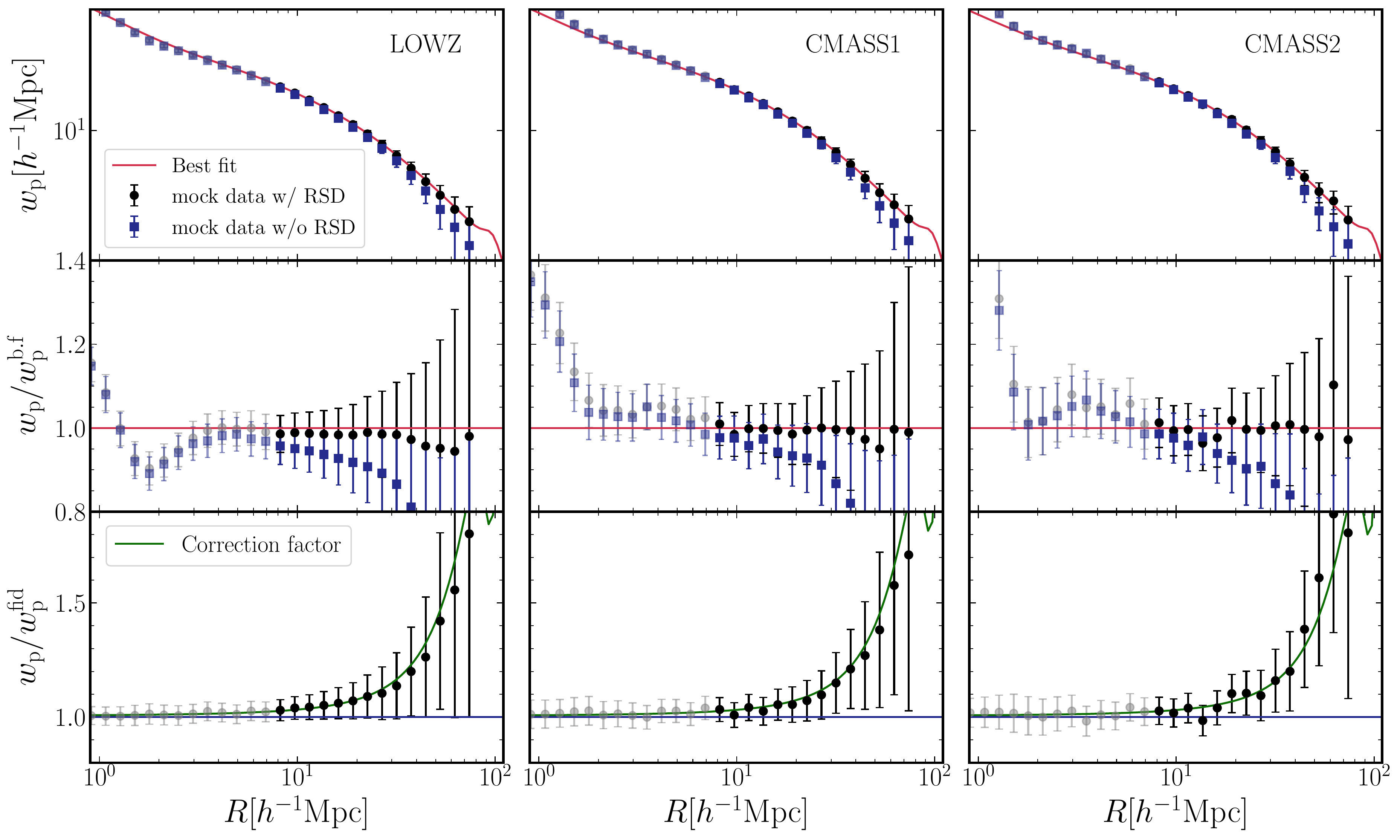}%
	\caption{
		Comparison of the best-fit model predictions of $\wproj$ (red-line) with the mock signals with the RSD effect, for each galaxy sample. For comparison, we also show the mock signals without the RSD effect. 
		In the bottom panels, we compare the best-fit model prediction of the linear Kaiser correction factor with the ratio of the mock signals with and without the RSD effect. The error bars in each separation bin are computed from the diagonal components of the covariance matrix of $\wproj$ for each sample. 
	}
	\label{fig:RSD-fit-signal}
\end{figure*}
In Figs~\ref{fig:fiducial-parameter-constraint} and \ref{fig:corner-RSD}, we also show the performance of the baseline setup when it is applied to the mock catalog including the RSD effect (the result, labeled as ``RSD'', in Fig.~\ref{fig:fiducial-parameter-constraint}). 
We here employ the method in Ref.~\cite{vandenboschCosmologicalConstraintsCombination2013} to include the RSD effect in the theoretical template of $\wproj$ (see around Eq.~47 in their paper). 
We multiply the ``RSD correction factor'' with our real-space model prediction in each $R$ bin, which is computed from the linear theory prediction assuming the Kaiser RSD effect.  The correction factor is not negligible for a finite projection length ($100~\hiMpc$ in our case) and for a large separation ($R$), and depends on redshift and a combination of the linear growth rate and the linear bias, usually denoted as $\beta\equiv f/b_1$, where $f\equiv \mathrm{d}\ln D_+/\mathrm{d}\ln a\simeq \Omegam(z)^{0.55}$. 
Hence including the RSD correction factor adds a sensitivity of the theoretical template to $\Omegam$ for flat $\Lambda$CDM model. 
This is the main reason why the RSD mock gives slightly smaller credible intervals (smaller error bars) of the cosmological parameters than those without the RSD effect. Note that a slight improvement in $\sigma_8$ is due to the fact that the parameter degeneracies are better broken by an improved constraint of $\Omegam$. 
We should note that the RSD mock catalog includes the full nonlinear effect; the nonlinear Kaiser effect and the finger-of-God effect due to the virial motions of galaxies \citep{kobayashiCosmologicalInformationContent2019}. 
How well does the linear Kaiser correction factor capture the RSD effect in the mock catalog? To answer this question, Fig.~\ref{fig:RSD-fit-signal} compares the best-fit model prediction with the mock signals for $\wproj$ for each galaxy sample. 
In particular, the bottom panel compares the correction factor for the best-fit model with the ratio of the mock signals with and without the RSD effect. The figure clearly shows that the linear Kaiser factor properly reproduces the simulation results within error bars. 
Hence we conclude that, for a projection length of $100~\hiMpc$, we can accurately model the RSD effect in the measured $\wproj$.

\subsection{Validation against scale cuts, PT methods, and bias models}
\label{subsubsect:result:fiducial:cutoff-and-bias}

We now show the performance against different scale cuts, different bias models, and different PT methods.
\ssrv{Fig.~\ref{fig:fiducial-parameter-constraint} gives the central value (blue circle symbol), and the 68\% and 95\% CL intervals 
\mtrv{(light- and dark-blue error bars),}
respectively, for each setup. We \mtrv{quantify a systematic bias in the cosmological parameter, estimated for each setup, by}
\begin{align}
	\Delta p \equiv \frac{|p^{\rm mode}-p^{\rm true}|}{68\%{\rm HDI}(p)/2}, \label{eq:relative-bias}
\end{align}
where $p^{\rm mode}$ is the mode value of parameter $p$, $p^{\rm true}$ is the true value, 
and $68\%{\rm HDI}(p)$ is the 68\% highest density interval (credible interval).  
Note that the systematic bias quantified in this way \mtrv{could arise from two contributions due to an inaccuracy in the model (theoretical template)
and due to the degeneracies with other parameter(s) or the effect of marginalization of asymmetric posterior distribution in multidimensional parameter space. For this reason, we think that a more useful question to quantify the validity of each method is to ask}
whether the true value of parameter is within the credible interval. 
Nevertheless, 
we will use the definition of Eq.~(\ref{eq:relative-bias}) 
to compare the results between different setups/analyses for a possible convenience.}

Fig.~\ref{fig:fiducial-parameter-constraint} shows that the smaller scale cut of $(6,4)$ appears be too aggressive in the sense that the central value of estimated parameters can be away from the true value by more than the 68\% CL interval. 
The choice of $(9,6)$ looks okay.
However, we confirmed that the $(9,6)$ scale cut generally leads to a larger bias in the cosmological parameter than the fiducial choice of $(12,8)$ for the different setups including the different mock signals, and in particular gives a more significant bias for the 
assembly bias mocks. 
In the following, we mainly consider the scale cut of $(12,8)$ as a conservative choice. 
For real data, we should explore how the parameter constraints vary with different scale cuts, e.g. from (12,8) to $(9,6)$, and the change in cosmological parameters, if observed, is consistent with what we expect for the {\it Planck}-like cosmology. 
On the other hand, while \mtrv{the} scale cuts larger than our baseline choice ($12,8$) safely recover the true value, the error bars become larger at the price of the conservative constraint. 
We note that, even if we employ a really conservative cut such as $(24,16)$, which are further in the linear regime, the central value does not necessarily recover the true value. 
This is due to the marginalization of parameters that have severe degeneracies as discussed in Sec.~\ref{subsubsect:result:fiducial:baseline-analysis}. 
In fact, we checked that, if we fix either of $\Omegam$ or $\sigma_8$ to the input value in the parameter fitting, we can recover the true value. 
Thus we conclude that the baseline choice of (12,8) is reasonable for the SDSS and HSC-Y1 dataset. 

In the lower half of \figref{fig:fiducial-parameter-constraint}, we show the performance against the different bias models, where we fixed the scale cuts  to those of our baseline setup, i.e. $(12,8)$. 
\mtrv{First let us discuss the results for the model including the $b_2$ bias parameter in addition to $b_1$.}
Naively we expect that including the higher-order bias parameters increases accuracy and flexibility of the model predictions, and enables a more accurate recovery of the cosmological parameters than the fiducial model. 
However, 
the figure shows that including the $b_2$ 
parameter does not improve the accuracy in parameter estimation; the central value has a larger shift from the true value, and the error bars do not shrink compared to those of the baseline setup. 
This is because 
our mock signals \mtrv{are not well captured by the model involving the $b_2$ parameter. In particular, recalling that our constraints are mainly 
from $\wproj$ due to its higher signal-to-noise ratio, the terms involving the $b_2$ coefficient cause an asymmetric change in the amplitude of $\wproj$, especially due to the term involving $b_2^2$ ($b_2^2P_{b_2^2}$ in Eq.~\ref{eq:Pgg}) that is always positive for both negative and positive values of $b_2$. Hence the model with $b_2$ terms tends to overpredict the amplitude of $\wproj$, and in turn leads to a lower $\sigma_8$ value to reproduce the observed amplitude of the mock signal \citep[also see Ref.][for the similar discussion]{saitoNeutrinoMassConstraint2011}. This is the main reason that a bias in 
$\sigma_8 $ becomes larger, on a negative side from its true value. On the other hand, 
\ssrv{since $\dSigma\propto \sigma_8^2\Omegam$ on scales of our interest, 
$\Omegam$ is biased to a negative side from the true value to compensate the lowered amplitude  due to the decrease in $\sigma_8$.
}}

\mtrv{The result for ``$(\dSigma_9,\wproj_6)$ /w $b_1,b_2$'' shows that the bias in cosmological parameter is not lifted even if applying the nonlinear bias model to the smaller scale cuts, $(9,6)$. On the other hand, 
the results for ``$(\dSigma_6,\wproj_4)$ /w $b_1,b_2$'' displays a good performance, and the cosmological parameter is apparently recovered to within the 
credible interval. However, we checked that the best-fit model for this model gives a rather bad $\chi^2$ value, compared to the fiducial method. 
Thus the good performance seems just a coincidence for the {\it Planck} cosmology. Because we do not know the true cosmology for an actual cosmological analysis, we are not able to tell that this model gives a good description of the actual observables unless the goodness of fit for the best-fit model is acceptable. Hence we conclude that it is not peferctly safe to use this method for actual data.} 
Very similarly, including further higher-order bias parameters, $b_{s_2}$ and/or $b_3$, does not improve the parameter estimation, but rather cause larger biases in the estimated parameters. Thus we conclude that, for the baseline scale cuts, the $b_1$ model 
with the nonlinear $P_{\rm NL}$ achieves a reasonably good performance to estimate the cosmological parameters, even if it is a phenomenological model.
\mtrv{Note that these results remain unchanged even if using the narrower prior ranges of the higher order bias parameters such as 
$b_2=[-10,10]$ than our default ranges in Table~\ref{tab:prior-range}.}


For comprehensiveness of discussion, we study whether the self-consistent PT based method can recover the cosmological parameters. 
Here we use the PT prediction for the nonlinear matter prediction, $P_{\rm SPT}$, instead of $P_{\rm NL}$, and then study the performance for different methods using different combinations of the bias parameters. As we stated, these methods are self-consistent in the PT framework in the sense that all the nonlinear corrections to nonlinear matter power spectrum and bias expansion are at the same order. The figure shows that the PT based methods are not better than our baseline method, i.e. the minimal bias method. With including more bias parameters, we find a larger bias in each cosmological parameter.

In Appendix~\ref{apdx:sigma-upsilon}, 
we also study the performance for the different observables, ADSD, or the surface mass density profile (see Eq.~\ref{eq:ADSD}) that can be reconstructed from the measured $\dSigma$. With these redefined observables, we can employ the same  scale cut for the weak lensing and the galaxy clustering to perform the parameter forecast. 
A brief summary of our finding is that we do not find any strong advantage for either of these redefined observables over the fiducial analysis of $\dSigma$ and $\wproj$, because it neither leads to a smaller bias in the estimated parameter nor a smaller credible interval than those for the 
fiducial analysis.

\subsection{Validation against different galaxy mocks}
\label{subsect:result:mock-challenge}

\begin{figure*}
	\centering
    \includegraphics[width=1.0\textwidth]{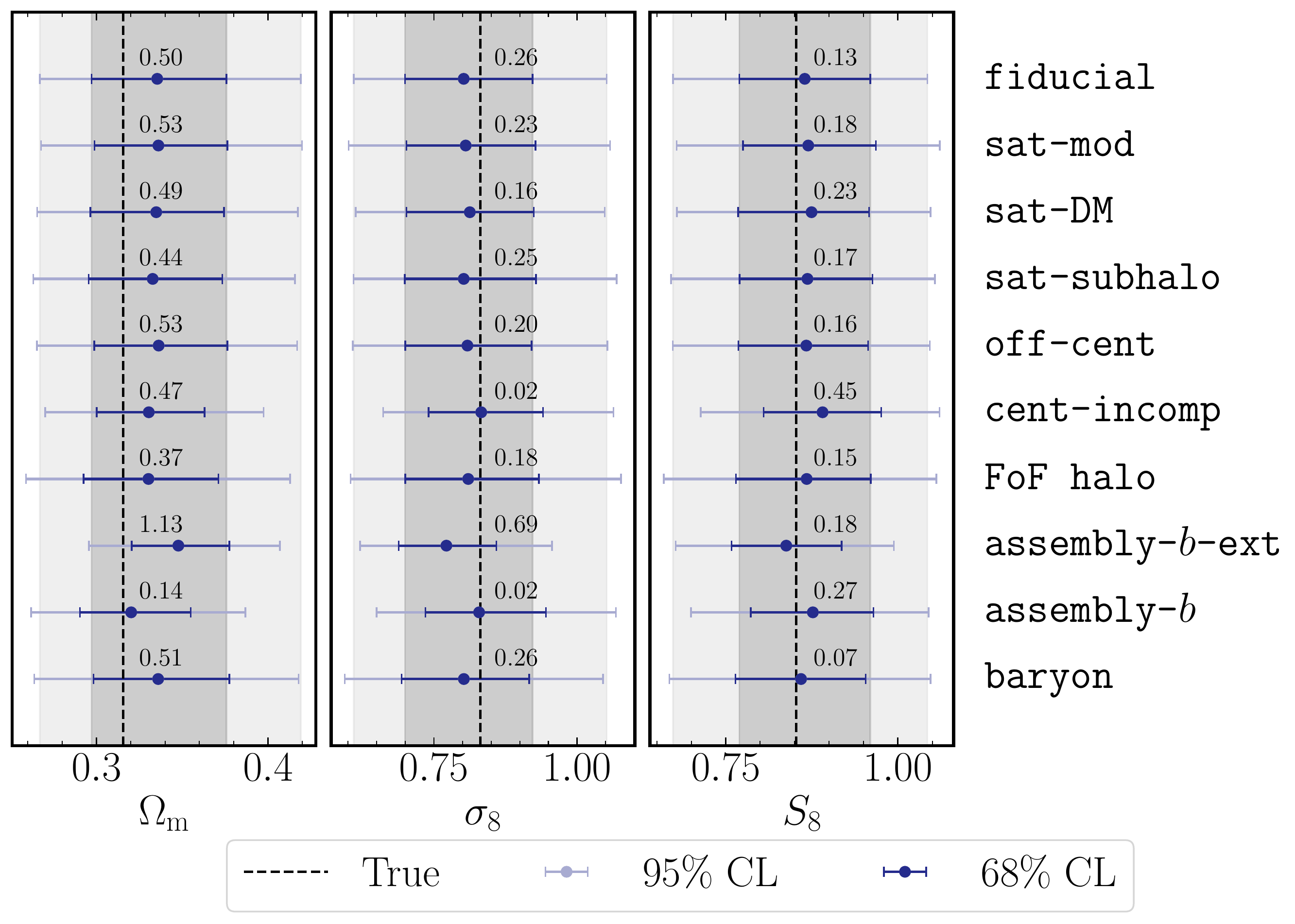}%
	\caption{
		Similar to the previous figure, but it shows the constraints against the different mock catalogs.
		Here we use the baseline analysis method of the $b_1$ bias and the scale cuts of $(12,8)~\hiMpc$.
		}
	\label{fig:mock-challenge-summary}
\end{figure*}
We now study how the minimal-bias method is robust against different mock galaxy catalogs. As long as we allow the bias parameter(s) to freely vary in the parameter estimation, we expect that the bias parameter(s) captures clustering properties of galaxies at large scales, and the method can recover the underlying cosmological parameters after marginalization over the bias parameter(s). Here we apply the baseline analysis method, the (12,8)-scale cuts  and $b_1$ model, to the mock signals measured from the 
different mock catalogs described in Sec.~\ref{sect:GalaxyMock}. 
Fig.~\ref{fig:mock-challenge-summary} shows the results. 
Encouragingly, the cosmological parameters, are recovered to within the 68\% CL interval for all mock catalogs, except for the $\Omegam$ result for the \mock{assembly-$b$-ext} mock.
This means that the baseline analysis is robust against details of the galaxy-halo connection.
We should stress that the minimal-bias method nicely recovers the true values of $\sigma_8$ and $\Omegam$ even for the worst-case mock, i.e. the \mock{assembly-$b$-ext} mock, which is one of the most dangerous effects, and predicts greater amplitudes in $w_\mathrm{p}$ and $\dSigma$, by up to a factor of 1.6, than that of the fiducial mock (Fig.~\ref{fig:galaxy-mock-signals}). 
If a strong prior between halo mass and the linear bias is employed, such as the scaling relation of halo bias with halo mass, this mock leads to a large bias in the estimated parameters (see Miyatake et al. in preparation).
A failure in $\Omegam$ for the \mock{assembly-$b$-ext} mock is likely to be due to the scale-dependent deviation from $r^{(w)}_{\rm cc}=1$ at scales around the scale cut in this mock (see Fig.~\ref{fig:galaxy-mock-r}), where a scale-dependent change in $\wproj$ due to a change in $\Omegam$ tends to compensate the deviation. In any case we should emphasize that the extreme assembly bias mock is the worst case scenario leading too large effect. 
A nice recovery of $\sigma_8$ and $S_8$ for this extreme mock is rather encouraging.
In summary,  combining $\dSigma$ and $\wproj$ allows us to robustly break the degeneracy of the cosmological parameters with the galaxy bias even if the spatial distribution of galaxies is affected by the assembly bias effect. 
The results are justified by 
Fig.~\ref{fig:galaxy-mock-r}, which shows that the cross-correlation coefficient is close to unity at scales greater than the scale cut we employed. 

\section{Conclusion}
\label{sect:conclusion}

In this paper we have studied the validity of the perturbation theory (PT) based method for estimating cosmological parameters from the galaxy-galaxy weak lensing ($\dSigma$) and the projected galaxy clustering ($\wproj$), expected for the HSC-Y1 and SDSS data. 
To do this, we used the mock signals measured from the mock catalogs of \mtrv{spectroscopic} SDSS 
galaxies that are constructed using a suite of high-resolution $N$-body simulations in Ref.~\cite{nishimichiDarkQuestFast2018}.

In the ``PT''-based method, we introduced a parameter(s) to model galaxy bias, allowing the parameter(s) to freely vary in parameter estimation, and then derived the cosmological constraints after marginalization over the bias parameter(s).
We quantified the validity and performance of the PT-based method by the shift or bias in the estimated cosmological parameter from the true value as well as by the size of the credible interval, considering that a method having a smaller parameter bias and a smaller credible interval as more desirable. 
First we carefully estimated a proper choice of the ``scale cut''; we use the clustering signals at scales above the scale cut for the parameter estimation, because the PT-based method is valid only for large scales, but at the same time we want to use the information down to scales as small as possible since the inclusion of smaller-scale signals leads to more constraining power. 
We found that the scale cuts of $12$ and $8~\hiMpc$ for $\dSigma$ and $\wproj$, respectively, are reasonable, as taken by the DES joint probes analysis \cite{2018PhRvD..98d3526A}. 

The baseline method we used throughout this paper is the ``minimal-bias'' method: we model the underlying 3D power spectra, $P_{\rm gm}(k)$ and $P_{\rm gg}(k)$, in terms of the linear bias parameter $b_1$ (for each galaxy sample at each of the three redshifts) and the fully nonlinear matter power spectrum, as $P_{\rm gm}(k)=b_1P_{\rm NL}(k)$ and $P_{\rm gg}(k)=(b_1)^2P_{\rm NL}(k)$. 
This model yields the cross-correlation coefficient $r^{(\xi)}_{\rm cc}=\xi_{\rm gm}/\sqrt{\xi_{\rm gg}\xi_{\rm mm}}=1$ at all scales by construction.
From all the cosmology challenges we performed, this simplest model was found to recover the true cosmological parameters in the sense that the 68\% credible interval for either of $\Omegam$, $\sigma_8$ or $S_8(\equiv \sigma_8(\Omegam/0.3)^{0.5}$) includes the true value. 
The observables $\dSigma$ and $\wproj$ from the HSC-Y1 and SDSS data seem to allow for about 10\% fractional accuracy of $S_8$ determination if the $\Lambda$CDM model is close to the underlying true cosmology. 
Even if we use the linear matter power spectrum or the perturbation theory prediction for the nonlinear matter power spectrum instead of $P_{\rm NL}$, the method fairly well recovers the true values, although the method of $P_{\rm NL}$ seems best, as shown in Appendix~\ref{apdx:linear_treePT}. 
Our results showed that the galaxy-galaxy weak lensing and the galaxy clustering, $\dSigma$ and $\wproj$, are complementary to each other, and combining the two enables one to break the parameter degeneracies, especially with the linear bias parameter, to constrain the cosmological parameters. 
Although we hoped that a model using the higher-order bias parameters can serve as a more accurate theoretical template of the clustering observables, we did not find that this is the case: all the results using the higher-order bias parameters lead to a larger parameter bias compared to the baseline method.
Thus the higher-order bias prescription does not appear promising for the parameter estimation, at least for $\dSigma$ and $\wproj$. 
The effective field theory of large-scale structure \cite{baumann12} might work better, and this is our future work. 

One of the most important results is that the baseline method successfully recovers the input cosmological parameters from all the mock catalogs including the ``extreme'' assembly bias mock that predicts boosted amplitudes in $\dSigma$ and $\wproj$ by up to a factor of 1.6 over scales used in the parameter estimation, as long as the scale cuts of $12$ and $8~\hiMpc$ are employed. 
This is contrasted to the results of our companion paper, Miyatake et al.: they use a halo model approach including the theoretical knowledge on the dependence of halo bias on halo mass and cosmology, and found that a recovery of the cosmological parameters fails for the assembly bias mock (although successful for other mocks), because the assumed relation between halo bias and halo mass is violated in the assembly bias mock. 
However, the price to pay for the PT-based method is that the credible interval is larger than that of the halo model method. 
We gave a physical justification of the success of the PT-based method, as given in Fig.~\ref{fig:galaxy-mock-r}, which shows that the cross-correlation coefficient is close to unity at scales above the scale cut for all the mock catalogs \citep[see Fig.~31 in Ref.~][for the halo-matter 
correlations]{nishimichiDarkQuestFast2018}
\citep[also see][for the similar result in hydrodynamical simulation including galaxy physics]{2018MNRAS.475..676S}. 
As long as physical processes involved in galaxy formation and evolution are confined to local scales, say smaller than a few Mpc at most, the large-scale clustering amplitudes are governed by gravitational interaction, and combining the large-scale information of $\dSigma$ and $\wproj$ can calibrate out the bias parameter, even with the simple $b_1$ model, and then recover the underlying matter power spectrum $P_{\rm NL}$, which in turn allows us to constrain the cosmological parameters. 

This is an advantage of the real-space clustering observables. Due to the nature of Fourier transform, the real-space clustering of compact small scales can affect the Fourier counterpart, the power spectrum, down to small $k$ scales. 
The easiest example is the shot noise; for real-space observables, the shot noise contributes to the measurement only at zero lag ($R=0$), while it affects the power spectrum over all scales, behaving like white noise. 
While we believe that the different mock catalogs we used include a fairly broad range of variations we can think of for the SDSS-like early-type galaxies, if any physics that affects galaxy formation on larger scales other than those we anticipated in this paper is identified, the PT-based method might cease to recover the true cosmology. 
Such an example would be the effect of reionization bubbles on galaxy formation at high redshift or the effect arising from the difference between the CDM and baryon velocity fields, which are both relevant on large scales \cite{dalal08,2016PhRvD..94f3508S}. 
However, at least for early-type galaxies such as the SDSS-like galaxies, these effects seem to be small or already washed out by the observed redshifts. Another example is the effect of the primordial non-Gaussianity, which generally causes a modification in the large-scale clustering. 
This effect itself is exciting, so any observational hint for a violation in the relation between $\dSigma$ and $\wproj$ at large scales would be worth exploring.

Another important, and non-trivial, result we found is the effect of marginalization when obtaining the posterior distribution of cosmological parameters. 
Perhaps obviously, well-known cosmological parameters $\Omegam$ and $\sigma_8$ are not necessarily the best choices for parameter estimation with the given observables. 
More precisely, those parameters are generally different from the eigenvectors or principal components of the parameter covariance obtained in the MCMC. 
For this reason, the central value of the marginalized posterior of $\Omegam$ or $\sigma_8$ is generally off from the true value, even if we use exactly the same data vector that is predicted by the model, as a result of the projection of the ``banana''-shaped posterior distribution in a multidimensional parameter space. 
This might lead to some confusion, e.g. the inferred central value might be seen as in tension compared to external constraints from other surveys. 
Thus, the community needs to carefully think about what and how to deliver their cosmological constraints from given surveys. 
A possible choice is to constrain the combination of parameters such as $S_8\propto \sigma_8\Omegam^{0.5}$, which is defined by the eigenvalue of the parameter covariance, e.g. in a linear combination of  the log-mapped space of parameters, so that the combinations of parameters map the original banana-shaped posterior distribution to a more ``straight'' posterior distribution in the 
mapped parameter space.
This requires a more careful study, and will be presented elsewhere. 

\acknowledgments
We would like to thank Elisabeth Krause and Surhud More for useful discussion. 
This work was supported in part by World Premier International Research Center Initiative (WPI Initiative), MEXT, Japan, and JSPS KAKENHI Grant Numbers JP15H03654,
JP15H05887, JP15H05893, JP15H05896, JP15K21733, JP17H01131, JP17K14273, JP18H04358, JP19H00677 and JP20H04723, by Japan Science and Technology Agency (JST) CREST JPMHCR1414, and JST AIP Acceleration Research Grant Number JP20317829, Japan.
SS is supported by International Graduate Program for Excellence in Earth-Space Science (IGPEES), World-leading Innovative Graduate Study (WINGS) Program, the University of Tokyo.

\appendix

\section{Power spectra in galaxy bias expansion}
\label{sect:power-spectra}

Eqs.~(\ref{eq:Pgm}) and (\ref{eq:Pgg}) give expressions for the 3D power spectra, $P_{\rm gm}$ and $P_{\rm gg}$, based on the Eulerian perturbation theory and the PT expansion of bias parameters. The expressions involve the following functions that are given in terms of the integrals of the linear matter power spectrum:
\begin{align}
	P_{b_2}(k) &= \phaseint{p} \PL(\vec{p}) \PL(\vec{k}-\vec{p}) F_2(\vec{p},\vec{k}-\vec{p})\\
    P_{b_{s_2}}(k) &= \phaseint{p} \PL(\vec{p}) \PL(\vec{k}-\vec{p}) \nonumber\\
    &\hspace{2em}\times F_2(\vec{p},\vec{k}-\vec{p}) S(\vec{p},\vec{k}-\vec{p})\\
	P_{b_3}(K) &= \sigma_3^2(k)\PL(k)\\
	P_{b_2^2}(k)&= \frac{1}{2}\phaseint{p} \PL(\vec{p})\left[\PL(\vec{p})-\PL(\vec{k}-\vec{p})\right]\\
	P_{b_{s_2}^2}(k)&= \frac{1}{2}\phaseint{p} \PL(\vec{p})\nonumber\\
    &\hspace{1em}\times\left[S^2(\vec{p},\vec{k}-\vec{p})\PL(\vec{k}-\vec{p})-\frac{4}{9}\PL(\vec{p})\right]\\
	P_{b_2b_{s_2}}(k)&= \frac{1}{2}\phaseint{p} \PL(\vec{p})\nonumber\\
    &\hspace{1em}\times\left[S(\vec{p},\vec{k}-\vec{p})\PL(\vec{k}-\vec{p})-\frac{2}{3}\PL(\vec{p})\right].
\end{align}
Here $\PL$ is the linear matter power spectrum. The Fourier kernels, $F_2$ and $S$ are given by
\begin{align}
	F_2(\vec{k},\vec{p}) &= \frac{5}{7} + \frac{1}{2}\frac{\vec{k}\cdot\vec{p}}{kp}\left(\frac{k}{p}+\frac{p}{k}\right) + \frac{2}{7}\left(\frac{\vec{k}\cdot\vec{p}}{kp}\right)^2\\
	S(\vec{k},\vec{p}) &= \left(\frac{\vec{k}\cdot\vec{p}}{kp}\right)^2-\frac{1}{3},
\end{align}
and $\sigma_3$ is the loop contribution,
\begin{align}
	\sigma_3(k) = \int\diff\ln r \Delta^2(kr) \left(I(r)+\frac{5}{6}\right),
\end{align}
where $\Delta^2(k)=k^3\PL(k)/2\pi^2$ and 
\begin{align}
	I(r) &= \frac{105}{32}\int_{-1}^1 \diff\mu \frac{2}{7}\left[S(-\vec{p},\vec{k})-\frac{2}{3}\right]\nonumber\\
    &\hspace{2em}\times S(\vec{p},\vec{k}-\vec{p})\bigg|_{r=p/k,~\mu=\vec{k}\cdot\vec{p}/kp}
\end{align}

\section{Linear theory,  PT one-loop corrections or $P_{\rm NL}$  for $P_{\rm mm}$ in the theoretical templates}
\label{apdx:linear_treePT}

\begin{figure}[h]
	\centering
    \includegraphics[clip,width=1.0\columnwidth]{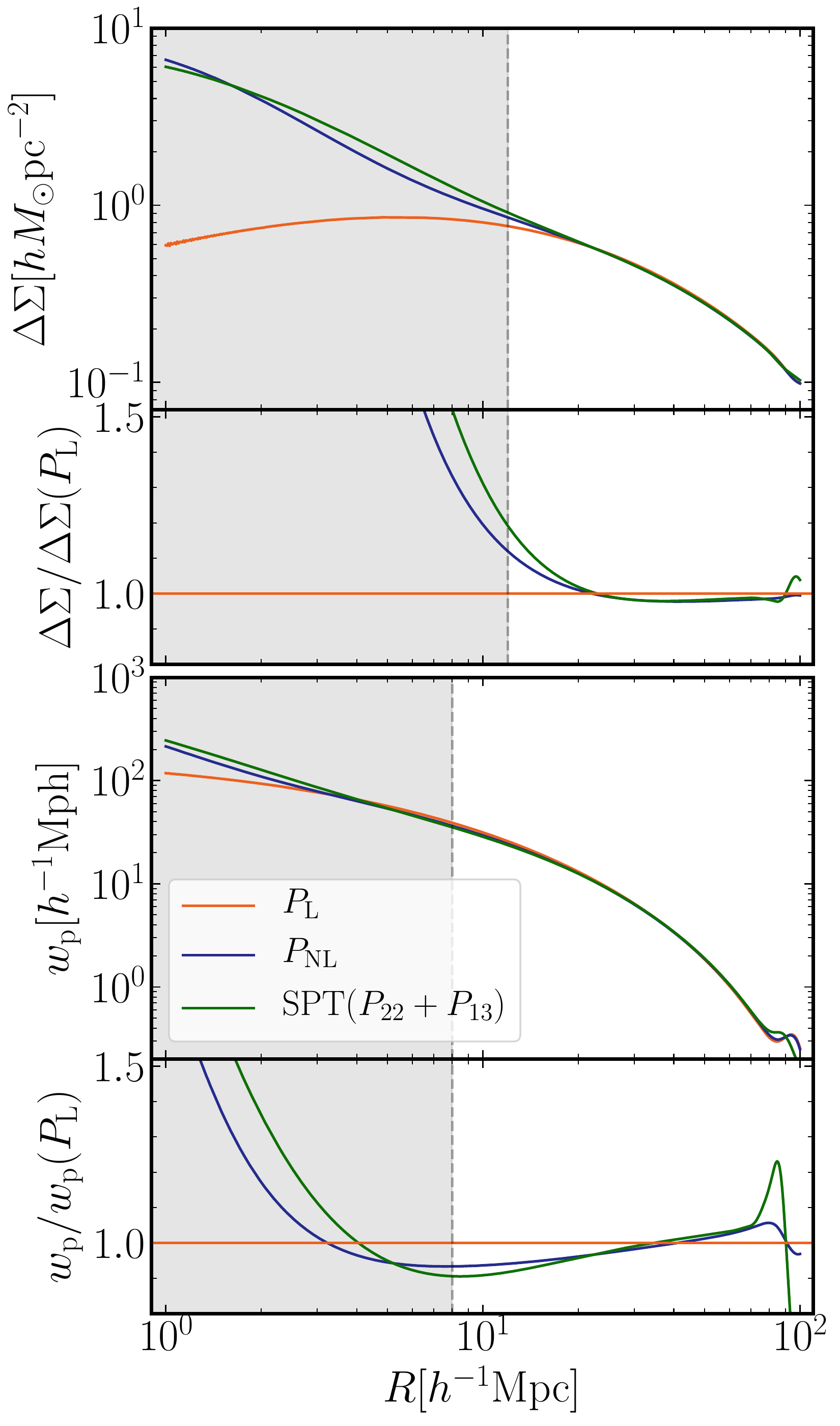}%
	\caption{
		Comparison of the model predictions of $\dSigma$ and $\wproj$ for the {\it Planck} cosmology, obtained by using the linear matter power spectrum ($P_{\rm L}$), the standard perturbation prediction ($P_{\rm SPT}$), and the halofit fitting formula ($P_{\rm NL}$) for the matter power spectrum entering into $P_{\rm gm}$ and $P_{\rm gg}$. Here we include the next leading-order \mtrv{(one-loop)} contribution in the perturbation theory for $P_{\rm SPT}$: $P_{\rm SPT}=P_{22}+P_{13}$. Note that we consider $z=0.251$ corresponding to the redshift of LOWZ sample, and include only the linear bias parameter by setting $b_1=1$ (that is, we set all the higher-order bias parameters to zero).
		Our default model is based on $P_{\rm NL}$. 
		The lower panel shows the ratio of the prediction relative to that for $P_{\rm L}$.
		}
	\label{fig:signal-plin-pnlin-pspt}
\end{figure}
\begin{figure}[h]
	\centering
	\includegraphics[clip,width=0.95\columnwidth]{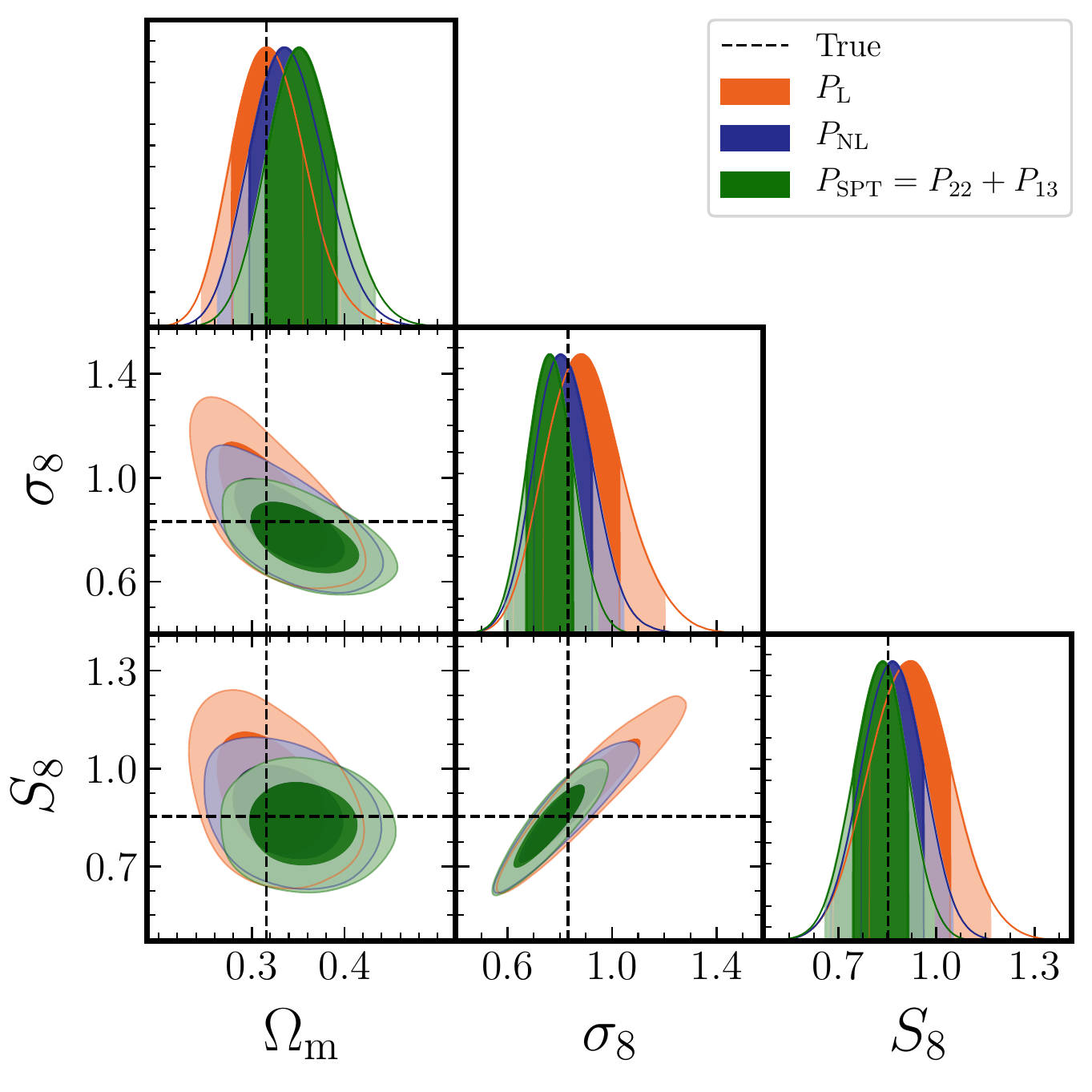}%
	\caption{
		Comparison of the marginalized posterior distributions of cosmological parameters, obtained by applying the theoretical templates of $P_{\rm L}$, $P_{\rm SPT}$ or $P_{\rm NL}$, as shown in the previous figure, to the fiducial mock.  
		The posterior distributions for $P_{\rm NL}$ are the same as those in Fig.~\ref{fig:corner-baseline}.
		}
	\label{fig:corner-plin-pnlin-pspt}
\end{figure}
Throughout this paper, we use the halofit fitting formula, $P_{\rm NL}(k)$, to model the nonlinear matter power spectrum entering into $P_{\rm gm}$ and $P_{\rm gg}$, with the linear bias parameter $b_1$ (Eqs.~\ref{eq:Pgm} and \ref{eq:Pgg}), which we call minimal-bias model.
However, this is a phenomenological approach and is not self-consistent in a perturbation theory manner because we only keep the linear bias parameter, but include fully nonlinear contributions in $P_{\rm NL}$ even including the non-perturbative effects after shell-crossing (halo/galaxy formation) as we discussed. 
Even if we use only the large-scale clustering information, this inconsistency might cause an unwanted bias in cosmological parameters. 
To study this concern, in this appendix we investigate how the results are changed if we use the standard perturbation theory (SPT) prediction including the next-to-leading order prediction, $P_{\rm SPT} (\equiv P_{22}+P_{13}$), instead of $P_{\rm NL}$, but still keeping the linear bias parameter ($b_1)$ alone. 
This method satisfies the cross-correlation coefficient $r^{(\xi)}_{\rm cc}=1$ at all scales.
As we described in the main text, we adopt the cutoff function $\exp(-k^2/k_{\rm off}^2)$ with $k_{\rm off}=10~\hiMpc$ to perform the Fourier transform of $P_{\rm SPT}$ to obtain the SPT prediction for $\xi_{\rm mm}(r)$.
In Sec.~\ref{subsubsect:result:fiducial:cutoff-and-bias} we showed the results further including the higher-order bias parameters in a self-consistent manner of the PT picture. Here we study the difference arising from $P_{\rm NL}$ and $P_{\rm SPT}$ in the minimum bias method. 
For comparison, we also study the result using the linear matter power spectrum ($P_{\rm L}$) instead of $P_{\rm NL}$.

Fig.~\ref{fig:signal-plin-pnlin-pspt} compares $\dSigma$ and $\wproj$ computed using $P_{\rm L}$, $P_{\rm SPT}$ and $P_{\rm NL}$, respectively. 
First of all, the predictions from $P_{\rm SPT}$ and $P_{\rm NL}$ are in good agreement with each other in the amplitudes over scales we use for the parameter estimation, although there is a subtle discrepancy around the BAO scales. In other words, the linear theory is not accurate enough even for these large scales that are naively well in the linear regime. 
From these results, we confirm the results of Fig.~\ref{fig:phalo-transition}, which shows that a change in $\sigma_8 $ alters $\dSigma$ and $\wproj$ over a window of $R\simeq [10,100]~\hiMpc$ due to the nonlinear clustering effect as predicted by SPT. 
However, the figure also shows that SPT prediction starts to deviate from the prediction of $P_{\rm NL}$ at scales below the scale cut, as it overpredicts the effect of the nonlinear clustering. 

In Fig.~\ref{fig:corner-plin-pnlin-pspt} we compare the posterior distributions obtained by using the model predictions of $P_{\rm L}$, $P_{\rm SPT}$ and $P_{\rm NL}$ in the baseline analysis. 
All the results are similar, but we find that the model with $P_{\rm NL}$ best recovers the input cosmological parameters; especially it gives smallest bias in $S_8$ from the true value. 
We also note that the model using $P_{\rm L}$ or $P_{\rm SPT}$ over- or under-estimates $\sigma_8$, because the two models under- or over-predict the clustering power at scales around the scale cut, respectively, as shown in Fig.~\ref{fig:signal-plin-pnlin-pspt}.
Thus we conclude that the model with $P_{\rm NL}$, although phenomenological, can work well in estimating the cosmological parameters for our baseline setup. 

\section{Performance of alternative observables for $\dSigma$: $\Upsilon$ and $Y$}
\label{apdx:sigma-upsilon}

\subsection{Definitions of $\Upsilon$ and $Y$}

The galaxy-galaxy lensing observable, $\dSigma$, is a non-local quantity reflecting the fact that it arise from the gravitational tidal field around lensing galaxies, rather than the surface mass density field that is a local quantity in contrast. 
Since it is difficult to model the matter and galaxy distributions at small scales in the deeply nonlinear regime, this non-locality forces us to use a scale cut for $\dSigma$ at relatively larger scale than in $\wproj$. For the baseline analysis, we employ the different scale cuts for $\dSigma$ and $\wproj$: 12 and $8~\hiMpc$, respectively. This choice is not rigorous, and somewhat arbitrary. 
To resolve this complexity, several works proposed an alternative observable for $\dSigma$: the Annular Differential Surface Density (ADSD), developed and used in \cite{baldaufAlgorithmDirectReconstruction2010,mandelbaumCosmologicalParameterConstraints2013}, and the reconstructed surface mass density, denoted as $Y$ \cite{parkLocalizingTransformationsGalaxyGalaxy2020}. 

The ADSD profile using a filtering scale $R_0$ is defined from the measured $\dSigma$ as
\begin{align}
	\Upsilon_{\rm gm}(R|R_0) = \dSigma(R) - \frac{R_0^2}{R^2}\dSigma(R_0)\label{eq:Upsilongm-def}.
\end{align}
With the second term in the above equation, the ADSD profile filters out the lensing contribution from $\xi_{\rm gm}$ at $R<R_0$. That is, ADSD is designed to explicitly filter out the small-scale contribution. 
Similarly, the ADSD profile for the galaxy clustering is defined from the measured $\wproj$ as
\begin{align}
    \Upsilon_{\rm gg}(R|R_0) &= \frac{2}{R^2}\int_{R_0}^{R}\diff R'R'\wproj(R') \nonumber\\
    &\hspace{2em} - \frac{1}{R^2}[R^2\wproj(R)-R_0^2\wproj(R_0)].\label{eq:Upsilongg-def}
\end{align}
$\Upsilon_{\rm gm}(R|R_0)$ and $\Upsilon_{\rm gg}(R|R_0)$ at a particular $R$ arise from the same Fourier modes by construction.

Another alternative choice for $\dSigma$ is the surface mass density profile that can be reconstructed from the observed $\dSigma$ as
\begin{align}
	\Sigma(R) = \Sigma(R_\mathrm{max})+\int_R^{R_{\rm max}} \diff \ln R' \left[ 2\dSigma(R')+ \diffrac{\dSigma}{\ln R'} \right] .
\end{align}
The integration constant, $\Sigma(R_\mathrm{max})$, is unknown and does not affect the parameter inference, because the relevant quantity is a difference vector between data and model, i.e. $\vec{D} - \vec{T}$. Hence we use as the new observable 
\begin{align}
    Y(R) &\equiv\Sigma(R)-\Sigma(R_\mathrm{max})\nonumber\\
    & = \int_R^{R_{\rm max}} \diff \ln R' \left[ 2\dSigma(R')+ \diffrac{\dSigma}{\ln R'} \right]\label{eq:Sigma-reconstruction-def}.
\end{align}

However, we comment that an estimation of the ADSD or $Y$ profiles from the measured $\dSigma$ and $\wproj$ involves a subtlety. As can be found from Eqs.~(\ref{eq:Upsilongm-def}), (\ref{eq:Upsilongg-def}), and (\ref{eq:Sigma-reconstruction-def}), the estimation involves an integration or derivative of the measured $\dSigma$ and $\wproj$, which is given at discrete data points and can be noisy due to the statistical errors. 
We have found, from the various tests using the mock data, that the estimations are not stable sometimes, especially when the measurement noise is present. 
For the ADSD reconstruction, we use the following method. We employ a particular point of the discrete bins of $R$ for $R_0$, rather than estimating $R_0$ from the interpolation between the discrete bins. 
If we refer $R_0=8$  or $12~\hiMpc$ in the following, it means that we take the nearest bin to the inferred $R_0$. For the $R$-integration in the estimation of $\Upsilon_{\rm gg}$, we use the trapezoidal integral method to perform the integration using the discrete data points. 
For the $Y$, we need to make a numerical derivative from the discrete data points of $\dSigma$ with respect to the discrete bins of $R$ as described in Ref.~\cite{parkLocalizingTransformationsGalaxyGalaxy2020}. 
We employ the two-side numerical derivative around a given bin of $R$. For the integration, we use the trapezoidal integral method. 
Since our mock signals of $\dSigma$ and $\wproj$ are computed from the average among about 20 realizations, corresponding to $20~(\hiGpc)^3$, and the $\dSigma$ signal does not include the shape noise errors, the mock signals are very smooth, and are least affected by the sample variance. 
This is an ideal situation. When we have statistical scatters in the signals in different $R$ bins, we found that the estimations of ADSD and $\Sigma$ profiles can be noisy. Hence, the following results are considered as an ideal case for these methods.

\subsection{Conversion matrices for $Y$, $\Upsilon$s}
\label{apdx:matrix-sigma-upsilon}

As described in the preceding section, we use a method to estimate the alternative observables, $\Upsilon_{\rm gm}$, $\Upsilon_{\rm gg}$ and $Y$, from the discrete data vector of $\dSigma$ and $\wproj$ given at discrete points of $R$ bins.
Hence the estimation is equivalent to an operation of linear-algebra conversion of the data vector to $\Upsilon$ and $Y$. 
This linear conversion allows us to obtain the covariance matrix for the redefined data vector from the original covariance matrix, as described below.

Suppose that the data vector of $\dSigma$ or $\wproj $ has $N_{\rm gm}$ or $N_{\rm gg} $ data points given in separation bins that are logarithmically-evenly spaced: $\dSigma(R_i)\equiv \dSigma_i$ ($i=1,2,\cdots, N_{\rm gm}$) or $\wproj(R_i)\equiv w_{{\rm p},i}$ ($N=1,2,\cdots,N_{\rm gg}$). 
An estimation of the alternative observable, $\Upsilon_{\rm gm}$, $\Upsilon_{\rm gg}$ or $Y$, is generally given as
\begin{align}
	A_{i} = (M_{AX}X)_{i} = \sum_{j}M_{AX,ij}X_{j},
\end{align}
where $A_i=\Upsilon_{\rm gm}(R_i)$, $\Upsilon_{\rm gg}(R_i)$ , or $\Sigma(R_i)$, $X_i=\dSigma_i$ or $w_{{\rm p},i}$, 
and $M_{AX,ij}$ is the transformation matrix. 
Thanks to property of the linear conversion we can compute the covariance matrix of the alternative observable from the original covariance 
matrix: 
\begin{align}
    \mathrm{Cov}_{A_i,A_j} &= (M_{AX}\mathrm{Cov}_{X,X}M_{AX}^{T})_{ij}\nonumber\\
    &= \sum_{k,l}M_{AX,ik}M_{AX,jl}\mathrm{Cov}_{X_k,X_l},
	\label{eq:cov_upsilon}
\end{align}
where $M^{T}$ is the transpose of matrix $M$, and ${\rm Cov}_{X_i,X_j}$ is the covariance matrix of our observables, $\dSigma$ or $\wproj$, as described in Sec.~\ref{subsect:measurement-of-covariance}. 
The conversion matrix, $M_{AX}$, is formally given as 
\begin{align}
	M_{\Sigma\dSigma} &= S+SD\\
	M_{\Upsilon_\mathrm{gm}\dSigma} &= I-U\\
	M_{\Upsilon_\mathrm{gg}\wproj} &= K-I+U,
\end{align}
where $S$ and $D$ are the numerical integration and differentiation matrices. The matrices $I$, $U$ and $K$ are
\begin{align}
	&I = 
	\begin{pmatrix}
		O&O\\
		O&\tilde{I}
	\end{pmatrix},\hspace{1em}
	U = 
	\begin{pmatrix}
		O&O\\
		O&\tilde{U}
	\end{pmatrix},\hspace{1em}
	K = 
	\begin{pmatrix}
		O&O\\
		O&\tilde{K}
	\end{pmatrix},
\end{align}
Here we divided the $N\times N (N=N_\mathrm{gm}$ or $N_\mathrm{gg})$ matrices into $i_0\times i_0$, $i_0\times(N-i_0)$, $(N-i_0)\times i_0$ and $(N-i_0)\times(N-i_0)$ block sub-matrices, where $i_0 = \mathrm{min}(\{i|R_i\geq R_0\})$. $\tilde{I}=\mathrm{diag}(1)$, $U=\mathrm{diag}((R_{i_0}/R_{i})^2)$ and $\tilde{K}=\mathrm{diag}(2/R_i^2)\tilde{S}\mathrm{diag}(R_i^2)$, where $\tilde{S}$ is numerical integration matrix of $(N-i_0)\times(N-i_0)$.

We use Eq.~(\ref{eq:cov_upsilon}) to compute the covariance matrix of $\Upsilon$ or $\Sigma$ from the original covariance matrices. 

\subsection{Performance of ADSD and $Y$ methods}
\label{apdx:sigma-upsilon-result}

\begin{table*}[htb]
	\centering
	\caption{
		Analysis setups using the alternative observables, the ADSDs of $\{\Upsilon_{\rm gm},\Upsilon_{\rm gg}\}$ or $\{Y,\wproj\}$ instead of our default observables $\{\dSigma,\wproj\}$.
		} 
	\begin{tabular}{l|cll}
		\hline\hline
		\multirow{2}{*}{Setup label} & Scale cuts  & \multirow{2}{*}{Sampled parameters} & \multirow{2}{*}{Notes}\\
		& in $\mathrm{Mpc}/h$ && \\
		\hline
		$(Y_{8},w_\mathrm{p,8})$ & (8,8) & \multirow{4}{*}{$\sigma_8,\Omegam,b_1(z_i)$} & \\
		$(Y_{12},w_\mathrm{p,12})$ & (12,12) & & Using $Y$\\
		$(Y_{8},w_\mathrm{p,8})$ w/ 0.1Cov${}_{Y}$ & (8,8) & & instead of $\dSigma$\\
		$(Y_{12},w_\mathrm{p,12})$ w/ 0.1Cov${}_{Y}$ & (12,12) & & \\
		\hline
		$(\Upsilon_\mathrm{gm,8},\Upsilon_\mathrm{gg,8})$ & (8,8) & \multirow{4}{*}{$\sigma_8,\Omegam,b_1(z_i)$} & \\
		$(\Upsilon_\mathrm{gm,12},\Upsilon_\mathrm{gg,12})$ & (12,12) & & Using ADSD\\
		$(\Upsilon_\mathrm{gm,8},\Upsilon_\mathrm{gg,8})$ w/ 0.1Cov${}_{\Sigma}$ & (8,8) & & instead of $(\dSigma,\wproj)$\\
		$(\Upsilon_\mathrm{gm,12},\Upsilon_\mathrm{gg,12})$ w/ 0.1Cov${}_{\Sigma}$ & (12,12)& &  \\
		\hline\hline
	\end{tabular}
	\label{table:sigma-upsilon-validation}
\end{table*}
\begin{figure*}
	\centering
    \includegraphics[clip,width=1.0\textwidth]{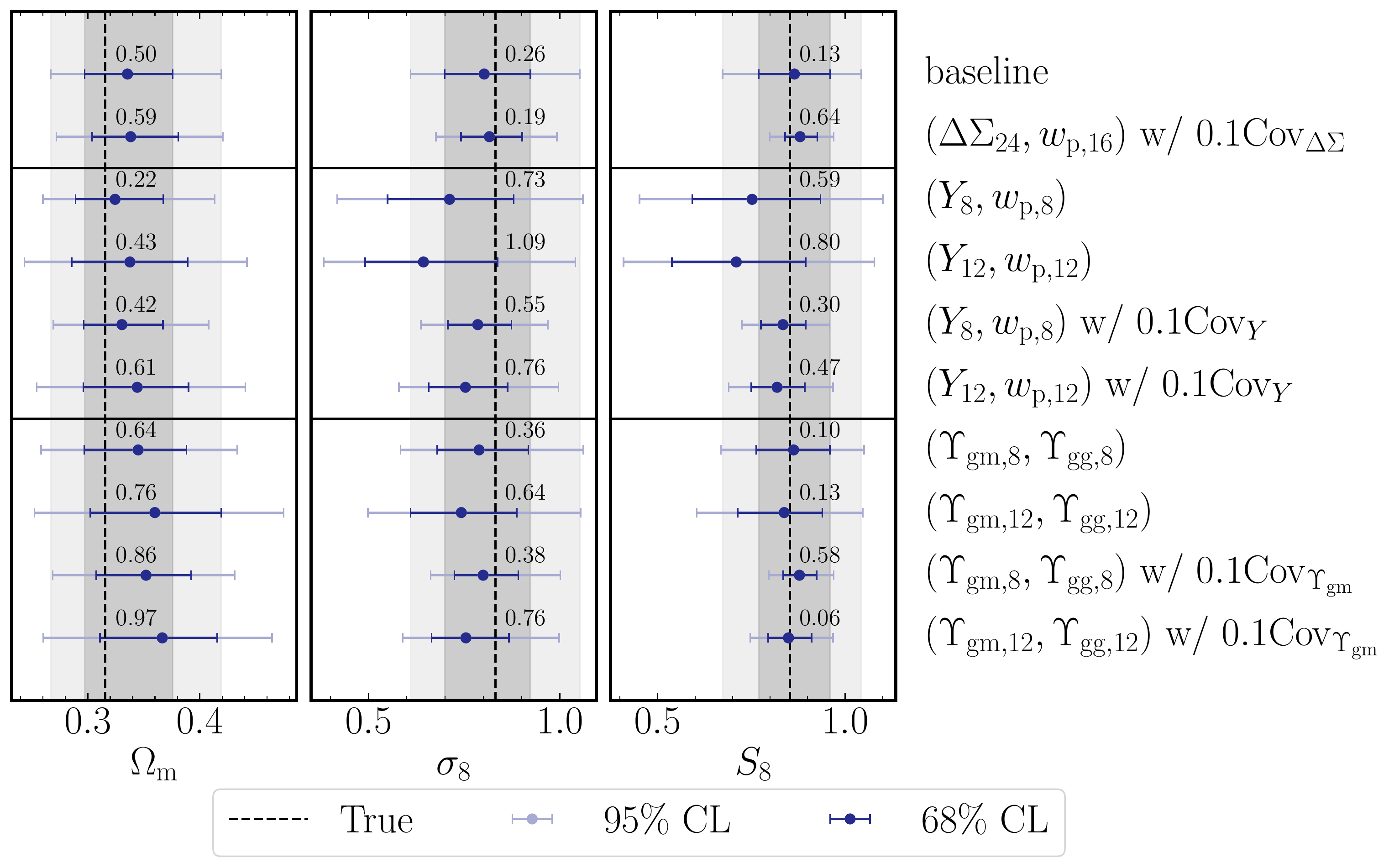}
	\caption{
		Similar to Fig.~\ref{fig:fiducial-parameter-constraint}, but the results obtained by applying the ADSDs ($\Upsilon$) or the $Y$ observables based on the setups in Table~\ref{table:sigma-upsilon-validation} to the fiducial mock. 
		}
	\label{fig:summary-sigwpSD-upup}
\end{figure*}
As in \sectref{fiducial-mock-validation}, we study performance of  the ADSD or $Y$ method, i.e. how these alternative observables, instead of the fiducial observables ($\dSigma, \wproj$), can recover the cosmological parameters. 
To do this, we adopt the analysis setups as given in Table~\ref{table:sigma-upsilon-validation}, and apply the analysis method to the fiducial mock catalog. 
Fig.~\ref{fig:summary-sigwpSD-upup} shows the results, compared to the results for the fiducial analysis of $\{\dSigma,\wproj\}$.
For the method using the observables $\{Y,\wproj\}$, we employ the same scale cut, $R=8~\hiMpc$, for both the observables in order to compare with the result for the fiducial analysis of $\{\dSigma,\wproj\}$, where we employed the scale cut  $8~\hiMpc$ for $\wproj$. 
A systematic bias in $\Omegam$ is slightly smaller than that for the fiducial analysis. However, the constraint for $S_8$ does not promising, because a bias looks larger and the credible interval is larger than those for the fiducial analysis. 

For the method using $\{\Upsilon_{\rm gm},\Upsilon_{\rm gg}\}$, it is not clear which filtering scale $R_0$ to employ. The ADSDs satisfy the condition $\Upsilon(R=R_0|R_0)=0$ by construction. Here we study the two cases of $R_0=8$ and 12~$\hiMpc$. 
The figure shows that the results for $\Upsilon$ using $R_0=8~\hiMpc$ look comparable with those for the fiducial analysis, although a bias in $\Omegam$ is larger than the 68\% credible interval. 
The figure also shows the forecast for these redefined observables expected for the full HSC data.   

In conclusion, we do not find a strong advantage for the alternative observables of $\Upsilon$ or $\Sigma$ over the fiducial analysis of $\{\dSigma,\wproj\}$ as long as the appropriate scale cuts, e.g. $12$ and $8~\hiMpc$, are employed. In addition, an estimation of the alternative observables seem to involve some uncertainty due to the numerical derivative and integration, and we found that the parameter estimation varies with how to implement the numerical calculations for the discrete data vector. 

\section{Sampler dependence}	
\label{apdx:sampler-check-emcee}

\begin{figure*}
	\centering
	\includegraphics[clip,width=1.0\textwidth]{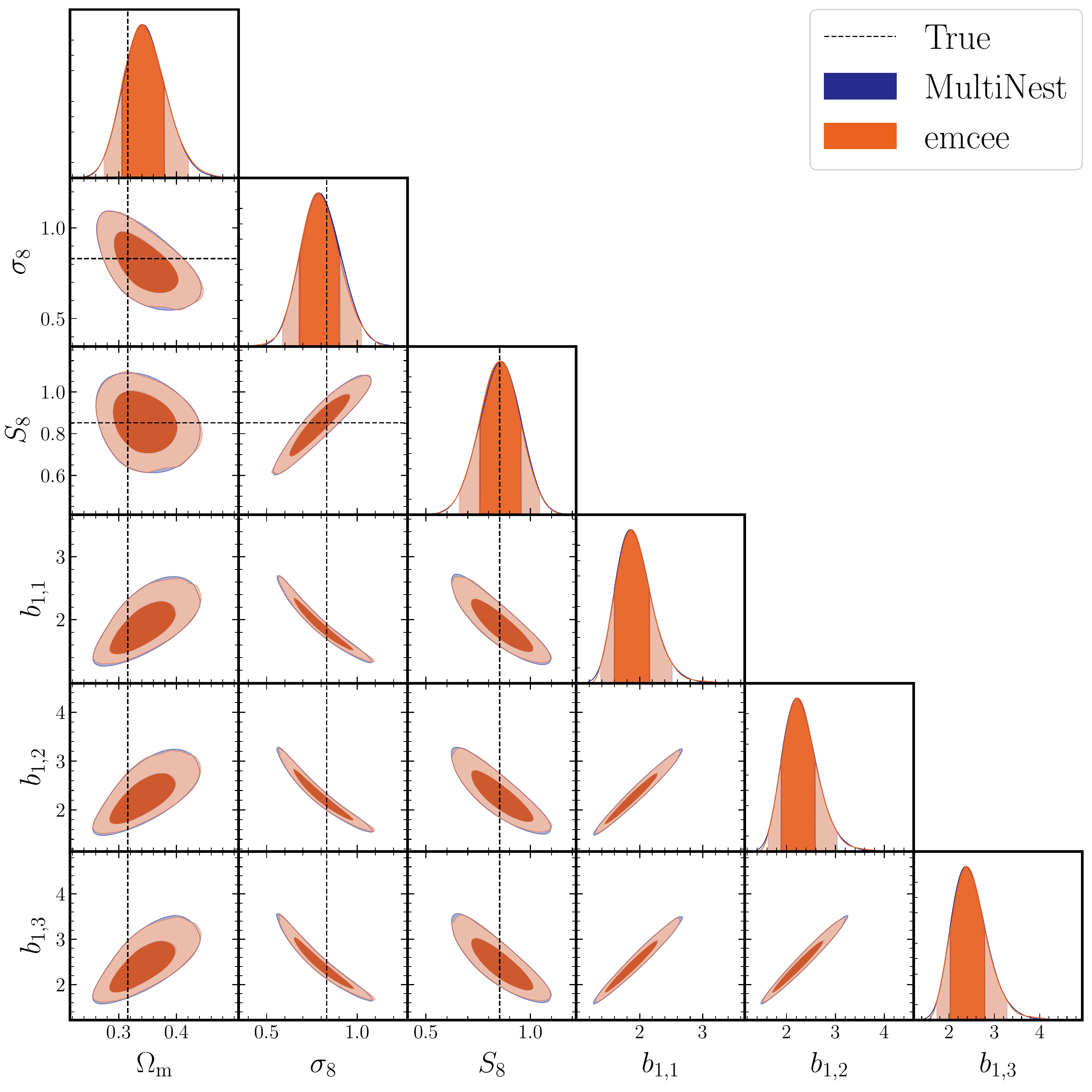}
	\caption{
		Comparison of the parameter inference results from the different MCMC samplers, \code{MultiNest}\cite{ferozMultiNestEfficientRobust2009} (our default sampler) and \code{emcee}\cite{foreman-mackeyEmceeMCMCHammer2013}. Here, we apply the analysis with baseline setup, \setup{$(\Delta\!\Sigma_{12},w_\mathrm{p,8})$}. (\tableref{table:fiducial-mock-setup}). Each panel shows the 2D or 1D posterior distribution of two parameter subspace or a given parameter, respectively. 
		}
	\label{fig:sampler-check-emcee}
\end{figure*}
Throughout this paper we use \code{MultiNest} for the parameter estimation. Here we study robustness of our results against the MCMC sampler. 
We compare the results obtained by using \code{MultiNest} and \code{emcee} in Fig.~\ref{fig:sampler-check-emcee}. Here we apply the baseline analysis to the fiducial mock as given in Tables~\ref{table:fiducial-mock-setup}. 
The figure shows that the results well agree with each other, and the difference looks negligible.


\bibliography{ssbib_old,lssref}

\end{document}